\documentclass[pdflatex, sn-apa]{sn-jnl} % APA Reference Style
\usepackage{amsmath}
\usepackage{xcolor}
\usepackage[caption=false]{subfig}
\jyear{2021}%

\theoremstyle{thmstyleone}%
\theoremstyle{thmstyletwo}%

\theoremstyle{thmstylethree}%

\begin{document}

\title[Injection-induced aseismic slip in tight fractured rocks]{Injection-induced aseismic slip in tight fractured rocks}

\author*[1]{\fnm{Federico} \sur{Ciardo}}\email{federico.ciardo@sed.ethz.ch}

\author[2]{\fnm{Brice} \sur{Lecampion}}\email{brice.lecampion@epfl.ch}

\affil[1]{\orgdiv{Swiss Seismological Service (SED)}, \orgname{ETH Zürich}, \orgaddress{\street{Sonneggstrasse 5}, \city{Zürich}, \postcode{CH-8092}, \state{Switzerland}}}

\affil[2]{\orgdiv{Geo-energy Laboratory - Gaznat Chair on Geo-energy}, \orgname{EPFL}, \orgaddress{\street{Rte Cantonale}, \city{Lausanne}, \postcode{CH-1015}, \state{Switzerland}}}

%%==================================%%
%% ABSTRACT %%
%%==================================%%
%Please provide an abstract of 150 to 250 words. 
%The abstract should not contain any undefined abbreviations or unspecified references.

\abstract{
We investigate the problem of fluid injection at constant pressure in a 2D Discrete Fracture Network (DFN) with randomly oriented and uniformly distributed frictionally-stable fractures. We show that this problem shares similarities with the simpler scenario of injection in a single planar shear fracture, investigated by \citet{BaVi19,Vi21} and whose results are here extended to include closed form solutions for aseismic moment as function of injected volume $V_{inj}$. Notably, we demonstrate that the hydro-mechanical response of the fractured rock mass is at first order governed by a single dimensionless parameter $\mathcal{T}$ associated with favourably oriented fractures: low values of $\mathcal{T}$ (critically stressed conditions) lead to fast migration of aseismic slip from injection point due to elastic stress transfer on critically stressed fractures. In this case, therefore, there is no effect of the DFN percolation number on the spatio-temporal evolution of aseismic slip. On the other hand, in marginally pressurized conditions ($\mathcal{T} \gtrsim 1$), the slipping patch lags behind the pressurized region and hence the percolation number affects to a first order the response of the medium. Furthermore, we show that the aseismic moment scales $\propto V_{inj}^2$ in both limiting conditions, similarly to the case of a single planar fracture subjected to the same injection condition. The factor of proportionality, however, depends on the DFN characteristics in marginally pressurized conditions, while it appears to be only mildly dependent on the DFN properties in critically stressed conditions.

}

%%==================================%%
%% KEYWORDS %%
%%==================================%%
% Please provide 4 to 6 keywords which can be used for indexing purposes.

\maketitle

%%==================================%%
%% INTRODUCTION %%
%%==================================%%

% I have added additional references in additionalBIB.bib
% FC: I have moved the new references in the Bibliography.bib file

\section{Introduction}
\label{sec1}
Field observations reveal that human activities associated with injection of fluids in the sub-surface can re-activate pre-existing fractures/faults and trigger micro-seismicity \citep{HeRu68, HaMe71, ZoHa97, El13, KeWe18}. Although with different purposes and different techniques, current industry practices, such as the ones operating in the field of deep geothermal energy or hydrocarbon extraction, utilize micro-seismic events to determine the extent of the stimulated rock volume and often correlate its propagation with a pore fluid diffusion process. As demonstrated by many observations, indeed, the migration of these ``wet" events is bounded by a power-law which grows proportional to the square-root of pressurization time \citep{ShHue97, ShRo02, PaRo03, HaFi12}. Although a possible explanation could be attributed to the presence of highly hydraulically conductive fractures at large scale, an increasing number of evidences show that seismic slip is not the only possible result of fluid injection (e.g. see \cite{Corn16} and references therein), and ``dry" micro-seismicity may be triggered and driven by elastic stress transfers to unstable patches caused by aseismic slip migrating away from injection point \citep{EyEa19}.

As demonstrated by in-situ large scale injection campaigns \citep{ScCo94, CornHe97, BoBe07} and recent field and laboratory experiments \citep{GuCa15,ScCo16,CaScu19}, aseismic slip is primarily induced by fluid injection and micro-earthquakes are only indirect effects associated with a stable aseismic rupture propagation.  Actually, the deformation that led to ground ruptures in the Baldwin Hills associated with injection waterflooding was shown to have a large aseismic component as early as 1971 \citep{HaMe71}. 
The  interest in  the spatio-temporal evolution of fluid-driven slip on predominantly frictionally-stable pre-existing discontinuities has prompted recent theoretical investigations. 
\cite{BaVi19} showed that fluid-induced aseismic slip on a critically stressed two-dimensional fault can outpace pore fluid diffusion if the ratio between the initial distance to failure of the fault over the injection strength is small. Similar results for a planar three-dimensional fault have been recently obtained by \cite{SaLe22}. 
In that ``critically stressed" limit, the fast migration of aseismic slip compared to the pore-pressure diffusion front may be the primary cause of observed (micro-)seismicity during in-situ injection experiments (e.g. \citep{DuDe17}).
In these studies, micro-seismicity is conceptually understood as the result of instabilities triggered by quasi-static stress perturbations on unstable patches around or on the main slipping fault plane of different sizes. Although this can clearly occur if a single main fault plane accommodate most of the slip, an injection in a deep fracture/fault zone most likely re-activates several pre-existing structural discontinuities present at various length scales, from meters to kilometres \citep{Wibberley20085, FauJa10}. Such heterogeneous discontinuities would plausibly serve as main fluid conduits for pore-pressure diffusion \citep{WhGa77} and may host aseismic slip that would cause elastic stress re-distribution in the fractured rock mass, leading  potentially to a more complex spatio-temporal evolution of micro-seismicity. 
Additionally, at a decameter scale, how does a highly fractured rock mass respond to fluid injection at pressures lower than the minimum stress (thus not promoting hydraulic fracturing) but sufficient to induce shear ruptures is an important topic in rock mechanics at large - from grouting to geothermal energy \citep{Corn21}.  

In this context, we study  how fluid-driven stable frictional ruptures propagate when a fluid is injected at constant pressure in a fractured rock mass. Notably, we model the fractured rock mass discretely accounting for fractures at multiple scales using a Discrete Fracture Network approach.  We focus on the case of randomly oriented and uniformly distributed pre-existing fractures. The rock matrix is assumed to be impermeable at the time scale of injection and hence fluid can flow only along the hydraulically connected pre-existing fractures.  

This paper is organized as follow. In Section \ref{sec2}, we present the mathematical formulation of the problem as well as a quick description of the methods used for its numerical resolution. As a limiting scenario of fluid injection into a fracture/fault that is not hydraulically connected to others, we revisit in Section \ref{sec3} the problem of fluid injection at constant pressure and aseismic slip propagation on a 2-dimensional planar shear fracture. The numerical results are thoroughly benchmarked against the analytical asymptotic solutions of \citet{BaVi19} and \cite{Vi21}. In Section \ref{sec4}, we extend the results of Section \ref{sec3} and investigate the problem of injection at constant pressure in a 2D Discrete Fracture Network, with randomly oriented and uniformly distributed fractures. Scaling considerations allow to identify the important governing parameters and drive the numerical experiments. 
Finally, in Section \ref{sec5} we discuss  the implications of our results to injection-induced micro-seismicity, as well as the limits of our modelling assumptions. 

\section{Problem formulation and methods}
\label{sec2}

We consider a homogeneous, isotropic and linear elastic infinite medium that hosts a set of inter-connected planar fractures, forming a so called Discrete Fracture Network (DFN). Under the assumption of plane strain conditions, these two-dimensional discontinuities reduce to one-dimensional entities in the $\mathbb{R}^2$ Euclidean space with a canonical global basis $\mathbf{e_x}=(1,0)$,  $\mathbf{e_y}=(0,1)$. The fractured medium is subjected to a uniform far-field stress state\footnote{with principal stress components denoted by $\sigma_{xx,o}$ and $\sigma_{yy,o}$, aligned along the principal $x-$ and $y-$axis, respectively. The subscript $_o$ refers to initial conditions} which, resolved on each fracture $^k$, leads to a uniform shear stress $t_{s,o}^k$ and effective normal stress $t_{n,o}^{\prime, k} = t_n^{k} - p_o$, with $p_o$ the in-situ uniform pore pressure field. We assume that frictional resistance of each pre-existing fracture is governed by Coulomb's friction with a constant friction coefficient $f$.
Initially, we assume that the frictional resistance of all fractures is sufficient to prevent any slip. Such an equilibrium  is perturbed by internal pore-fluid pressurization due to fluid injection into one fracture. Since we assume that the host medium is impermeable at the time scale of injection (i.e. tight rock such as granite), the induced pressure gradient generates fluid flow that propagates inside the hydraulically connected permeable fractures, all characterized by a constant and uniform hydraulic aperture $w_h \left[ L\right]$ and longitudinal permeability $k_f \left[ L^{2}\right]$. 
%Fluid flow propagation, therefore, is strongly affected by inter-connectivity degree of the fracture network.\\ 
The local reduction of effective normal stresses due to internal pore fluid pressurization may cause the shear resistance to drop below the applied stresses (in absolute terms). This activates slip, manifested with a local shear stress drop, that forces the applied stress to be equal (or lower) to the available frictional resistance, here assumed to obey the Mohr-Coulomb yielding criterion without cohesion. In other words, for all fractures, at any time
\begin{equation}
\lvert t_s^{k}\rvert \le f \left( t_{n,o}^{\prime, k} - \bar{p}(\mathbf{x}, t) \right),
\label{eq:interfacial_law}
\end{equation}
where $\bar{p}(\mathbf{x}, t) = p(\mathbf{x}, t) - p_o$ is the internal pore fluid over-pressure.
Upon frictional yielding (when $\lvert t_s^{k}\rvert = f \left( t_{n,o}^{\prime, k} - \bar{p}(\mathbf{x}, t) \right)$), the fracture(s) can slip following a non-associated plastic flow rule \citep{CiLe20}. We assume here that slip occurs at critical state (without any volumetric change), such that
the slip component $\delta$ of the displacement discontinuity is the only unknown, while its opening part $w$ always remain zero.

The conservation of momentum under quasi-static approximation leads to a linear relationship between tractions and slip on the pre-existing fracture network. This relation is represented mathematically by the following boundary integral equations \citep{Mogi2014}
\begin{equation}
t_i^k \left(\mathbf{x},t \right) = t^k_{i,o}\left(\mathbf{x}\right) + \int_{\Gamma} K_{is} \left( \mathbf{x} - \mathbf{\xi}, E^\prime, \nu \right) \delta \left(\mathbf{\xi}, t \right) \, \text{d}\xi, \quad \mathbf{x} \in \Gamma, \quad i,j = s,n
\label{eq:elasticity_equations}
\end{equation}
where $\Gamma$ denotes the boundary region in the medium traced by the DFN and locus of slip $\delta$, $t_i^k$ is the current total traction vector on the generic fracture $^k$ and $t^k_{i,o}$ its initial value under the in-situ stress state. %characterized by a second order tensor $\Sigma$ with components $\sigma_{ij}$.
Finally, $K_{is}$ is the singular elastostatic traction kernel function of both plane-strain elastic modulus $E^\prime$ and Poisson's ratio $\nu$. Its mathematical expression is known in closed form for an infinite 2-dimensional medium (see e.g. \citep{Ge1891, HiKe96}).
The assumption of zero dilatancy as the frictional slip develops entails 
that on all fracture the opening displacement discontinuity $w$ remains zero.
The non-local elastic kernel $K_{is}$, therefore, maps the effects of slip distribution on the normal and shear component of the traction vector along $\Gamma$. \\
%The conservation of momentum under quasi-static approximation leads to a linear relationship between tractions and displacement discontinuities (slip $d_s$ and opening $d_n$) on the pre-existing fracture network. This relation is represented mathematically by the following boundary integral equations \cite{Mogi2014} 
%\begin{equation}
%t_i^k \left(\mathbf{x},t \right) = t^k_{i,o}\left(\mathbf{x}\right) + \int_{\Gamma} K_{ij}\left( \mathbf{x} - \mathbf{\xi}, E^\prime, \nu \right) d_j, \text{d}\xi, \quad \mathbf{x} \in \Gamma, \quad i,j = s,n
%\label{eq:elasticity_equations}
%\end{equation}
%where $\Gamma$ denotes the boundary region in the medium traced by the DFN  which corresponding to the locus of displacement discontinuities $d_j$, $t_i^k$ is the current total traction vector on the generic fracture $^k$ and $t^k_{i,o}$ its initial value under the in-situ stress state. %characterized by a second order tensor $\Sigma$ with components $\sigma_{ij}$.
%Finally, $K_{ij}$ is the singular elastostatic traction kernel function of both plane-strain elastic modulus $E^\prime$ and Poisson's ratio $\nu$. Its mathematical expression is known in closed form for an infinite 2-dimensional medium (see e.g. \citep{Ge1891, HiKe96}).
%The assumption of zero dilatancy as the frictional slip develops entails 
%that on all fracture the opening displacement discontinuity $d_n$ remains zero.
%As a result, the non-local elastic kernel $K_{is}$  maps the effects of slip distribution on  normal and shear tractions along $\Gamma$. \\
Isothermal single-phase fluid flow inside the permeable fracture network is governed by the width-averaged fluid mass conservation equation over the fractures hydraulic thickness, whose expression in terms of the pore fluid over-pressure reads
\begin{equation}
w_h \beta \frac{\partial \bar{p}}{\partial t} + \nabla_{\lvert \rvert} q_{\lvert \rvert} = 0, \label{eq:diffusion_equation}
\end{equation}
where $\beta \left[ M^{-1} T^2 L\right]$ is a coefficient sums of the fluid and pore compressibility, and $q_{\lvert \rvert}$ is the longitudinal fluid flux inside the fracture(s) $\Gamma$ given by Darcy's law
\begin{equation}
q_{\lvert \rvert} = - \frac{w_k k_f}{\mu} \nabla_{\lvert \rvert}\bar{p}, \label{eq:Darcy_flux}
\end{equation}
with $\mu$ $\left[ M L^{-1} T^{-1}\right]$ is 
 the fluid dynamic viscosity. In this contribution, we investigate sustained fluid injection into a single fracture at the center of the DFN and we denote this fracture as $^{k_*}$. Notably, we assume that the injection is performed at constant over-pressure:
\begin{equation}
\bar{p}\left( x^{k_*}_{inj}, y^{k_*}_{inj}, t\right) = \Delta P, \quad \forall t
\label{eq:injection_condition}
\end{equation}
where the injection over-pressure $\Delta P$ remains always below the minimum principal effective stress in order to avoid the creation of an hydraulic fracture.\\ 
As previously stated, shear-induced dilatancy is neglected and all the fractures have a uniform and constant hydraulic transmissivity $k_f w_h$. The hydro-mechanical problem therefore uncouples such that the flow 
Equations (\ref{eq:diffusion_equation}) and (\ref{eq:Darcy_flux}) reduce to a simple diffusion equation
\begin{equation}
\frac{\partial \bar{p}}{\partial t} - \alpha \nabla_{\lvert \rvert}^2 \bar{p} = 0
\label{eq:diffusion_equation2}
\end{equation}
where $\alpha = \dfrac{k_f}{\mu \cdot \beta}$ is the constant hydraulic diffusivity $\left[L^2T^{-1}\right]$.\\ 
In the following sub-section, we discuss the numerical methods used to solve numerically the system of equations (\ref{eq:interfacial_law}-\ref{eq:injection_condition}) in terms of spatial-temporal distribution of shear displacement discontinuities $\delta$ and pore fluid over-pressure $\bar{p}$.

\subsection{Numerical methods}
\label{sec2.1}

\begin{figure}[t]
\centering
\includegraphics[width=0.55\textwidth]{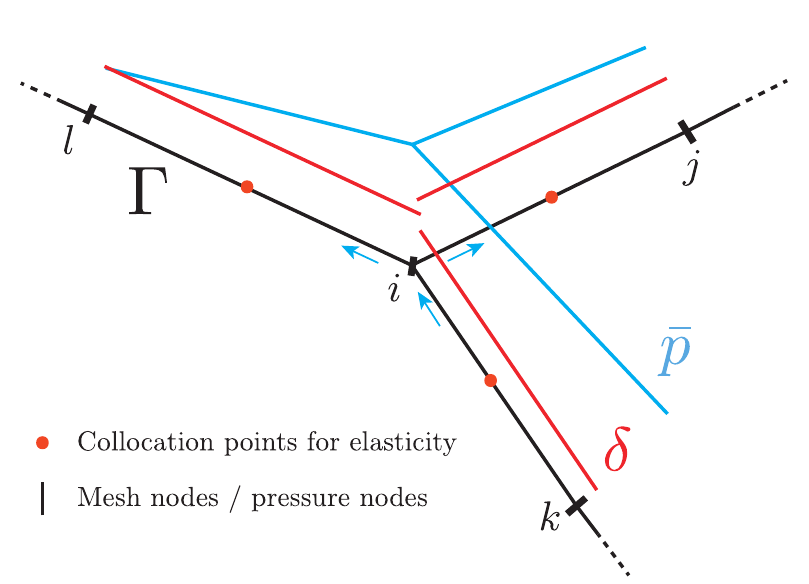}
\caption{Sketch of spatial discretization of displacement discontinuities (slip - red) and pore fluid over-pressure (blue) field along three mesh elements converging in one generic node $i$. The former is based on a piece-wise constant variation between different segments, with one collocation point per element for the evaluation of elasticity equations (\ref{eq:elasticity_equations}) (orange dots). The latter, instead, varies linearly and continuously between adjacent elements.}
\label{fig:sketch_numerics}
\end{figure}

The hydro-mechanical problem described in the previous section uncouples thanks to the assumption of constant fracture(s) permeability and negligible dilatancy. The fluid flow problem can thus be solved separately first, and the resulting pressure input into the  mechanical problem (i.e. quasi-static elastic equilibrium and Mohr's Coulomb criterion).\\
We use a continuous Galerkin Finite Element method for the spatial discretization of the fluid flow problem and solve for the unknown pore fluid over-pressure field $\bar{p}(\bf{x},t)$ at nodes. The set of fractures $\Gamma$ is discretized into a number of finite 1D linear iso-parametric elements. The nodal (pressure) values coincide with mesh nodes as depicted in Figure \ref{fig:sketch_numerics}, and thus continuity between elements and fractures intersection are automatically handled. 
Using a backward Euler time integration scheme, Equations (\ref{eq:diffusion_equation}-\ref{eq:injection_condition}) upon discretization reduces to a linear system of equations
\begin{equation}
    \bf{A} \bf{\bar{p}^{n+1}} = \bf{b},
    \label{eq:system_pressure}
\end{equation}
where $\bf{A}$ is a sparse matrix (combining the storativity and conductivity matrices) and $\bf{b}$ is the right hand side that contains the injection condition (\ref{eq:injection_condition}). 
The sparse system of equations (\ref{eq:system_pressure}) is solved numerically using a Lower-Upper decomposition method \citep{QuaSa2000} for the over-pressure distribution at  time $t^{n+1}$. The latter is then entered as change of effective normal tractions to a boundary-element-based solver with an elasto-plastic like constitutive interfacial law for the simultaneous resolution of the quasi-static elastic equilibrium and elasto-plastic non-associated Mohr-Coulomb interface law. 
Such a solver is similar to the one reported in \citet{SaLe22} (albeit its usage was only showed for 3D problems) where additional details can be found. The only difference compared to that contribution is the problem dimension, the type of boundary elements adopted and their interpolation order for displacement discontinuity field. 
Notably, we employ piece-wise constant boundary elements (straight segments) with only one collocation point located at their middle point (see Figure \ref{fig:sketch_numerics}). This results in having one degree of freedom (slip) for each boundary element in the computational mesh.

Note also that we adopt an automatic adaptive time-stepping scheme  \citep{SoWa03} based on an estimate of the local truncation error obtained via an explicit predictor implicit corrector. 
 
\section{Injection in a planar shear fracture}
\label{sec3}
Before investigating the problem of injection into a discrete fracture network, we present the case of an injection in an isolated planar fracture. This represents the limiting case where a pre-existing fracture network is loose, poorly connected and fluid is injected into a fracture/fault which is not hydraulically connected to others. The large inter-distances between pre-existing fractures further avoid elastic stress interactions, owing to the spatial decay of the elastic kernel $K_{is}$, which in 2-dimension scales as $\sim 1/r^2$ (with $r$ being the Euclidean distance).
Since we assume the fracture plane to be planar, the elastic equation uncouple ($K_{ns} = 0$ in Equation (\ref{eq:elasticity_equations})) such that slip does not induce any variation of normal traction along the planar fracture.
Fluid flow is confined inside the conductive fracture characterised by constant and uniform hydraulic transmissivity $k_f w_h$. Equation (\ref{eq:diffusion_equation}), or equivalently Equation (\ref{eq:diffusion_equation2}), together with injection condition ($\ref{eq:injection_condition}$) and the following boundary and initial conditions
\begin{equation}
\bar{p}(x, 0) = 0, \quad \bar{p}(\pm \infty, t) = 0,    
\end{equation}
can be solved analytically for the spatial and temporal evolution of pore-fluid over-pressure \citep{CaJa59}
\begin{equation}
  \bar{p}(x, t) = \Delta P \cdot \text{Erfc} \left( \frac{\lvert x \rvert}{\ell_d(t)}\right),
  \label{eq:analytical_flow}
\end{equation}
where $x$ is the fracture longitudinal coordinate, %(with origin located at mid-point),
Erfc is the complementary error function and $\ell_d(t) = \sqrt{4 \alpha t}$ is the fluid diffusion length-scale (function of pressurization time $t$). Plastic flow, manifested as a symmetric shear crack propagating away from injection point, occurs when 
\begin{equation}
    \Delta P > t^{\prime}_{n,o} - t_{s,o}/f
\end{equation}
Note that the superscript $^k$ has been dropped for clarity for this single fracture case.

This single planar fracture problem has been solved by \citet{BaVi19} and \citet{Vi21} in details. In the next sub-sections, we recall their main findings that - on one hand - will provide benchmark solutions for our numerical solver, and - on the other hand - will pave the way for later investigations. 

\subsection{Self-similar shear crack propagation}
\label{subsec1}

Under the specific conditions previously defined, fluid injection activates a shear crack that propagates symmetrically from injection point paced by pore fluid diffusion (and we denote its half-length as $a(t)$). Its propagation, however, is always stable (aseismic) due to the assumption of a constant frictional coefficient. No dynamic instabilities can occur. 
\cite{BaVi19} and \cite{Vi21} showed that, in this particular case, the one-way coupled hydro-mechanical problem is self-similar:  the shear crack front propagates strictly proportional to the pore-pressure diffusion front
\begin{equation}
\frac{a(t)}{\ell_d(t)} = \lambda,
\label{eq:self_sim}
\end{equation}
where $\lambda$ is simply the ratio between the shear crack length and the pore-pressure diffusion front. Values of $\lambda$ lower or greater than 1 means that the crack lags or outpaces the diffusion of pore-pressure, respectively.
Using scaling analysis, they also showed that the solution of the problem is governed by a single dimensionless parameter
\begin{equation}
%T = \left(1 - \frac{t_{s,o}}{f t_{n,o}^{\prime}} \right) \frac{t_{n,o}^{\prime}}{\Delta P},
\mathcal{T} = \left(\frac{ f t_{n,o}^{\prime} - t_{s,o}}{f \Delta P} \right),
\label{eq:T_parameter}
\end{equation}
which can be obtained by scaling shear tractions, spatial variables and slip in the shear elasticity equation (\ref{eq:elasticity_equations}) respectively with ambient fracture shear strength $f t_{n,o}^{\prime}$, fracture half-length $L$ and $l_*=\dfrac{f \Delta P}{E^\prime} L$ (the latter being a length-scale that results from the normalization of elasticity equation and yielding criterion (\ref{eq:interfacial_law})).\\
It is interesting to note that the $\mathcal{T}$ parameter includes i) the distance to failure $f t_{n,o}^{\prime} - t_{s,o}$
that quantifies how far the fracture/fault stress state is from failure prior fluid injection and ii) the destabilizing effect due to pressurization $f \Delta P$. For this reason we call this parameter as \textit{fracture-stress-injection} parameter.

Following the nomenclature of \cite{GaGe12}, we define the condition when the crack lags well behind the fluid front ($\lambda \ll 1$) as \textit{marginally pressurized} condition. Indeed, this occurs when injection over-pressure is just enough to activate slip, i.e. when $f \Delta P \simeq f t_{n,o}^{\prime} - t_{s,o}$ and hence when $\mathcal{T} \to 1$. On the other hand, in the limit when $\mathcal{T} \to 0$, the crack rapidly outpaces the fluid front upon fluid injection ($\lambda \gg 1$). 
Whilst such a condition can be attained for increasing values of normalized injection over-pressure, the ratio $\Delta P/t_{n,o}^\prime$ must be below 1 in order to avoid tensile opening. This implies that the limit $\mathcal{T}\to0$ can be achieved only when the fault is critically stressed prior fluid injection, i.e. when $t_{s,o}\simeq f t_{n,o}^\prime$. For this reason, we define this limit as \textit{critically stressed} limit.\\
By considering these two limiting behaviors, \cite{BaVi19} derived two corresponding asymptotic solutions that link the $\mathcal{T}$ parameter with the self-similarity factor $\lambda$:
\begin{equation}
\begin{split}
&\mathcal{T} = 1-\lambda \frac{4}{\pi^{3/2}} \qquad \text{(Marginally Pressurized)} \\
&\mathcal{T} = \frac{2}{\pi^{3/2} \lambda} \qquad \text{(Critically Stressed)}
\end{split}
\label{eq:viesca_solution}
\end{equation}
For a given value of  $\mathcal{T}$, Equation (\ref{eq:viesca_solution}) can be easily inverted to estimate the $\lambda$ parameter, which can then be used to evaluate the corresponding slip distribution obtained in \citep{Vi21}.
Notably, the analytical solutions of the slip distributions follow 
\begin{equation}
\delta(\bar{x}) \simeq \frac{\lambda^2 \ell_d(t) f \Delta P}{E^\prime /4} \frac{2}{\pi^{3/2}} \left( \sqrt{1-\bar{x}^2} - \bar{x}^2\text{Atanh}\sqrt{1-\bar{x}^2}\right)
\label{eq:viesca_solution_slip}
\end{equation}
in the marginally pressurized limit, and
\begin{equation}
\begin{split}
&\delta(\bar{x}) \simeq \frac{\ell_d(t) f \Delta P}{E^\prime /4} \frac{2}{\pi^{3/2}} \left( \text{Atanh}\sqrt{1 - \bar{x}^2} - \sqrt{1-\bar{x}^2}\right) \quad \text{(Outer solution)}\\
&\delta(x/\ell_d(t)) \simeq \delta(0) - \frac{\ell_d(t) f \Delta P}{\mu} \int_0^{x/\ell_d(t)} \left[ \frac{1}{\pi} \int_{-\infty}^{\infty} \frac{\textrm{erfc}(\hat{s})}{\hat{x}- \hat{s}} \textrm{d} \hat{s}\right] \textrm{d}\hat{x} \quad \text{(Inner solution)}
\end{split}
\label{eq:viesca_solution_slip2}
\end{equation}
in the critically stressed limit\footnote{The slip at the center is obtained by matching the inner with the outer solution at intermediate distances and is given by $\delta(0) \simeq \dfrac{\ell_d(t) f \Delta P}{\mu} \frac{2}{\pi^{3/2}} \left( \textrm{ln}(2 \lambda) + \gamma/2\right)$, where $\gamma$ is the Euler-Maraschoni constant \citep{Vi21}.}. Note that the slip distribution in the critically stressed limit is decomposed into an outer asymptotic solution that is valid on distances comparable to the rupture distance $a(t)$ (i.e. when $\lvert \bar{x} \rvert = \lvert x/a(t)\rvert \sim 1$), and an inner asymptotic solution valid on distances comparable to the diffusion length scale $\ell_d (t)$ (see \citep{Vi21} for the complete details of the solution). 

From Equations (\ref{eq:self_sim}) and  (\ref{eq:viesca_solution_slip2}), we observe that rupture velocity is $\lambda \sqrt{\alpha}/\sqrt{t}$ and the slip rate scales as $\sqrt\alpha f\Delta P/\left(E^\prime \sqrt t\right)$. Taking realistic values $f=0.6$, $E^\prime$ = 50 GPa, $\Delta P$ =10 MPa and $\alpha={10}^{-3} \text{m}^2/s$, after few seconds, we obtain micro meters per second for the slip rate and, assuming $\lambda \sim 100$ (critically stressed condition), a rupture speed of 1 m/s (clearly below the Rayleigh wave speed). This kind of rupture is therefore clearly aseismic.

\subsection{Aseismic moment}
\label{subsec2}

Using the asymptotic slip solutions previously reported, we can take a further step and derive closed-form expressions of the scalar aseismic moment $M_o(t) = E/\left( 2 ( 1+\nu)\right) \int_{-a(t)}^{a(t)} \delta\left( x\right) \text{d}x$ \citep{Aki66} for the two limiting regimes. In the marginally pressurized limit, we integrate the analytical solution (\ref{eq:viesca_solution_slip}) and obtain
\begin{equation}
M_o(t) \simeq \frac{4}{3 \sqrt{\pi}} \cdot \left( \frac{a\left(t\right)^3 f \Delta P \left(1 - \nu\right)}{\ell_d(t)}  \right)   
\label{eq:aseismic_moment_marginally_press}
\end{equation}
In the critically stressed limit, the pressurization region is always very confined near injection point ($\ell_d \gtrsim 0$), while the slipping patch continuously accelerates and outpaces pore-fluid front ($a(t) \gg \ell_d(t)$). The pressure distribution can be effectively replaced by an equivalent “point-force” distribution and the second term of the inner asymptotic solution (\ref{eq:viesca_solution_slip2})-b can be neglected. The main contribution to the spatial slip accumulation is thus given by the outer asymptotic solution and, for this reason, we use its analytical expression (\ref{eq:viesca_solution_slip2})-a to obtain the expression of the scalar aseismic moment in the critically stressed limit:
\begin{equation}
M_o(t) \simeq \frac{2}{\sqrt{\pi}}\cdot \left( a\left(t\right) f \Delta P \left(1 - \nu\right) \ell_d(t)  \right)
\label{eq:aseismic_moment_critically_stress}
\end{equation}
%\begin{figure}[t!]
 %   \centering
  %  \begin{tabular}{cc}
   % \begin{minipage}{0.5\textwidth} \includegraphics[width=0.95\textwidth]{Plot_HalfCrack_vs_Time_SingleFault(2).pdf} \\ \includegraphics[width=0.95\textwidth]{Plot_T_vs_lambda_SingleFault.pdf} \end{minipage}&
    %\begin{minipage}{0.5\textwidth} \includegraphics[width=0.95\textwidth]{RelErrPlot_MP_HalfCrack_vs_Time_SingleFrac.pdf} \\ \includegraphics[width=0.95\textwidth]{RelErrPlot_CS_HalfCrack_vs_Time_SingleFrac.pdf} \end{minipage} 
    %\end{tabular}
    %\caption{Top left: evolution of normalized crack half-length $a(t)/L$ (where $L$ is half-length of the fault) with normalized time $\sqrt{4 \alpha t}/L$, for different values of $\mathcal{T}$ parameter. 
    %Two of them are representative of critically stressed conditions (i.e. $\mathcal{T}=0.1, 0.15$), while the others two are representative of marginally pressurized conditions (i.e. $\mathcal{T}=0.8, 0.9$). 
    %The blue dashed line represents the scenario in which fluid front and shear crack tips coincide (i.e. $\lambda=1$). The obtained self-similarity parameter $\lambda$ for \textcolor{red}{two numerical simulations corresponding to marginally pressurized ($\mathcal{T}$=0.9, 0.8) and critically stressed ($\mathcal{T}$=0.1, 0.15)  limiting conditions} matches the one retrieved from the asymptotic analytical solutions (\ref{eq:viesca_solution}) (see bottom left panel). Right column: relative error in terms of time evolution of crack half-length for one marginally pressurized (top) and critically stressed (bottom) scenario.}
    %\label{fig:res_viesca1}
%\end{figure}
\begin{figure}[t!]
    \centering
       \subfloat[]{
      \includegraphics[width=.45\textwidth]{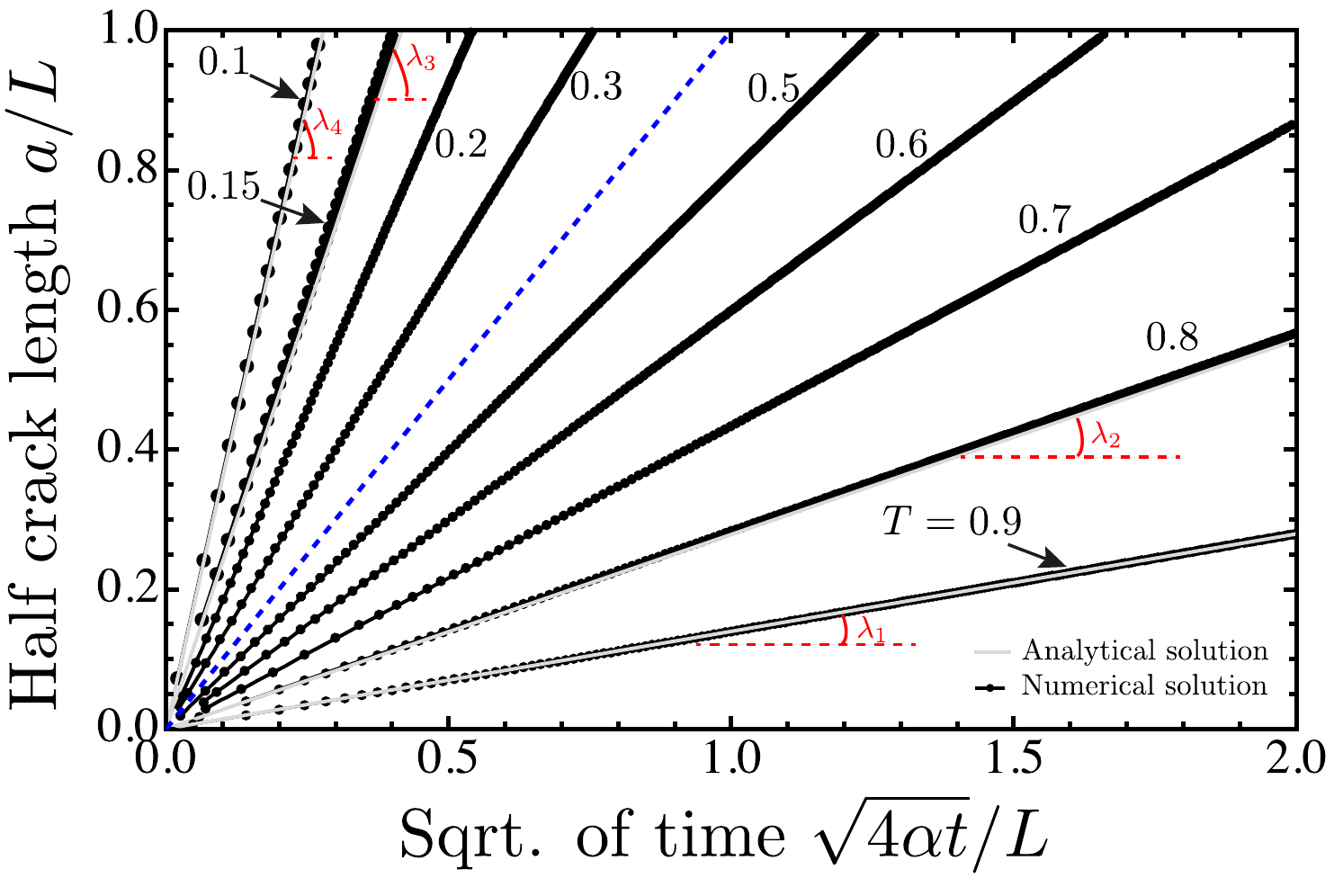}}
~
   \subfloat[]{
      \includegraphics[width=.48\textwidth]{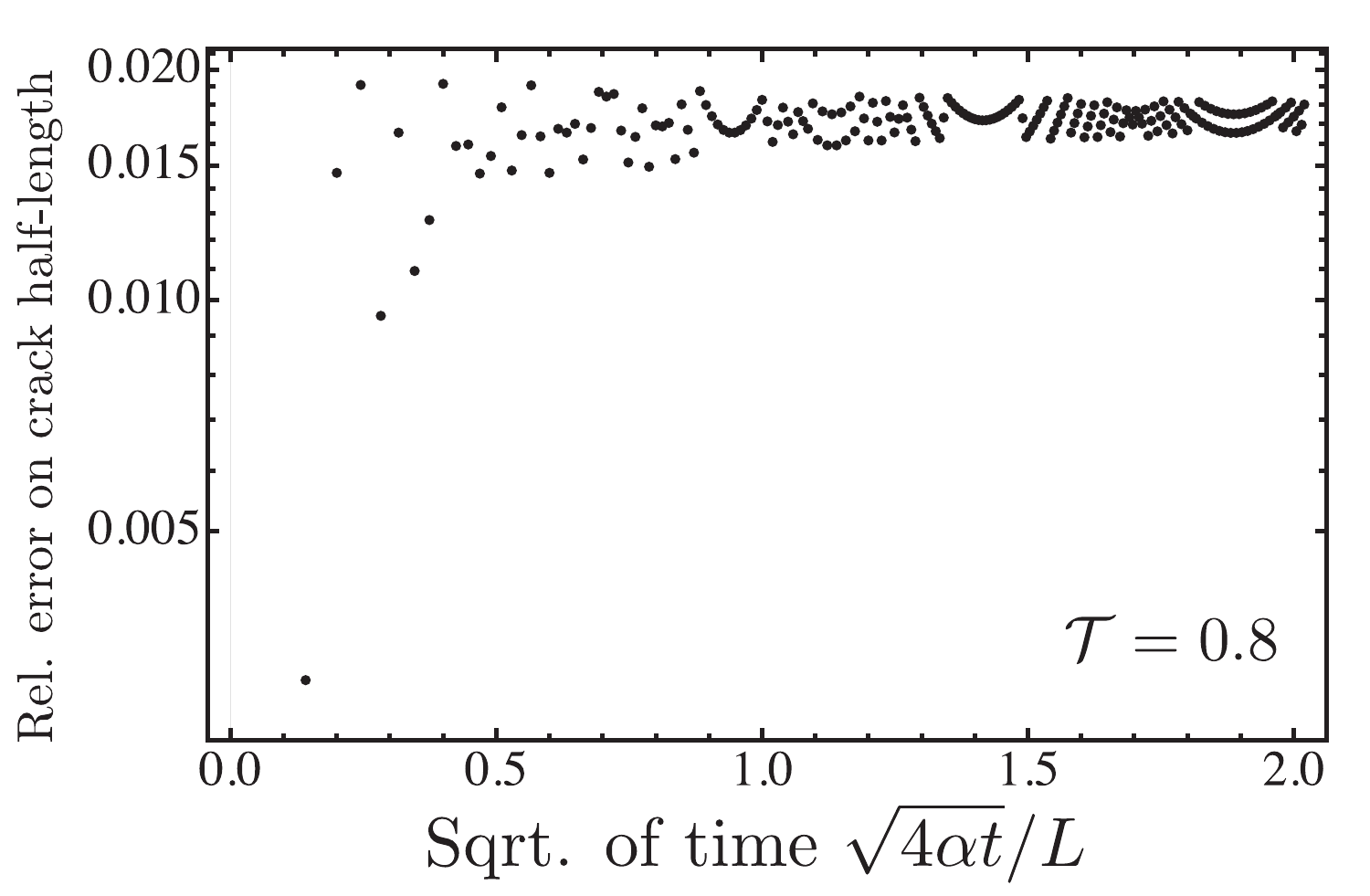}}
\\[-0.5cm]
   \subfloat[]{
      \includegraphics[width=.45\textwidth]{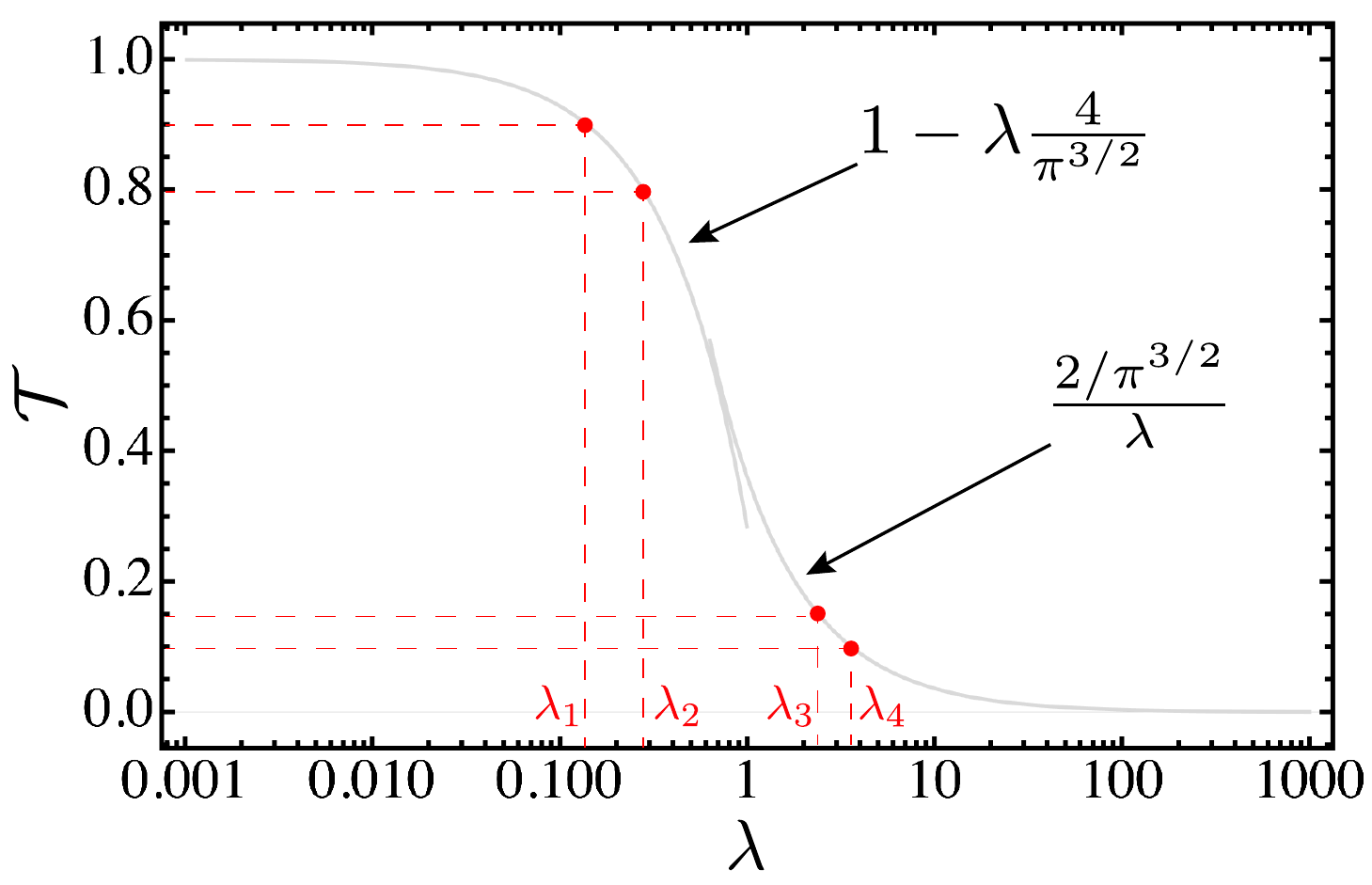}}
~
   \subfloat[]{
      \includegraphics[width=.48\textwidth]{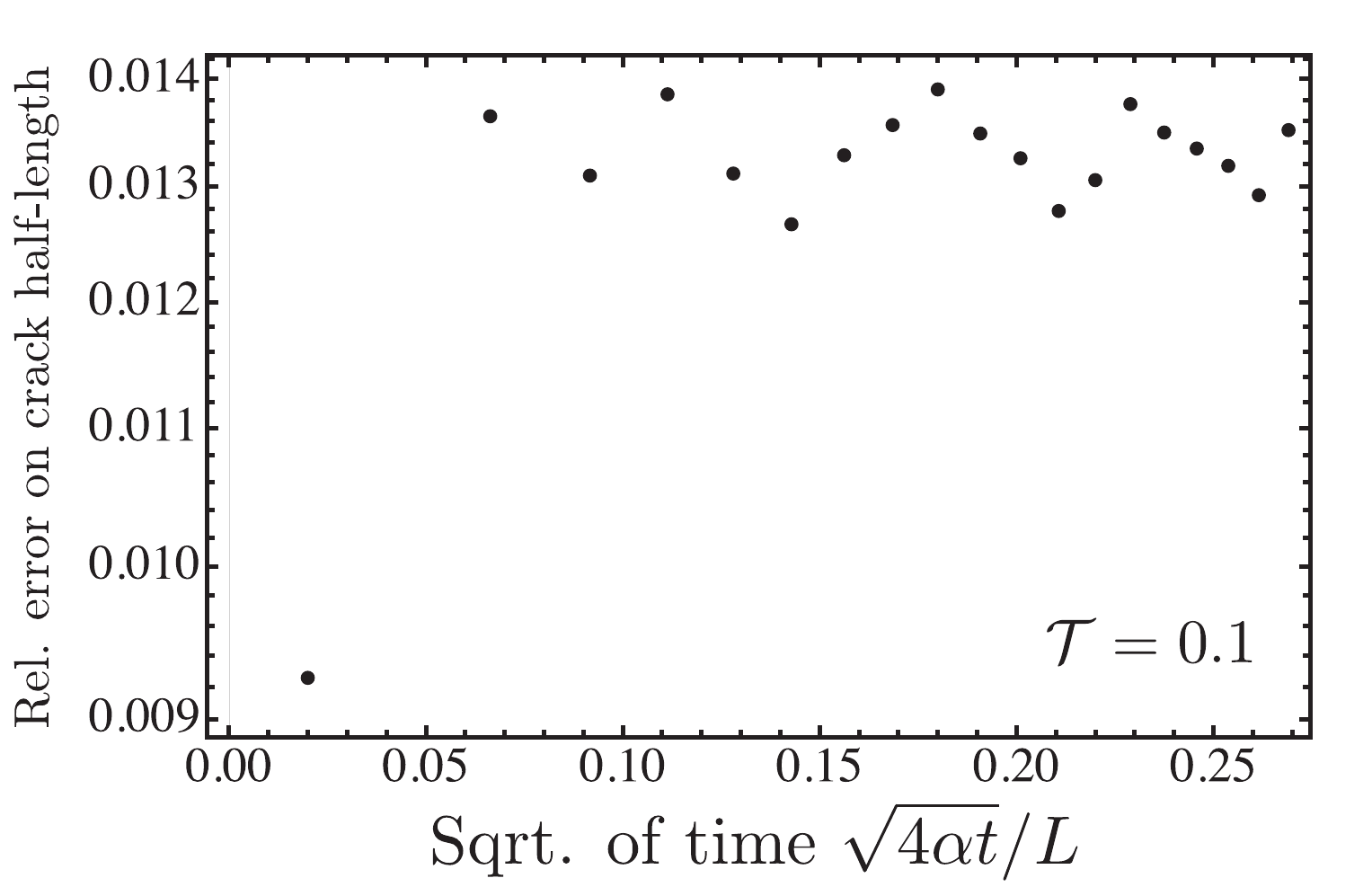}}
    \caption{Panel (a): evolution of normalized crack half-length $a(t)/L$ (where $L$ is half-length of the fault) with normalized time $\sqrt{4 \alpha t}/L$, for different values of $\mathcal{T}$ parameter. 
    %Two of them are representative of critically stressed conditions (i.e. $\mathcal{T}=0.1, 0.15$), while the others two are representative of marginally pressurized conditions (i.e. $\mathcal{T}=0.8, 0.9$). 
    The blue dashed line represents the scenario in which fluid front and shear crack tips coincide (i.e. $\lambda=1$). The obtained self-similarity parameter $\lambda$ for two numerical simulations corresponding to marginally pressurized ($\mathcal{T}$=0.9, 0.8) and critically stressed ($\mathcal{T}$=0.1, 0.15)  limiting conditions matches the one retrieved from the asymptotic analytical solutions (\ref{eq:viesca_solution}) (see panel (c)). Panels (b) and (d): relative error in terms of time evolution of crack half-length for one marginally pressurized (b) and critically stressed (d) scenario.}
    \label{fig:res_viesca1}
\end{figure}
Interestingly, we find that the aseismic moment scales linearly to pressurization time $t$ in both limits (since $a(t) = \lambda \ell_d(t)$ and $\ell_d(t) = \sqrt{4 \alpha t}$). If we define the injected volume as $V_{inj} = Q_{inj} \cdot t$, with $Q_{inj} =  \left. - \dfrac{w_h k_f}{\mu}\dfrac{\partial \bar{p}}{\partial x} \right \rvert_{x=0^{\pm}}$ being the fluid flux entering into the fault at injection point $x=0$, using Equation (\ref{eq:analytical_flow}) we can obtain its analytical expression 
\begin{equation}
V_{inj}\left( t\right) = \frac{2 k_f w_h \Delta P \sqrt{t}}{\sqrt{\pi} \mu \sqrt{\alpha}},
\label{eq:injected_volume}
\end{equation}
and thus come to the observation that the aseismic moment, in both marginally pressurized and critically stressed limit, scales \textit{quadratically} to the injected volume. 
%\begin{figure}[t!]
%    \centering
%    \begin{tabular}{cc}
%    \begin{minipage}{0.5\textwidth} \includegraphics[width=0.95\textwidth]{Slip_profiles_MP.pdf} \\ \includegraphics[width=0.95\textwidth]{PlotErrSlipMP.pdf} \end{minipage}&
%    \begin{minipage}{0.5\textwidth} \includegraphics[width=0.95\textwidth]{Slip_profile_CS.pdf} \\ \includegraphics[width=0.95\textwidth]{PlotErrSlipCS.pdf} \end{minipage} 
 %   \end{tabular}
 %   \caption{Left column: comparison between normalized slip profiles for marginally pressurized conditions (which collapse into one black curve due to scaling adopted) and asymptotic solution (\ref{eq:viesca_solution_slip}), together with the corresponding relative and absolute errors (bottom). Right column: comparison between normalized slip profile for $\mathcal{T}=0.1$ (critically stressed condition) and asymptotic inner and outer solutions (\ref{eq:viesca_solution_slip2}). The corresponding relative and absolute errors are displayed in the bottom, with the former calculated with respect to the inner solution (\ref{eq:viesca_solution_slip2}-b) and the latter with respect to the outer solution (\ref{eq:viesca_solution_slip2}-a).}
 %   \label{fig:res_viesca2}
%\end{figure}
\begin{figure}[t!]
    \centering
    \subfloat[]{
      \includegraphics[width=.45\textwidth]{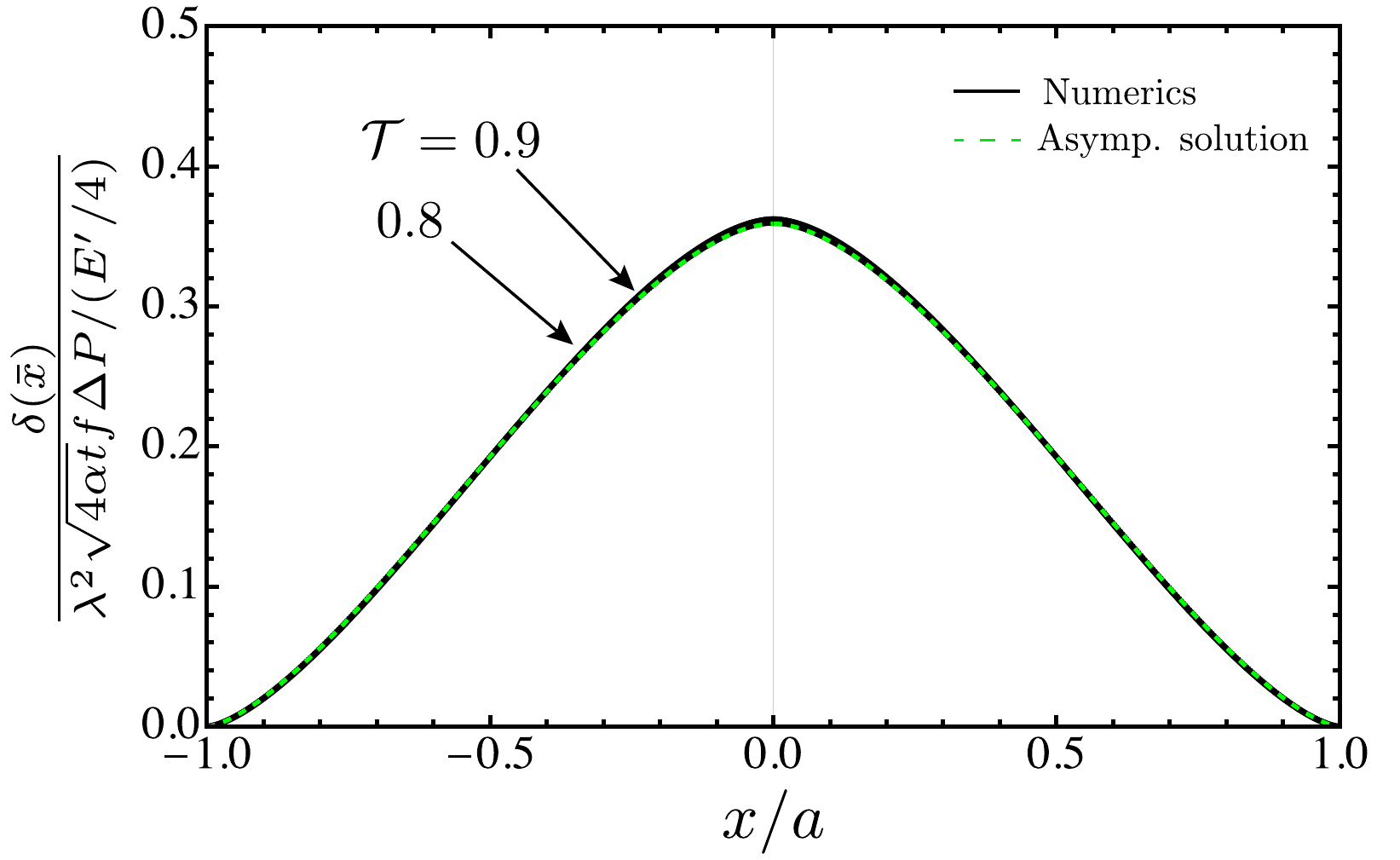}}
~
   \subfloat[]{
      \includegraphics[width=.45\textwidth]{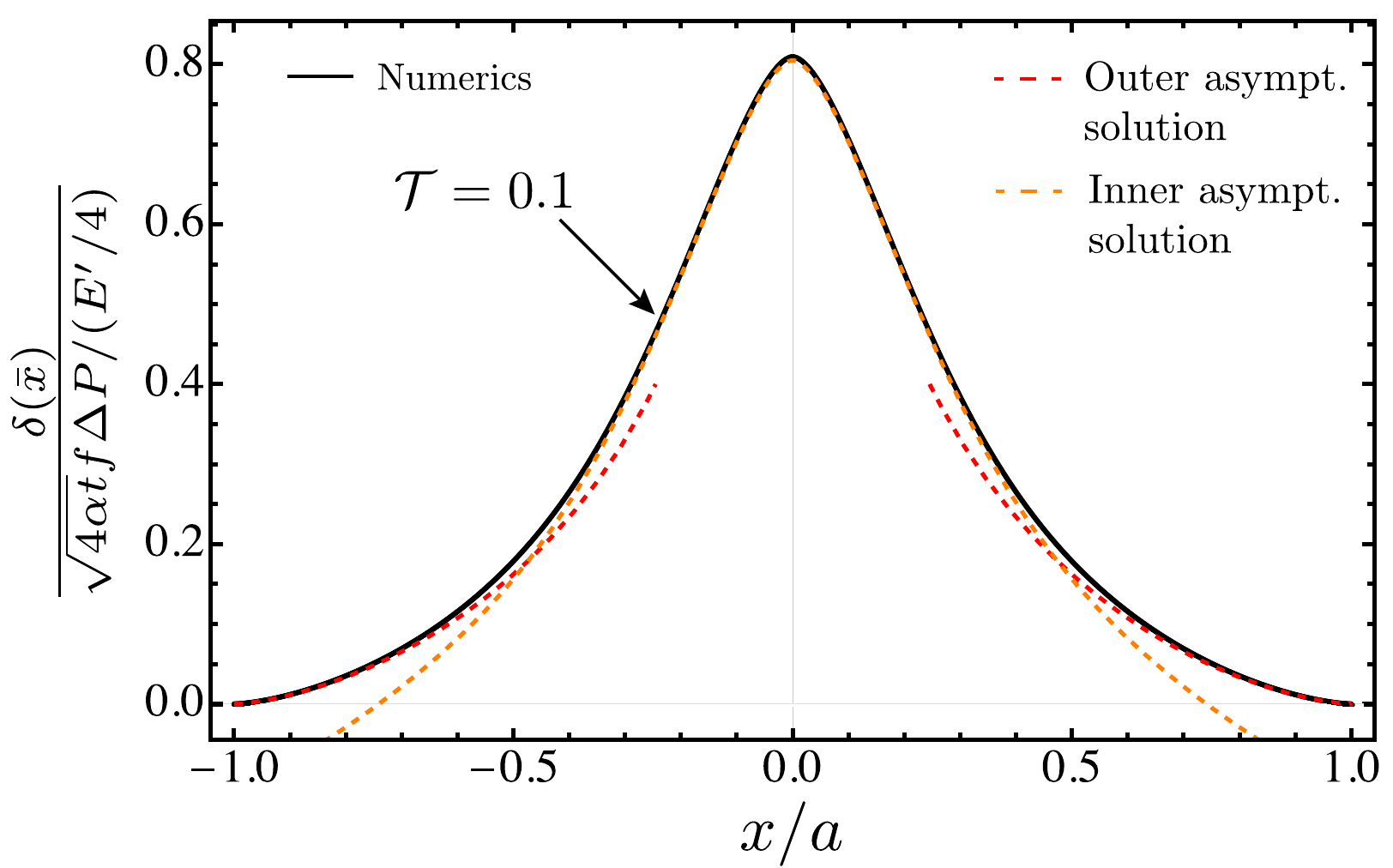}}
\\[-0.2cm]
   \subfloat[]{
      \includegraphics[width=.47\textwidth]{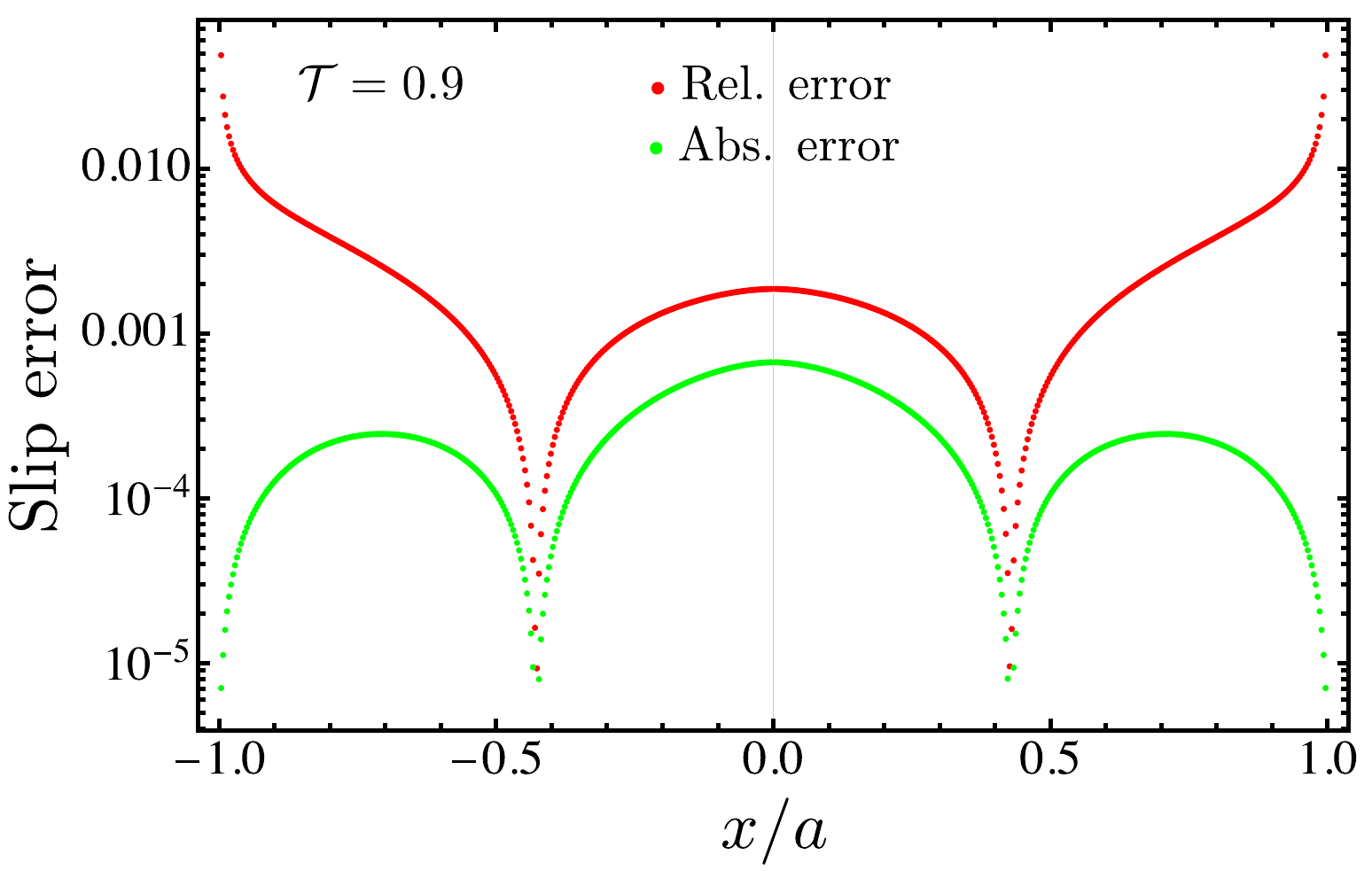}}
~
   \subfloat[]{
      \includegraphics[width=.48\textwidth]{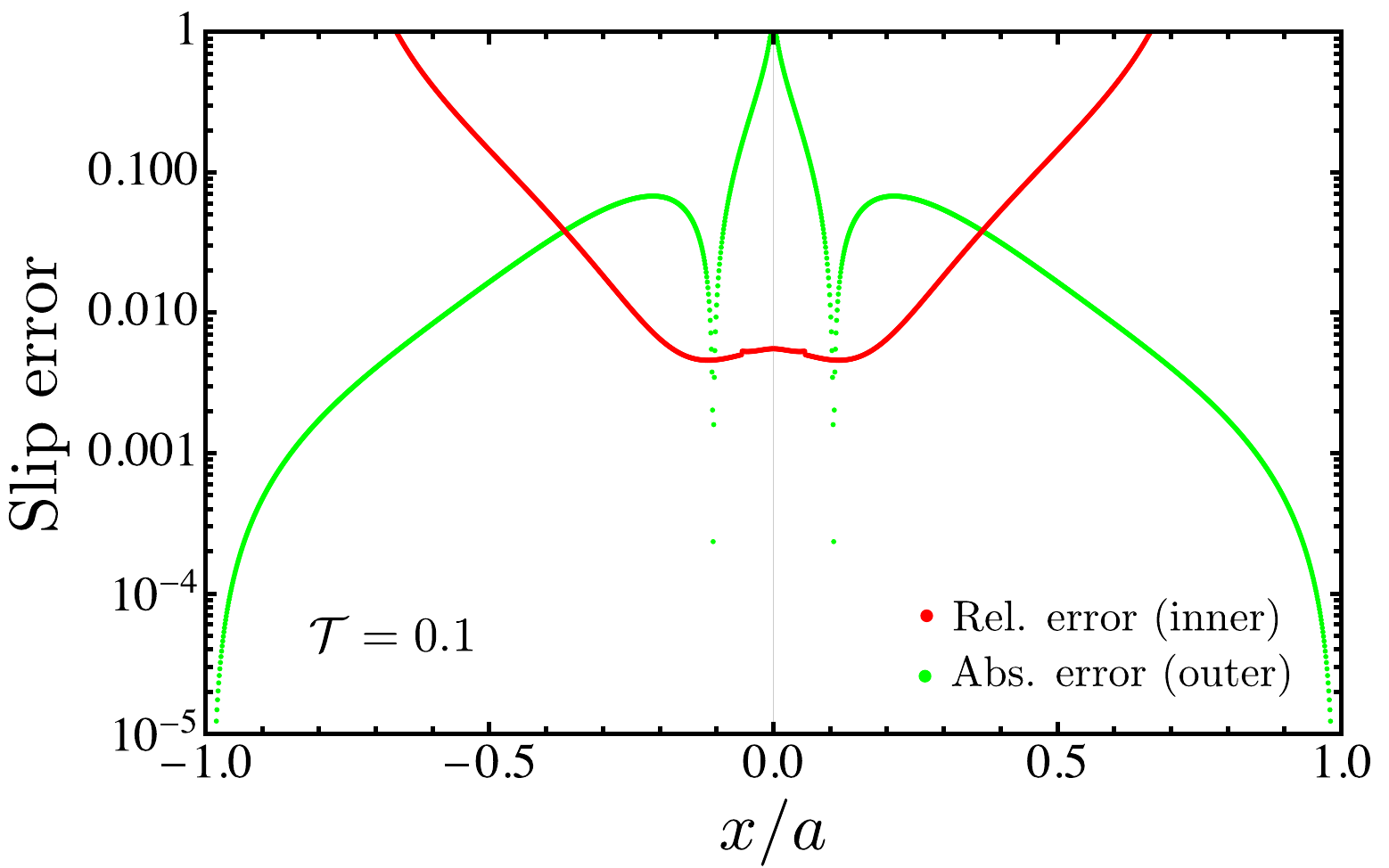}}
\caption{Panels (a) and (c): comparison between normalized slip profiles for marginally pressurized conditions (which collapse into one black curve due to scaling adopted) and asymptotic solution (\ref{eq:viesca_solution_slip}), together with the corresponding relative and absolute errors. Panel (b): comparison between normalized slip profile for $\mathcal{T}=0.1$ (critically stressed condition) and asymptotic inner and outer solutions (\ref{eq:viesca_solution_slip2}). The corresponding relative and absolute errors are displayed in (d), with the former calculated with respect to the inner solution (\ref{eq:viesca_solution_slip2}-b) and the latter with respect to the outer solution (\ref{eq:viesca_solution_slip2}-a).}
    \label{fig:res_viesca2}
\end{figure}
Indeed, by replacing time $t$ in Equations (\ref{eq:aseismic_moment_marginally_press}) and (\ref{eq:aseismic_moment_critically_stress}) with the corresponding expression obtained from Equation (\ref{eq:injected_volume}), after few manipulations we find that the aseismic moment can be expressed as   
\begin{equation}
M_o(t)  \simeq \frac{4}{3}\sqrt{\pi} \left( \frac{\lambda^3 f \left( 1-\nu\right)}{\beta^2 w_h^2 \Delta P}\right) V_{inj}^2
\label{eq:aseismic_moment_marginally_press_vs_volume}
\end{equation}
in the marginally pressurized limit, and as
\begin{equation}
M_o(t)  \simeq 2 \sqrt{\pi} \left( \frac{\lambda f \left( 1-\nu\right)}{ \beta^2 w_h^2 \Delta P}\right) V_{inj}^2
\label{eq:aseismic_moment_crticially_stress_vs_volume}
\end{equation}
in the critically stressed limit. If we introduce the following characteristic scales for the aseismic moment and the injected volume
\begin{equation}
    \frac{V_{inj} (t)}{w_h \beta \Delta P L} \rightarrow \mathcal{V}_{inj} \qquad \frac{M_{o} (t)}{f \Delta P L^2 (1-\nu) / 2} \rightarrow \mathcal{M}_o,
\end{equation} 
we can express these two scalar quantities in function of only the dimensionless fracture-stress-injection parameter:
\begin{equation}
\begin{split}
&\mathcal{M}_o  \simeq \frac{1}{24} \pi^5 \left( 1 - \mathcal{T}\right)^3 \cdot \mathcal{V}_{inj}^2  \qquad \text{(Marginally Pressurized)} \\
&\mathcal{M}_o  \simeq \frac{8}{\pi \mathcal{T}} \cdot \mathcal{V}_{inj}^2  \qquad \text{(Critically Stressed)}
\label{eq:aseismic_moment_vs_volume_dimensionless}
\end{split}
\end{equation}
Not surprisingly, the aseismic moment, in the marginally pressurized limit, decreases non-linearly with increasing values of $\mathcal{T}$ parameter and vanishes when $\mathcal{T} \to 1$.  
Conversely, in the critically stressed limit, the aseismic moment is unbounded with respect to the parameter $\mathcal{T}$ and a singularity appears when $\mathcal{T} \to 0$, which corresponds for example to the case where prior to injection the system is on the verge of failure.

\subsection{Verification}
\label{subsec3}

\begin{figure}[t]
\centering
    \subfloat[]{
      \includegraphics[width=.5\textwidth]{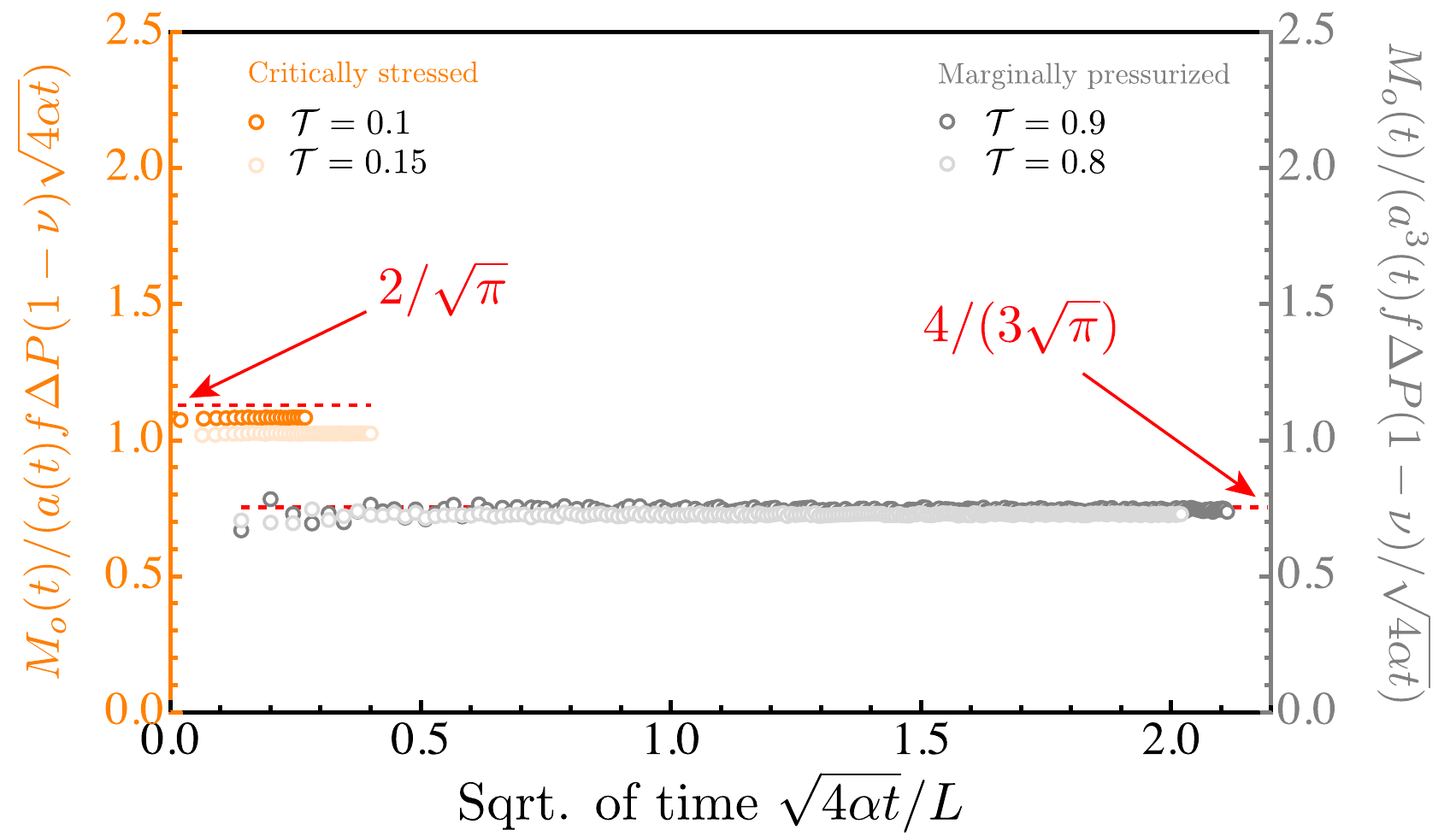}}
~
    \subfloat[]{
     \includegraphics[width=.47\textwidth]{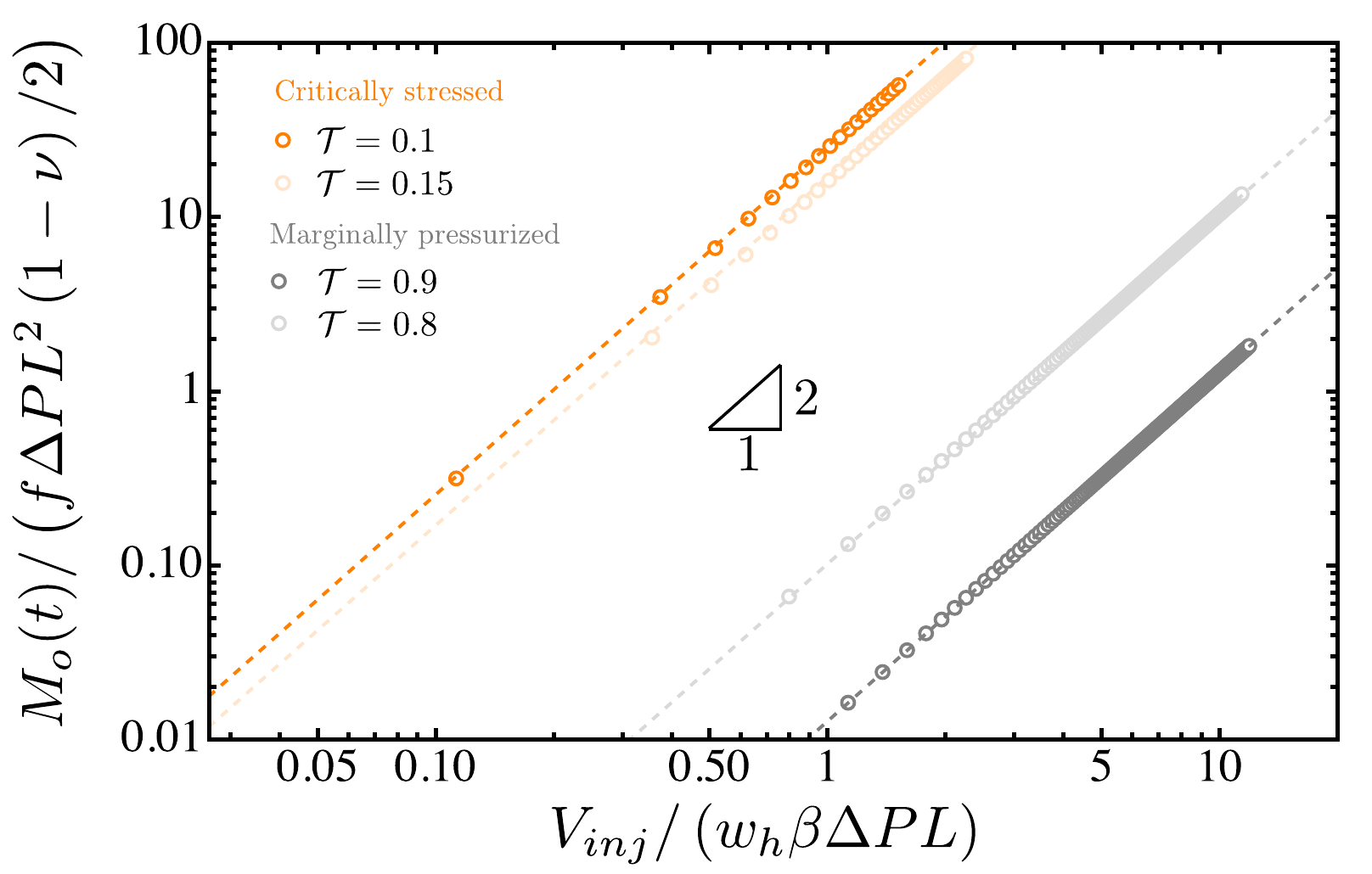}}
\caption{Panel (a): Time evolution of normalized aseismic moment in both critically stressed conditions (orange) and marginally pressurized conditions (gray). The scaling adopted for the two limits is directly obtained from analytical expressions (\ref{eq:aseismic_moment_marginally_press}) and (\ref{eq:aseismic_moment_critically_stress}). Panel (b): Evolution of normalized aseismic moment with normalized injected volume in a log-log plot, for different values of fracture-stress-injection parameters $\mathcal{T}$. Dashed lines refer to the corresponding analytical solutions (\ref{eq:aseismic_moment_vs_volume_dimensionless}).}
\label{fig:Aseismic_moment_single_fracture}
\end{figure}
The theoretical considerations previously recalled provide important solutions against which we verify our numerical solver. We consider a 1-dimensional planar fracture and discretize its total length $2L$ with $2000$ straight finite elements. 
We fix the injection over-pressure $\Delta P$ as well as the friction coefficient $f$, and vary the initial effective stress state on the fracture plane, in order to obtain different values of $\mathcal{T}$ parameter to simulate the marginally pressurized and critically stressed limit.   
Figure \ref{fig:res_viesca1}-a displays the temporal evolution of crack half-length $a(t)/L$ for different values of fracture-stress-injection parameter. We can note that the slipping patch grows in time always quasi-statically for each value of $\mathcal{T}$ parameter considered, with its tips located inside or ahead of the pressurized region. These different scenarios do depend on the value of $\mathcal{T}$ and it can be easily grasped by comparing the numerical results with the blue dashed line that represents the case in which the fluid front coincides with crack tips ($\lambda=1$). In the marginally pressurized limit (large values of $\mathcal{T}$), the slipping patch lags well behind the pressurized region, while the opposite occurs for low values of $\mathcal{T}$  at all times. It is worth mentioning that the coefficient of proportionality $\lambda$ obtained with different values of $\mathcal{T}$ parameter in the two limiting conditions, i.e. $\mathcal{T}$=0.9, 0.8 and $\mathcal{T}$=0.1, 0.15, matches perfectly with the analytical expressions (\ref{eq:viesca_solution}) (see Figure \ref{fig:res_viesca1}-c). Furthermore, panels (b) and (d) in Fig.~\ref{fig:res_viesca1} show that the relative error in terms of temporal evolution of crack half-length is always below $2\%$, under both marginally pressurized and critically stressed conditions.

A very good agreement with the analytical solutions (\ref{eq:viesca_solution_slip}) and (\ref{eq:viesca_solution_slip2}) is also obtained in terms of slip distributions. As demonstrated in Figure \ref{fig:res_viesca2}-a and Figure \ref{fig:res_viesca2}-c, the numerical results in the marginally pressurized limit collapse in one curve due to the scaling adopted and match perfectly with the asymptotic solution (\ref{eq:viesca_solution_slip}). The absolute error as well as the relative error are very small everywhere within the crack, with the latter that increases near fracture tips (due to the fact that slip goes to zero such that the absolute error provides a true comparison near the tip). Similar results are obtained  in the critically stressed limit: the numerical result matches very well with the outer asymptotic solution for $\lvert x/a(t) \rvert \lesssim 1$, as well as with the inner asymptotic solution near injection point $\lvert x/a(t) \rvert \sim 0$, i.e. within the pressurization region (see Figure \ref{fig:res_viesca2}-b). The relative error in panel (d) is everywhere below $1\%$ in the center region around the injection point, and the absolute error drops to very small values near fracture tips. 

Finally, Figure \ref{fig:Aseismic_moment_single_fracture} shows the aseismic moment evolution as function of normalized time (a) and normalized injected volume (b), in both marginally pressurized (gray) and critically stressed (orange) limit. On the left panel, we can observe that our numerical results (empty circles) are in very good agreement with the analytical expressions (\ref{eq:aseismic_moment_marginally_press}) and (\ref{eq:aseismic_moment_critically_stress}) under both limit conditions. Note, however, that the lower is the $\mathcal{T}$ parameter, the closer is the numerical aseismic moment to the analytical expression (\ref{eq:aseismic_moment_critically_stress}) (due to the lower spatial effect of pore-fluid pressurization).
Similarly, the log-log plot on the right panel shows a good match between our numerical results and the analytical expressions (\ref{eq:aseismic_moment_vs_volume_dimensionless}) (denoted by dashed lines). The quadratic scaling between normalized injected volume and aseismic moment can be grasped by comparing the slopes of all the analytical and numerical curves with the one of the hypothenuse of the small black triangle (which is equal to $2$).

%%%%%%%%%%%%%%%%%%%%%%%%%%%%%%%%%%%%%%%%%%%%%%%%%%%%%%%%%%%%%%%%%%%%%
% DFN
\section{Injection in a Discrete Fracture Network}
\label{sec4}
Now, we extend the previous findings to the case in which fluid is injected in one fracture that is hydraulically connected to a large number of pre-existing fractures, forming a Discrete Fracture Network. In the next sub-sections, we first describe the DFN model adopted in this contribution and then present the key dimensionless parameters that will guide our numerical investigations.

\subsection{DFN model}  
Formally, a DFN model provides the number of fractures in any given volume, with given lengths and orientations \citep{DaBo06}. 
%Restricting to the 2-dimensional case, this can be quantitatively defined with the following general mathematical expression 
%\begin{equation}
%    N_{2d} \left( L, l, \theta, \psi \right) \textrm{d}l \textrm{d}\theta \textrm{d}\psi \textrm{d}\dots ,
%    \label{eq:DNF_definition}
%\end{equation}
%which represents the number of fractures contained in an area of typical size $L$, with length between $l$ and $l + \textrm{d}l$, orientations in $\theta$ and $\textrm{d}\theta$, positions in $\psi$ and \textrm{d}$\psi$, and a set of other properties (denoted by the dots $\dots$). 
Several fracture distributions models have been introduced in literature, 
%for the variables in (\ref{eq:DNF_definition}), 
such as lognormal distribution, gamma law, exponential law among others (see \cite{BoBo01,LeLa17} for reviews).
%, leading to a non-unique choice of the expression for $N_{2d}$.
Each distribution model, however, must contain a scaling law that assembles scattered real data at different scales into an unified scaling model.

In this contribution, we adopt a distribution model that contains two scaling laws: a fractal density (given by the fractal dimension $D_{2d}$) and a power-law distribution for fracture length generation (characterised by an exponent $b$) with cut-off for minimum and maximum fracture lengths (denoted by $l_{min}$ and $l_{max}$, respectively). This choice has been used in numerous studies at different scales and in different tectonic settings \citep{HaMa94,SorDa93,AndWil94,Kra83,WalSor96}. Assuming fracture lengths, positions (or density) and orientations independent entities, we can write the DFN model as
\begin{equation}
N_{2d} \left( L,l,\theta,\psi\right) = \rho(\theta, \psi) L^{D_{2d}} l^{-b}, \quad \text{for} \quad l\in\left[ l_{min}, l_{max}\right]
\label{eq:DFN_model}
\end{equation} 
where $\rho(\theta, \psi)$ is the fracture density term, which depends on orientations $\theta$ and positions $\psi$. Note that the only intrinsic characteristic length scales in this model are the smallest $l_{min}$ and the largest $l_{max}$ fracture lengths. The exponents $D_{2d}$ and $b$ quantify the scaling aspects of the DFN model \citep{LeiWa16,LeiGao18}: the former governs the fracture density, whereas the latter governs the lengths distribution. According to extensive outcrop data, $D_{2d}$ typically varies between 1.5 and 2.0, whereas $b$ ranges between 1.2 and 3.5 \citep{BoBo01}. In this investigation, we assume that fractures are uniformly distributed within the region of interest $L\times L$, with random locations and orientations (albeit this may not reflect most of the real cases, in which discrete sets of joints with given orientations are observed). This assumption implies that the fractal dimension $D_{2d}$ equals the Euclidean dimension and hence $D_{2d}=2$ (homogeneous and isotropic case). 

In order to build a DFN model representative of systems of various sizes, we normalize (\ref{eq:DFN_model}) with $L^{D_{2d}}$ and 
$\dfrac{(l_{max} l_{min})^{-b} (l_{max}^b l_{min} - l_{max} l_{min}^b) \cdot \rho\left( \theta, \psi\right)}{-1+b}$, the latter being the number of fractures with length greater than $l_{min}$ but lower than $l_{max}$. We obtain
\begin{equation}
n_{2d}(l) = \frac{(-1 + b) (l_{max} l_{min})^b}{l_{max}^b l_{min} - l_{max} l_{min}^b} \cdot l^{-b}, \quad \text{for} \quad  l\in\left[ l_{min}, l_{max}\right]
\label{eq:pdf_lengths}
\end{equation}
which represents the scaling law for fracture lengths for a given DFN. Typically, this law must span at least two orders of magnitude. It is worth mentioning that the choice of a given scaling law strongly impacts fluid flow organization, especially if flow is confined only inside the DFN. A parameter that assesses the geometrical connectivity of a DFN is the \textit{percolation} parameter, which is commonly denoted in literature as $p$. For non-fractal discrete fracture networks (like the ones used in this contribution), such a dimensionless parameter is defined as \citep{BoDa97}
\begin{equation}
p(l) = \int_{l_{min}}^{l_{max}} \frac{n_{2d}(l) l^2}{L^2} \text{d}l
\label{eq:percolation_parameter}
\end{equation}
The larger is the percolation parameter, the more connected is the system and thus the more homogeneous is the fluid diffusion inside the DFN. Typically, a DFN is \textit{statistically connected} if $p$ is greater than a percolation threshold $p_c $, with $p_c \sim 5.6$ for 2D networks \citep{BoDa97}.

\subsection{Governing dimensionless parameters}
The scaling analysis of the governing equations (elasticity equations (\ref{eq:elasticity_equations}), yielding criterion (\ref{eq:interfacial_law}) and fluid mass conservation equation (\ref{eq:diffusion_equation})) is similar to the one proposed by \citet{BaVi19} for the case of injection in a planar shear fracture (and discussed in Section \ref{subsec1}). However, for a Discrete Fracture Network, fluid flow is not known analytically, but is part of the numerical solution of the coupled hydro-mechanical problem. Without loss of generality, we can express the pore fluid over-pressure field as 
\begin{equation}
    \bar{p}(\mathbf{x},t) = \Delta P \times \mathcal{H}\left( \theta^k, \alpha, r, p, t \right),
    \label{eq:overpressure_DFN}
\end{equation}
where $\mathcal{H}$ is a dimensionless function that depends on local fracture orientation $\theta$ of fracture $^k$, pressurization time $t$, constant hydraulic diffusivity $\alpha$, percolation parameter $p$, and distance to injection point $r$. Following the same scaling procedure qualitatively described in Section \ref{subsec1}, but now considering the generic over-pressure field (\ref{eq:overpressure_DFN}), it is straightforward to show that the hydro-mechanical problem in the case of injection in a DFN is governed by a family of dimensionless parameters, namely one for each pre-existing fracture
\begin{equation}
    \mathcal{T} = \frac{f t_{n,o}^{\prime, k} \left( \theta^k, \sigma^{\prime}_{xx,o}, \sigma^{\prime}_{yy,o} \right)- t_{s,o}^{k}\left( \theta^k, \sigma^{\prime}_{xx,o}, \sigma^{\prime}_{yy,o} \right)}{f \Delta P},
    \label{eq:T_parameter_DFN}
\end{equation}
on top of the percolation parameter $p$ (\ref{eq:percolation_parameter}) that characterizes the geometrical connectivity of the DFN.\\
The initial effective tractions on each fracture can be obtained by projecting the principal effective stress tensor $\Sigma = \begin{pmatrix} \sigma^{\prime}_{xx,o} & 0 \\ 0 & \sigma^{\prime}_{yy,o} \end{pmatrix} $ onto each fracture plane. Assuming $\theta$ the angle between the trace of fracture $^k$ and the direction of the minimum far-field principal effective stress $\sigma^{\prime}_{yy,o}$ (in absolute terms), we get 
\begin{equation}
\begin{split}
    & t_{n,o}^{\prime, k} = \sigma^{\prime}_{xx,o} \cos (\theta^k)^2 + \sigma^{\prime}_{yy,o} \sin (\theta^k)^2 \\
    & t_{s,o}^{k} = \sigma^{\prime}_{xx,o} \cos (\theta^k) \sin (\theta^k) -  \sigma^{\prime}_{yy,o} \sin (\theta^k) \cos (\theta^k)
\end{split}
\label{eq:generic_tractions}
\end{equation}

For a given pressurization scenario, large values of $\mathcal{T}$ parameter suggest that the fractures are not favourably oriented with respect to the in-situ stress field. On the contrary, low values of $\mathcal{T}$ suggest that the fractures are critically stressed at ambient conditions and thus prompt to fail upon stress perturbation. Given an initial effective stress state, the critical orientation at which the $\mathcal{T}$ parameter reaches a minimum value or, equivalently, at which the distance to failure is minimum, can be obtained knowing the in-situ friction coefficient and reads
\begin{equation}
    \theta_c = \pm \left( \frac{\pi}{4} + \frac{\textrm{ArcTan} (f)}{2}\right)
    \label{eq:critical_angle}
\end{equation}
The expression of $\mathcal{T}$ parameter at critical angle $\theta_c$, therefore, can be retrieved by replacing (\ref{eq:critical_angle}) into (\ref{eq:T_parameter_DFN}-\ref{eq:generic_tractions}) and reads \footnote{Here we consider only the positive angle.}
\begin{equation}
    \mathcal{T}_c = \frac{\sigma^{\prime}_{yy,o}}{2 f \Delta P } \left(f (\kappa +1)-\sqrt{f^2+1} (\kappa -1)\right)
    \label{eq:critical_T}
\end{equation}
where $\kappa=\dfrac{\sigma^{\prime}_{xx,o}}{\sigma^{\prime}_{yy,o}} \geq 1$ is the in-situ effective stress anisotropy ratio. It is interesting to note that $\mathcal{T}_c$ is given by a product of two functions: one strictly related to in-situ conditions (namely friction coefficient and far-field stress state), while the other is related to the pressurization scenario. This is an unique parameter that characterizes each DFN and is the complete analogy of the $\mathcal{T}$ parameter introduced by \cite{BaVi19} for the case of injection in a single shear fracture (see Section \ref{sec3}).

\begin{figure}[t]
\centering
   \subfloat[]{
      \includegraphics[width=0.5\textwidth]{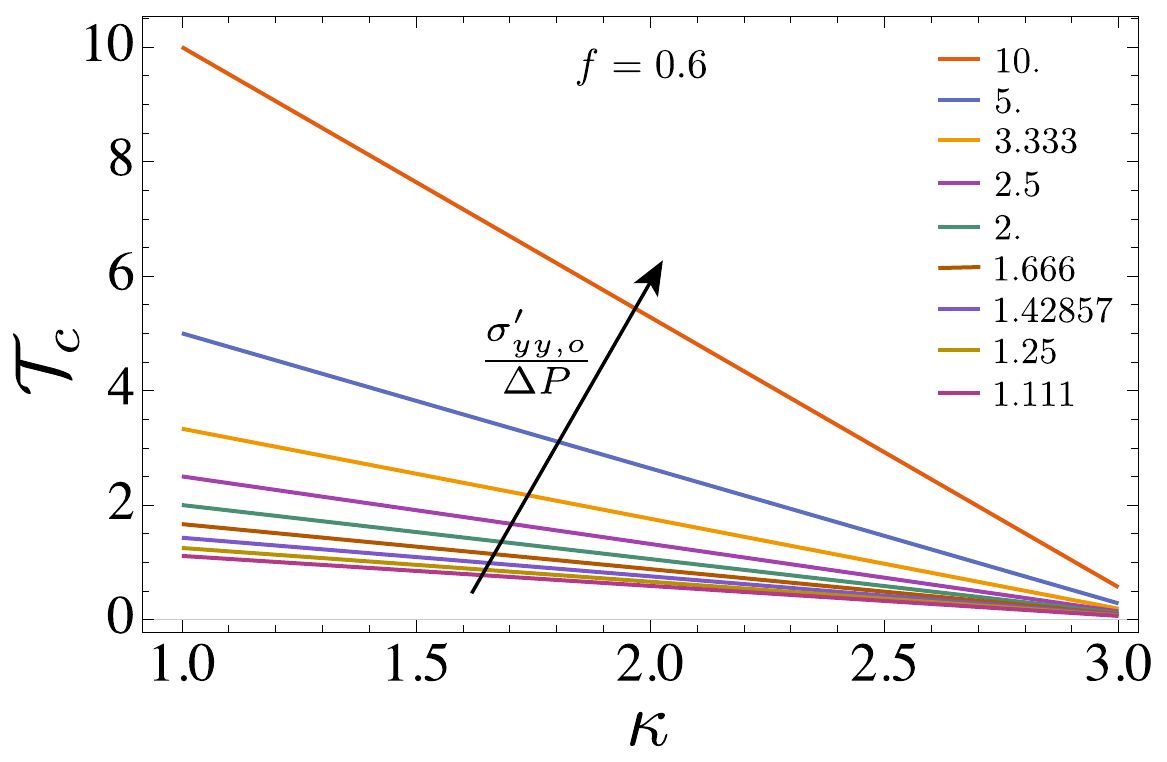}}
~
   \subfloat[]{
      \includegraphics[width=0.53\textwidth]{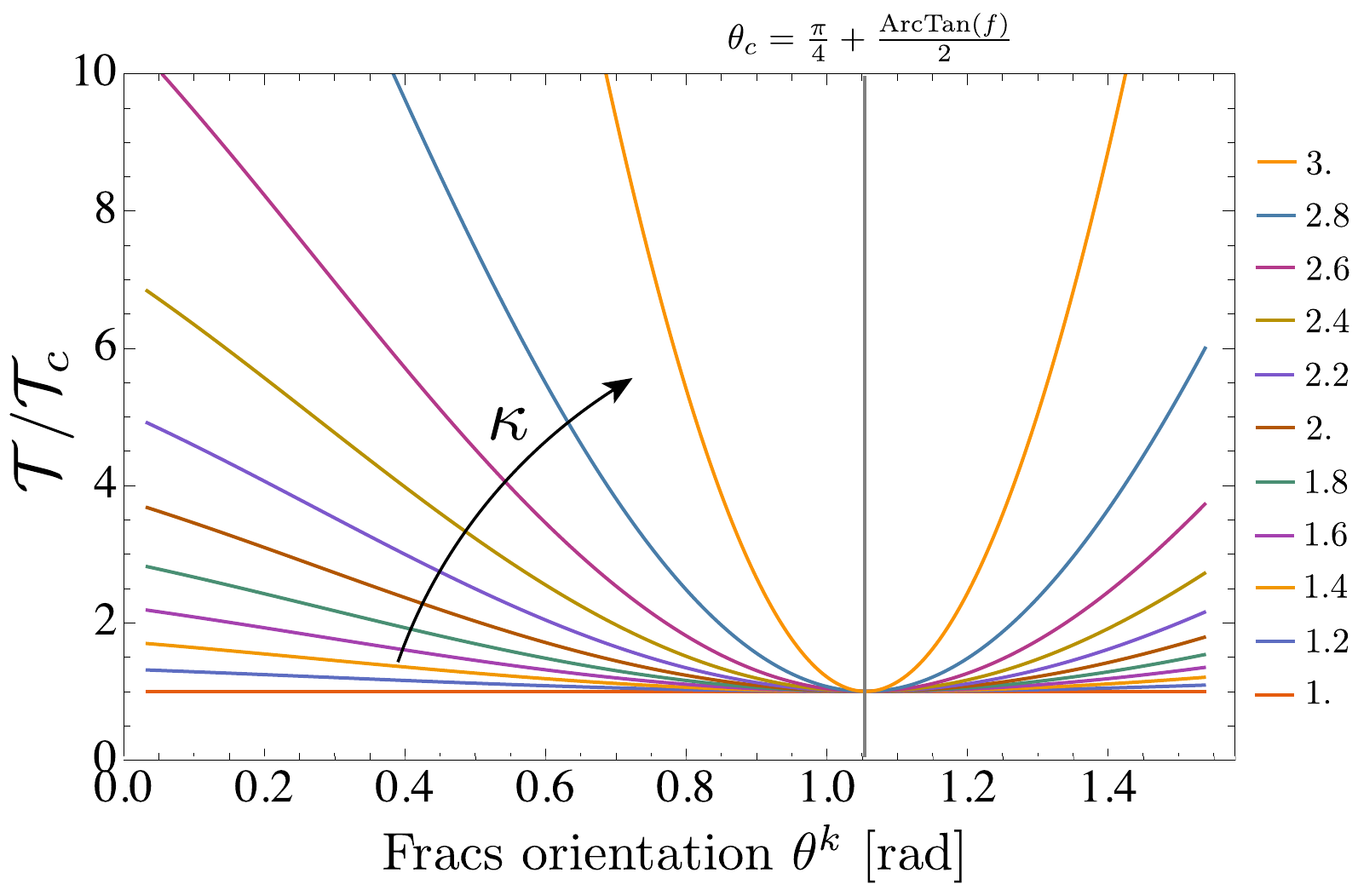}}
\caption{Panel (a): Variation of the critical $\mathcal{T}$ parameter in function of effective stress anisotropy ratio $\kappa$ and increasing values of the inverse of normalized injection over-pressure $\dfrac{\sigma^{\prime}_{yy,o}}{\Delta P}$, for a fixed value of friction coefficient $f=0.6$. Panel (b): Variation of $\mathcal{T}/\mathcal{T}_c$ in function of local fracture orientation $\theta^k$ and increasing values of effective stress anisotropy ratio $\kappa$, for a fixed value of friction coefficient $f=0.6$. The grey vertical line corresponds to the critical fracture orientation $\theta_{c}=\pi/4+\dfrac{\textrm{ArcTan}(f)}{2}\simeq 1.056$ radians.}
\label{fig:SC_function_theta}
\end{figure}

As displayed in Figure \ref{fig:SC_function_theta}-a, for a given pressurization scenario $\Delta P/\sigma^{\prime}_{yy,o}$ (or its inverse), large values of $\kappa$ lead to low values of $\mathcal{T}_c$ (with the maximum value of $\kappa$ achieved in the limit when $\mathcal{T}_c\to0$). This means that the fractures oriented at $\theta_c$ are critically stressed and thus prone to fail upon fluid injection. 
Vice-versa, lower values of $\kappa$ lead to increasing values of $\mathcal{T}_c$. Hence, even if the fractures are oriented at the critical angle $\theta_c$, the far-field effective stress state induces tractions that well satisfy the activation criterion (\ref{eq:interfacial_law}). The maximum value of $\mathcal{T}_c$ is reached in the limit when $\kappa\to1$, for which $\mathcal{T}_c = \dfrac{\sigma^{\prime}_{yy,o}}{\Delta P}$ (always greater than one). Finally, it is worth mentioning that large values of normalized injection over-pressure lead to relatively low values of $\mathcal{T}_c$ for any in-situ effective stress states, similarly to the case of injection in a single  fracture.   

By taking the ratio between the generic $\mathcal{T}$ parameter (\ref{eq:T_parameter_DFN}-\ref{eq:generic_tractions}) and its value at critical angle $\mathcal{T}_c$ (\ref{eq:critical_T}), we can uncouple the effect of in-situ conditions and pressurization scenario. Indeed, after few manipulations we obtain 
\begin{equation}
    \frac{\mathcal{T}}{\mathcal{T}_c} = \frac{2 \sqrt{f^2+1} \left(f \kappa  \cos ^2(\theta^k)-(\kappa -1) \sin (\theta^k) \cos (\theta^k) +f \sin ^2(\theta^k)\right)}{1 -\kappa + f \left(\sqrt{f^2+1} (\kappa +1)+ f ( 1 -\kappa )\right)}
\end{equation}

Figure \ref{fig:SC_function_theta}-b displays the variation of $\mathcal{T}/\mathcal{T}_c$ as function of fracture orientation $\theta^k$ (expressed in radians) and increasing effective stress anisotropy ratios $\kappa$ (denoted by the arrow), for a given value of friction coefficient $f$. % (specifically $f=0.6$). 
Specifically, we consider a friction coefficient of $f=0.6$ that is a typical value for granitic rocks under confinements between 10 to 80 MPa \citep{Byerlee78}, with this range being plausible at few kilometer depths.
We can readily observe that the orientation at which the $\mathcal{T}$ parameter is minimum always correspond to the critical value $\theta_c$ (see grey vertical line), regardless of the value of stress anisotropy ratio $\kappa$. Furthermore, for increasing values of $\kappa$, only the fractures with orientations similar to $\theta_c$ are critically stressed and thus characterized by a ratio of $\mathcal{T}/\mathcal{T}_c$ that tends to one, with $\mathcal{T}_c$ being very close to zero (see panel (a)). The remaining fractures are instead characterized by initial stress states that are far from failure prior fluid injection (large values of $\mathcal{T}/\mathcal{T}_c$). We refer to this specific condition as critically stressed condition due to the presence of some highly critically stressed fractures in the DFN.\\
On the other hand, the lower is the value of $\kappa$, the more uniform is the distribution of $\mathcal{T}/\mathcal{T}_c$ over all the fracture angles, implying that more and more fractures are far from being critically stressed. In the limit case of $\kappa =1$, all the fractures of the DFN are characterized by $\mathcal{T}/\mathcal{T}_c=1$, with $\mathcal{T}_c$ that has achieved its maximum value equal to $\dfrac{\sigma^{\prime}_{yy,o}}{\Delta P}$ (see left panel). Therefore, we define this condition as marginally pressurized condition, in complete analogy with what presented in Section \ref{sec3}.\\

These considerations suggest that a critically stressed DFN (low value of $\mathcal{T}_c$) subjected to fluid injection exhibits fast aseismic slip regardless of the percolation value that characterises its fluid inter-connectivity as well as the value of injection over-pressure (as long as it is sufficient to activate slip). This is because the main driving force for the slipping patch propagation is the stress-interactions between the pre-existing critically stressed fractures. On the other hand, the scenario may change for marginally pressurized or slightly critically stressed DFN characterized by a large/moderate $\mathcal{T}_c$. The aseismic slip propagation in this case is directly affected by pore pressure diffusion inside the DFN and as a result the corresponding percolation parameter plays an important role. Low percolation values may lead to fluid localization and possibly contained ruptures.

In order to verify these fore-thoughts, we present several numerical investigations using a number of DFN realizations with different degrees of fluid percolation (see Appendix \ref{app:appendix1}). Each DFN realization includes $1000$ fractures whose lengths are sampled from a power law distribution (\ref{eq:pdf_lengths}), using different power law coefficients $b$ and maximum fracture length $l_{max}$ (see Appendix \ref{app:appendix1} for more details). 
For a given value of friction coefficient $f=0.6$ and a given value of normalized injection over-pressure $\dfrac{\Delta P}{\sigma^{\prime}_{yy,o}} = 1.0$, we vary the effective stress anisotropy ratio $\kappa$ in order to reproduce  critically stressed and marginally pressurized conditions. 

\subsection{Marginally pressurized conditions}
\label{subsec:MP}
\begin{figure}[t!]
\centering
   \subfloat[]{
      \includegraphics[width=.5\textwidth]{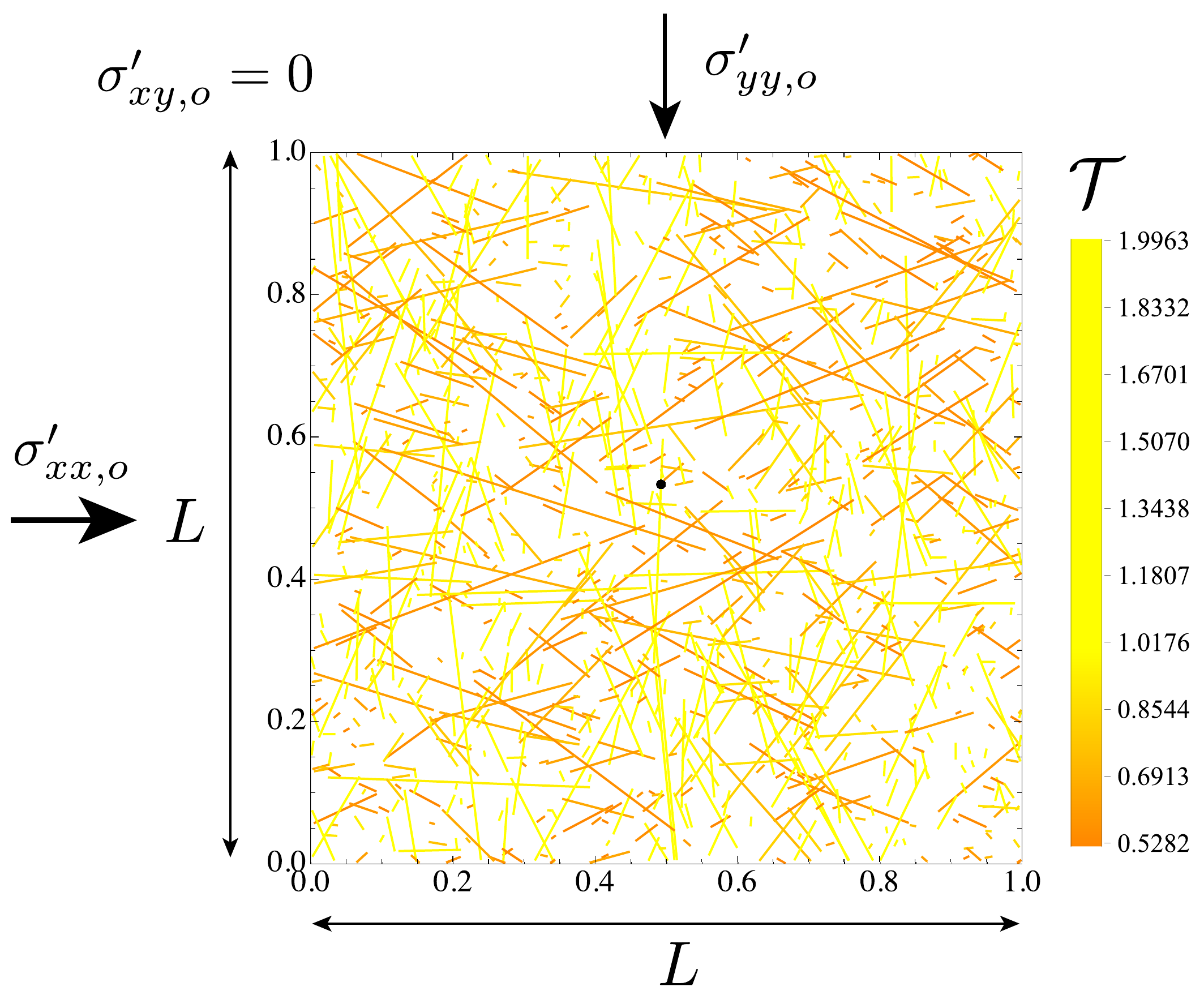}}
~
   \subfloat[]{
      \includegraphics[width=.4\textwidth]{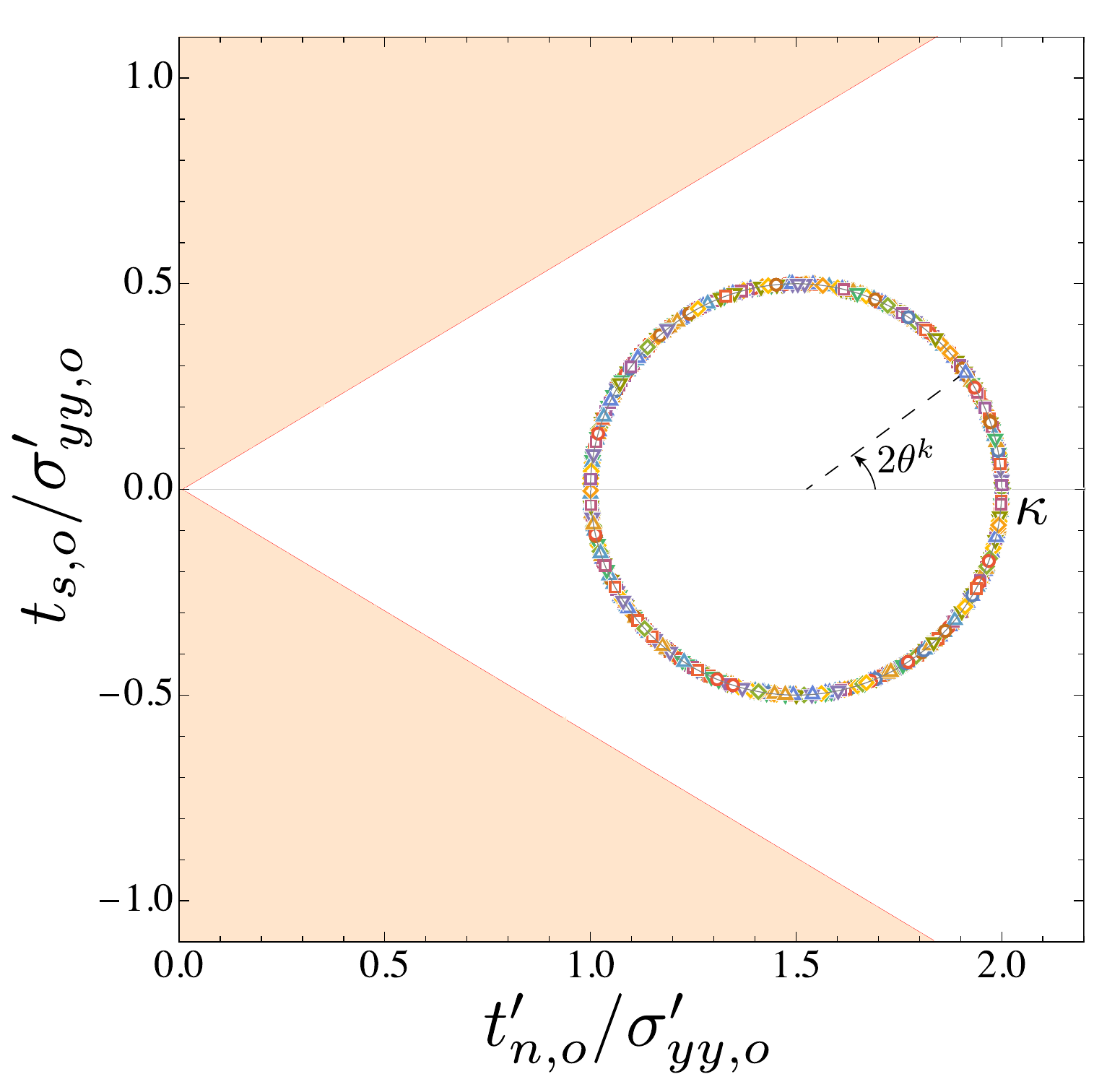}}
      
   \caption{Panel (a): marginally pressurized Discrete Fracture Network with percolation parameter $p =  12.21$, subjected to an in-situ principal effective stress state. The color of each fracture denotes the corresponding magnitude of the fracture-stress-injection parameter $\mathcal{T}$ prior fluid injection. The black dot, instead, denotes the location of injection point. Panel (b): initial effective stress state in the Mohr-Coulomb plane. Note that the effective tractions are normalized with the minimum far-field principal effective stress $\sigma_{yy,o}^{\prime}$.  
   Along the Mohr-Coulomb circle, whose diameter equals the value of stress anisotropy ratio considered (i.e. $\kappa=2$), all the uniform stress states of each pre-existing fracture are reported. Since all the pre-existing fractures are randomly oriented within the region $L\times L$, the whole circle is uniformly ``covered'' by the fractures' initial stress states.}
   \label{fig:SC1}
\end{figure}
 We fixed the effective stress anisotropy ratio at a relative low value $\kappa=2$, obtaining a moderate value of $\mathcal{T}_c$ (specifically $\mathcal{T}_c\simeq 0.528$) representative of marginally pressurized conditions (for which $\lambda=a(t)/\sqrt{4\alpha t} \ll 1 $ for the single fracture case). 
 In each DFN realization considered (see Figure \ref{fig:DFN_realizations}), all the pre-existing fractures within the region of interest $L\times L$ are characterized by a relatively low stress criticality $(f t^{\prime, k}_{n,o})/t^k_{s,o}$ or a relative large $\mathcal{T}$ parameter.
Fluid is injected approximately around the centre of the domain area, i.e. $\sim \left(L/2, L/2 \right)$, in a point crossed by a pre-existing fracture that is hydraulically connected to at least another one and with an orientation $\theta$ different than $\pi/2$ (in order to avoid tensile opening due to the injection condition $\dfrac{\Delta P}{\sigma^{\prime}_{yy,o}} = 1.0$).
All the simulations are stopped when the normalized over-pressure $\bar{p}/\Delta P$ reached the boundaries of the simulation box.\\ 
For sake of space, we do not report all the numerical results in terms of spatial and temporal evolution of pore fluid over-pressure and aseismic rupture extent. As a representative example, we display only those associated with the DFN with percolation parameter $p=12.21$ (see Figure \ref{fig:DFN_realizations}).\\

Figure \ref{fig:SC1} displays the spatial variation of the $\mathcal{T}$ parameter (\ref{eq:T_parameter_DFN}) on each pre-existing fracture as a result of the applied far-field effective stress state (see panel (b) for its representation in the Mohr-Coulomb plot). As one can observe, the effective stress state, which is represented by a circle whose diameter is equal to the difference between the two principal effective stress at ambient conditions (i.e. $\sigma_{xx, o}^{\prime}$ and $\sigma_{yy,o}^{\prime}$), is relatively far from the yielding failure line (\ref{eq:interfacial_law}) (denoted by red lines). All the pre-existing fractures that are randomly oriented and uniformly distributed within the region $L\times L$ are thus characterised by a $\mathcal{T}$ parameter not lower than its critical value $\mathcal{T}_c\simeq 0.528$ (see panel (a)).\\
%a stress criticality $\Lambda$ not higher than $\sim 0.575$, or $T$ parameter not below $\sim 0.85$.\\
\begin{figure}[th!]
\centering
   \subfloat[$\sqrt{4\alpha t}/L \simeq 0.203 $]{
      \includegraphics[width=.48\textwidth]{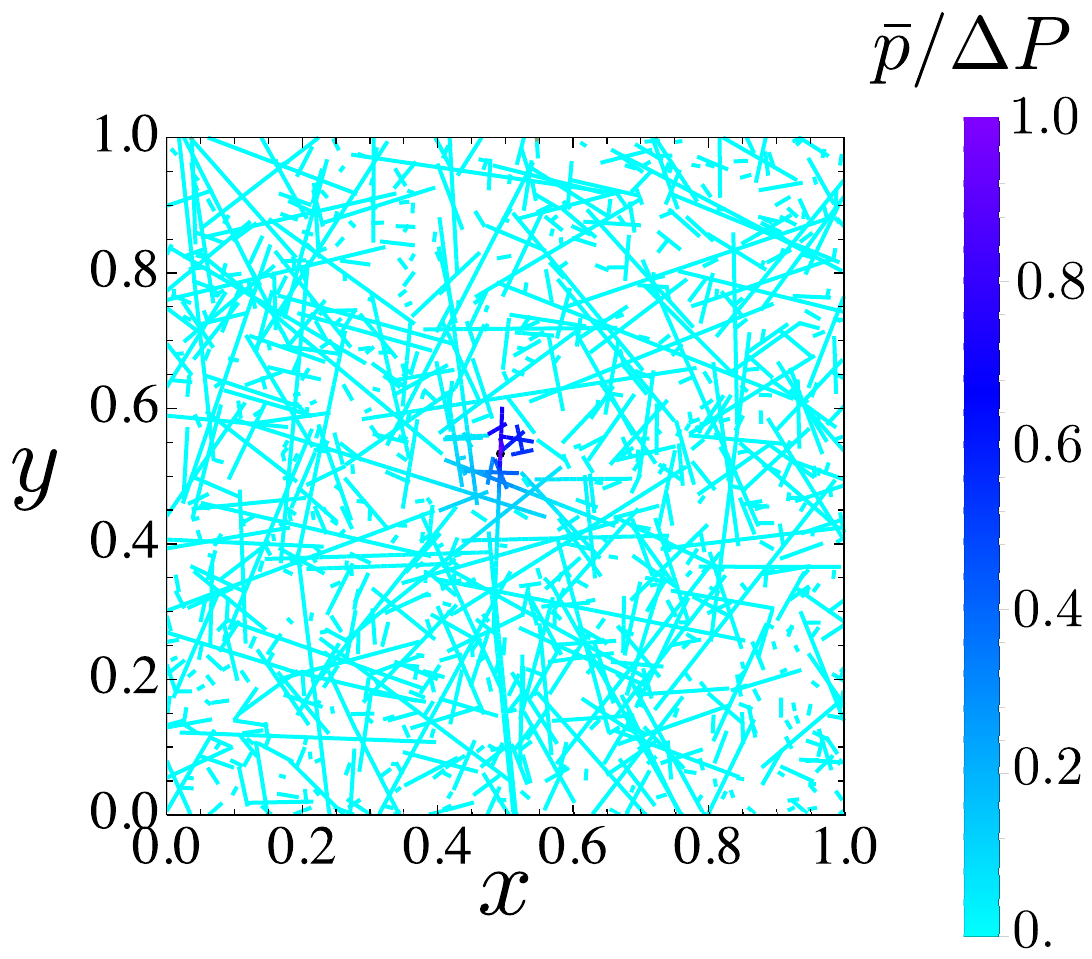}}
~
   \subfloat[$\sqrt{4\alpha t}/L \simeq 0.203 $]{
      \includegraphics[width=.4\textwidth]{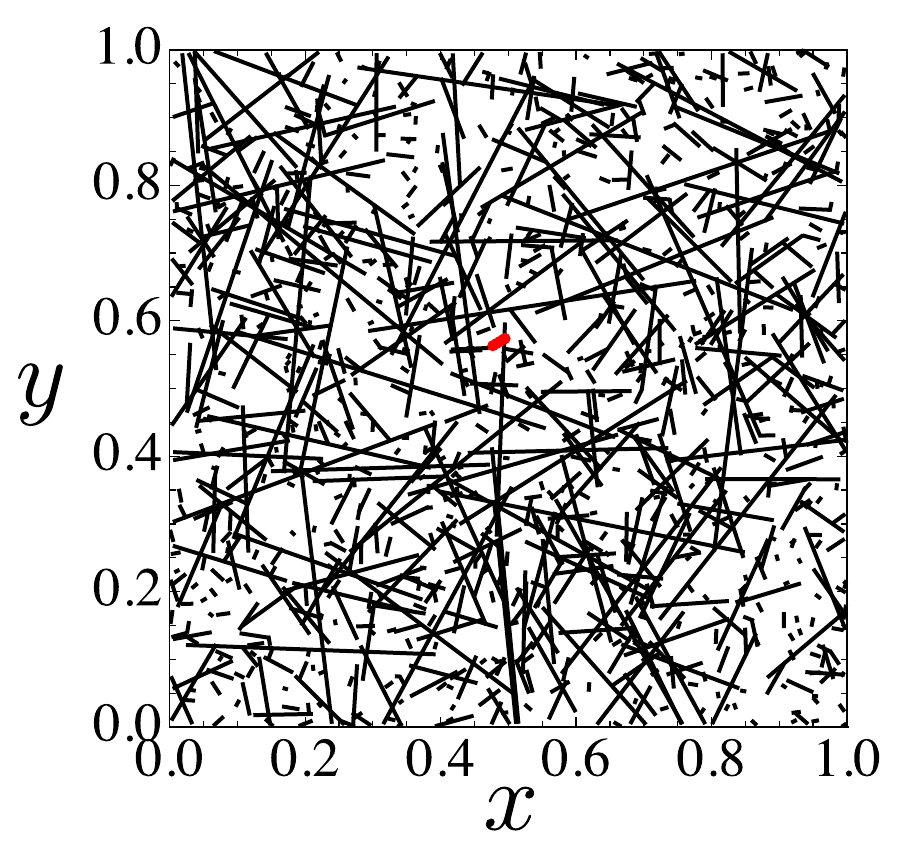}}
\\
   \subfloat[$\sqrt{4\alpha t}/L \simeq 0.47 $]{
      \includegraphics[width=.48\textwidth]{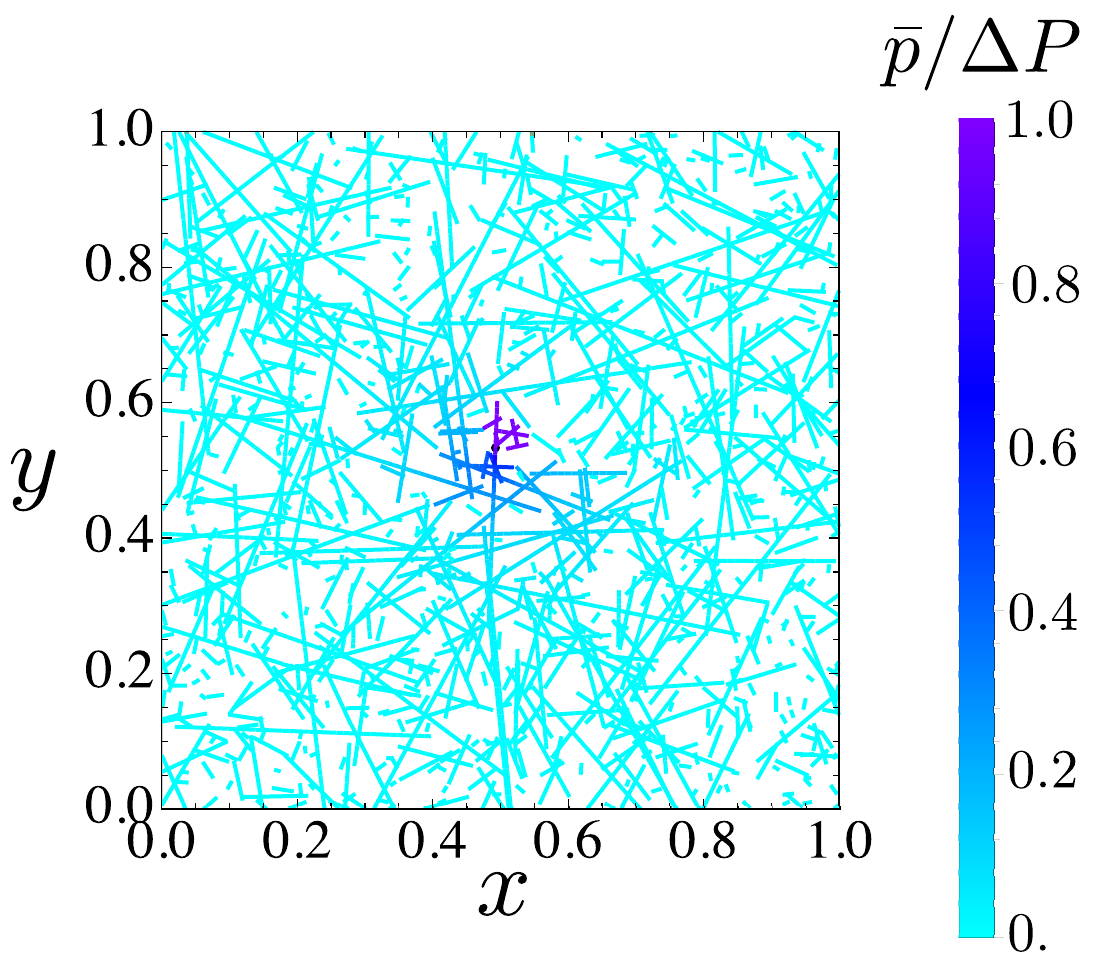}}
~
   \subfloat[$\sqrt{4\alpha t}/L \simeq 0.47 $]{
      \includegraphics[width=.4\textwidth]{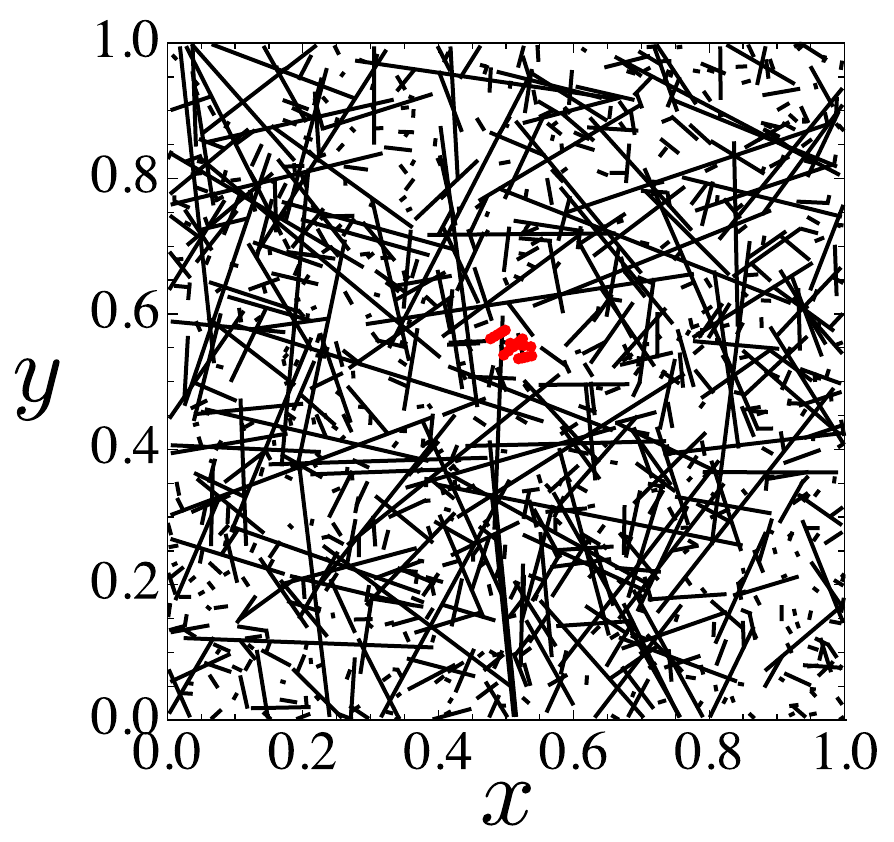}}
\\
   \subfloat[$\sqrt{4\alpha t}/L \simeq 1.045 $]{
      \includegraphics[width=.48\textwidth]{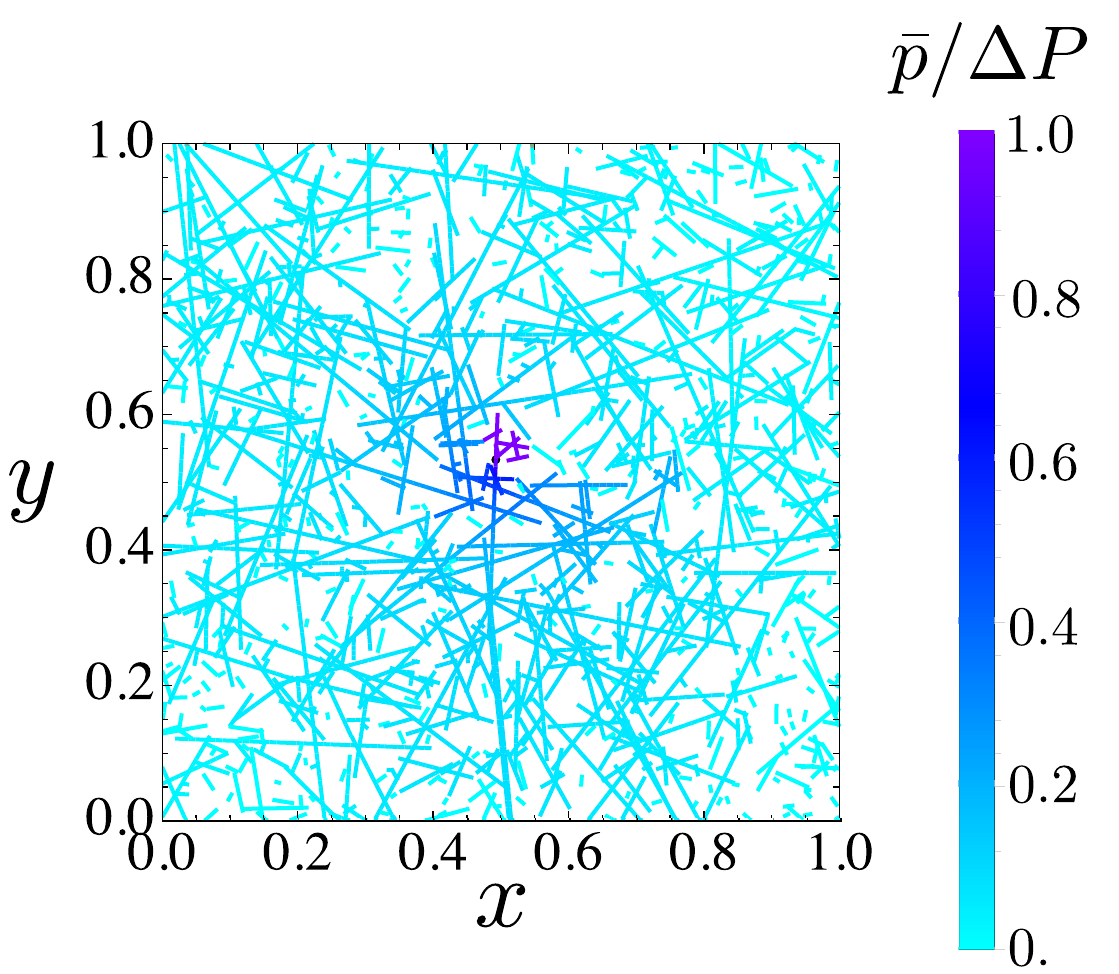}}
~
   \subfloat[$\sqrt{4\alpha t}/L \simeq 1.045 $]{
      \includegraphics[width=.4\textwidth]{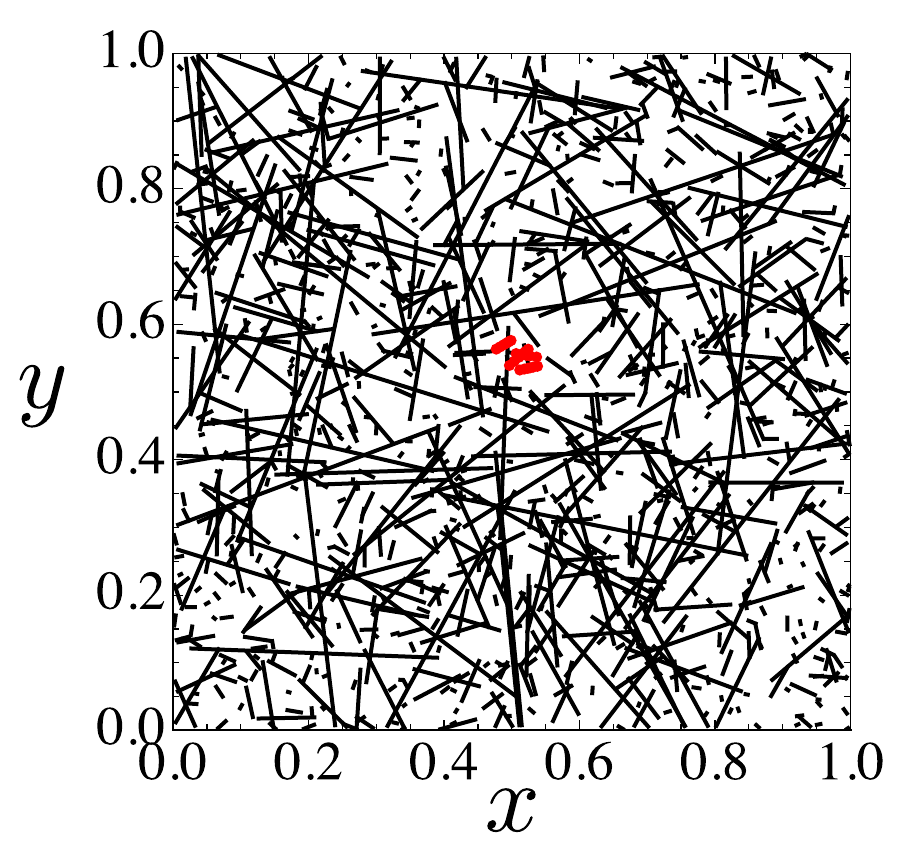}}

   \caption{Evolution of normalized over-pressure $\bar{p}/\Delta P$ (a-c-e) and aseismic rupture extent (denoted by red color - (b-d-f)) along the marginally pressurized DFN with $p=12.21$, at different normalized time snapshots $\sqrt{4 \alpha t}/L$. Fluid is injected at $\sim\left( 0.493, 0.533 \right)$, in one fracture that is hydraulically connected to many others.}
   \label{fig:overpress&aseismic_patch_MP}
\end{figure}
We discretized the DFN with 11690 finite straight segments (mean mesh size $0.005 L$) and used the numerical methods described in Section \ref{sec2.1} to solve the coupled hydro-mechanical problem. % Due to the large number of unknowns, 
We used a hierarchical approximation of the fully populated elasticity matrix for computational efficiency that resulted in a memory compression ratio of $c_r = 0.114$ with a relative accuracy of $10^{-4}$  for the approximated far-field contributions (see \citep{CiLe20} and references therein for more details). This was  sufficient to be able to run the simulation with a 2.8 GHz Quad-Core Intel Core i7 processor with 16 GB of memory.

Upon fluid injection in a point located at $\sim \left(0.493, 0.533 \right)$ (see black dot in Fig. \ref{fig:SC1}-a), the slipping patch starts to expand paced by pore fluid diffusion. Because of the in-situ marginally pressurized conditions, the aseismic slipping patch always lags 
%is located well inside 
the pressurized region, even at large pressurization times. This can be clearly grasped from Figure \ref{fig:overpress&aseismic_patch_MP} in which the over-pressurized region (left column) is compared to the shear rupture extent (right column), at different normalized time snapshots. At $\sqrt{4 \alpha t}/L \simeq 1.045$, the normalised fluid over-pressure $\bar{p}/\Delta P$ has nearly reached the external boundaries of the region $L\times L$, while the slipping patch is confined near the injection point (where the over-pressure is maximum and equal to $\bar{p}/\Delta P = 1.0$). Not surprisingly, such limited extent of aseismic rupture occurs in all the different DFNs considered. 
\begin{figure}[t!]
\centering
\includegraphics[width=0.75\textwidth]{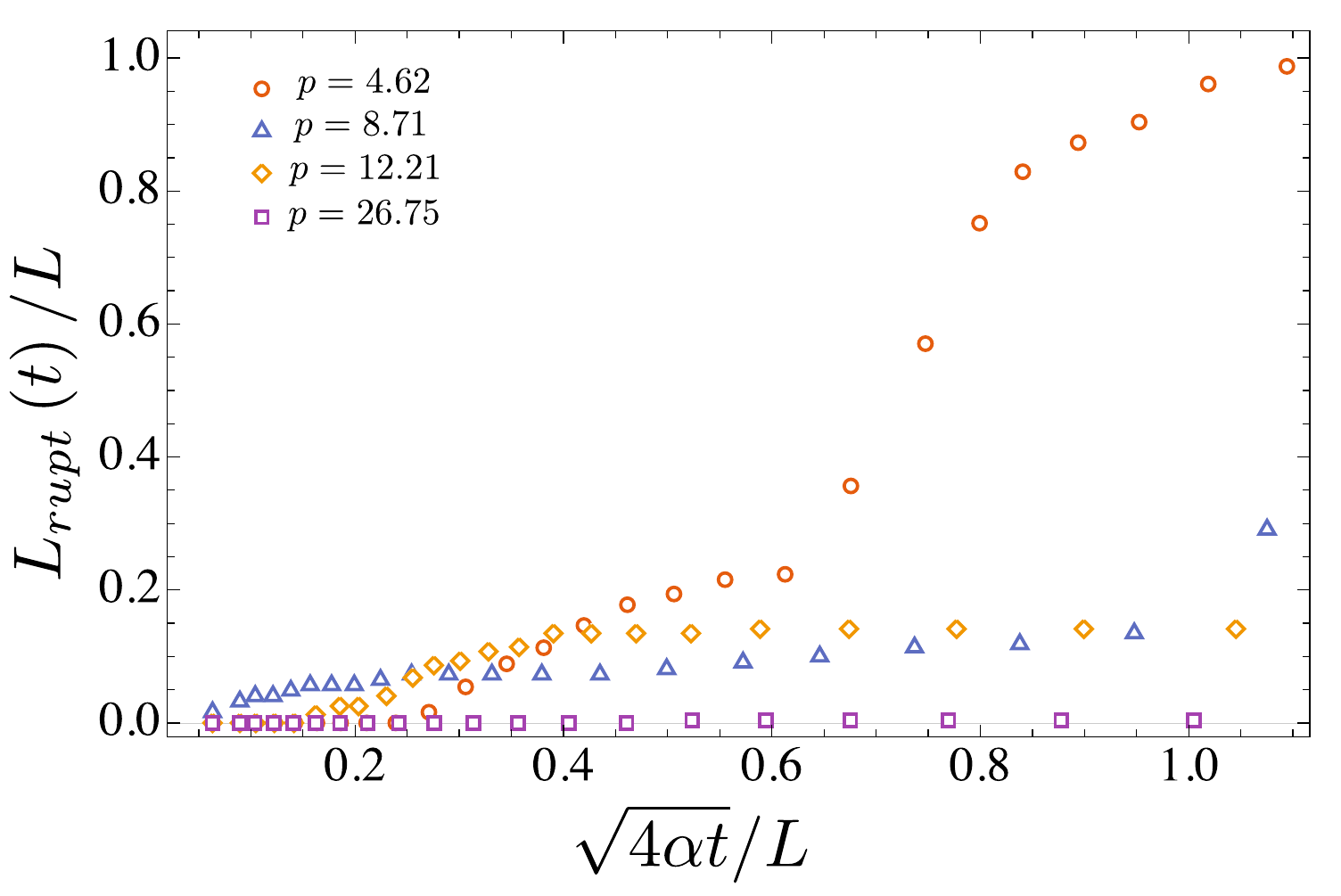}
\caption{Time evolution of the normalized total rupture length $L_{rupt}\left(t \right)/L$ for each DFN realization considered (see Appendix \ref{app:appendix1}), in marginally pressurized conditions. $p$ denotes the percolation parameter (\ref{eq:percolation_parameter}).}
\label{fig:rupture_length_MP}
\end{figure}
Figure \ref{fig:rupture_length_MP}, indeed, shows the time evolution of the total rupture length $L_{rupt}\left( t\right)/L$, defined as the total sum of yielded elements during pressurization, for each DFN realization. We can observe that throughout pressurization the total rupture length is always limited for every value of percolation parameter $p$. Interestingly, at large pressurization time, the rupture length is larger for lower values of DFN percolation. This is because low percolation values promote zones of flow localization where pore fluid over-pressure can increase  during sustained injection. On the contrary, a high percolation degree facilitates the dissipation of pore fluid over-pressure via a rather homogeneous radial diffusion from injection point.
The rupture lengths evolution reported in Figure \ref{fig:rupture_length_MP} actually confirm these considerations: for $p=26.75$, the rupture extent is really localized near injection point (i.e. $\text{max}\left(L_{rupt}\left( t\right)/L \right) \simeq 0.004$) throughout the whole pressurization history, whereas for $p=4.62$ the rupture extent increases in time with the local over-pressure accumulation.

Obviously, these results depend on the level of effective stress anisotropy adopted and the resulting value of $\mathcal{T}_c$. Lower values of $\kappa$ would lead to larger values of $\mathcal{T}_c$ and thus to even more confined aseismic ruptures or - ultimately - to the absence of any ruptures (i.e. when the injection over-pressure is not enough to activate slip). Finally, it is worth mentioning that the only scenario in which the rupture extent might be similar to the extent of the over-pressurization region is when the percolation parameter of the DFN is very very low. In such a case, in fact, fluid flow is most likely confined in a sub-region of the DFN where, upon over-pressure accumulation, it may lead to its complete rupture (if the stress state is critical enough).   

\subsection{Critically stressed conditions}
\label{subsec:CS}
\begin{figure}[t!]
\centering
   \subfloat[$p=4.62$]{\label{fig:angles_ruptures1}
      \includegraphics[width=.4\textwidth]{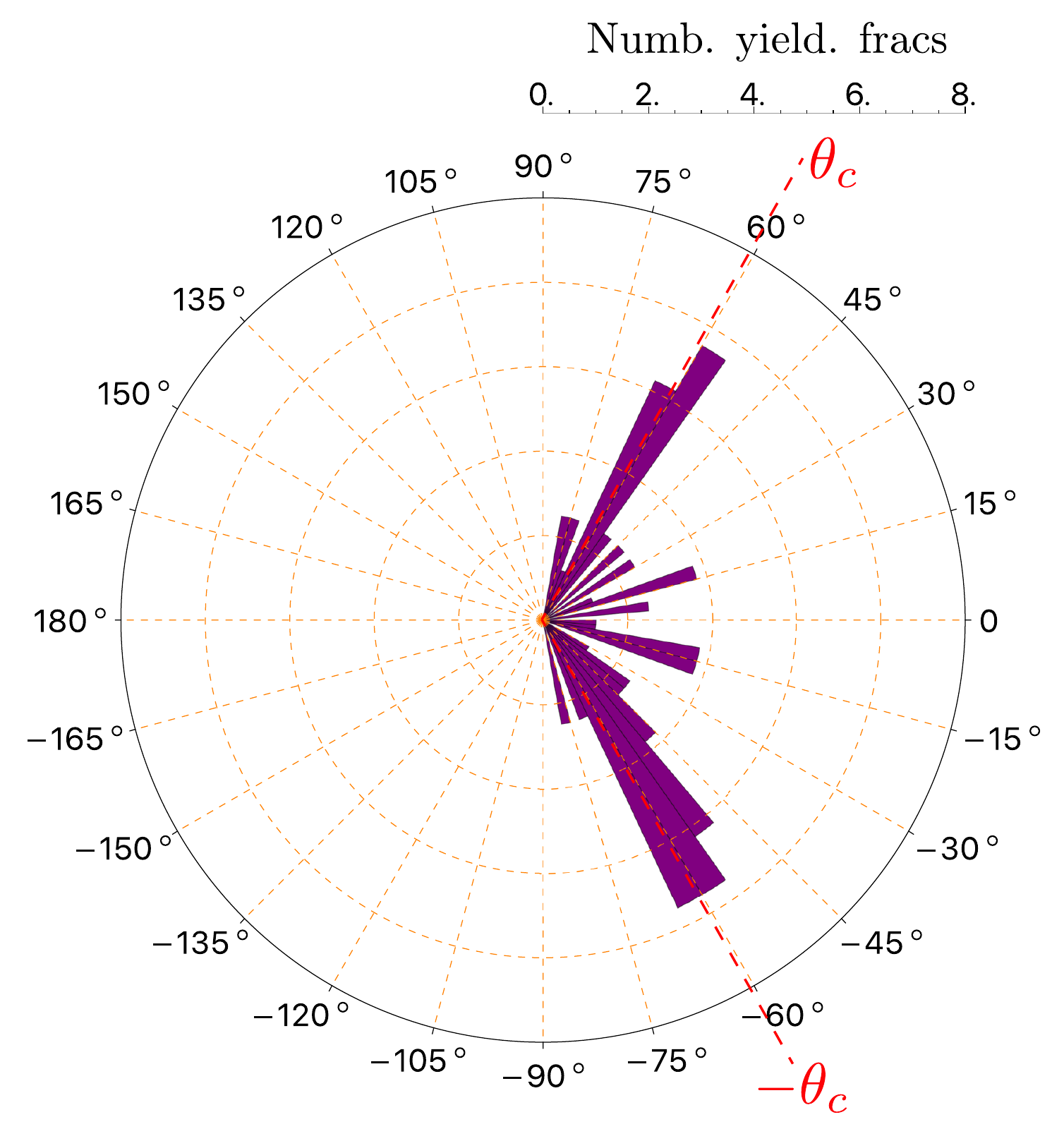}}
~
   \subfloat[$p=8.71$]{\label{fig:angles_ruptures2}
      \includegraphics[width=.4\textwidth]{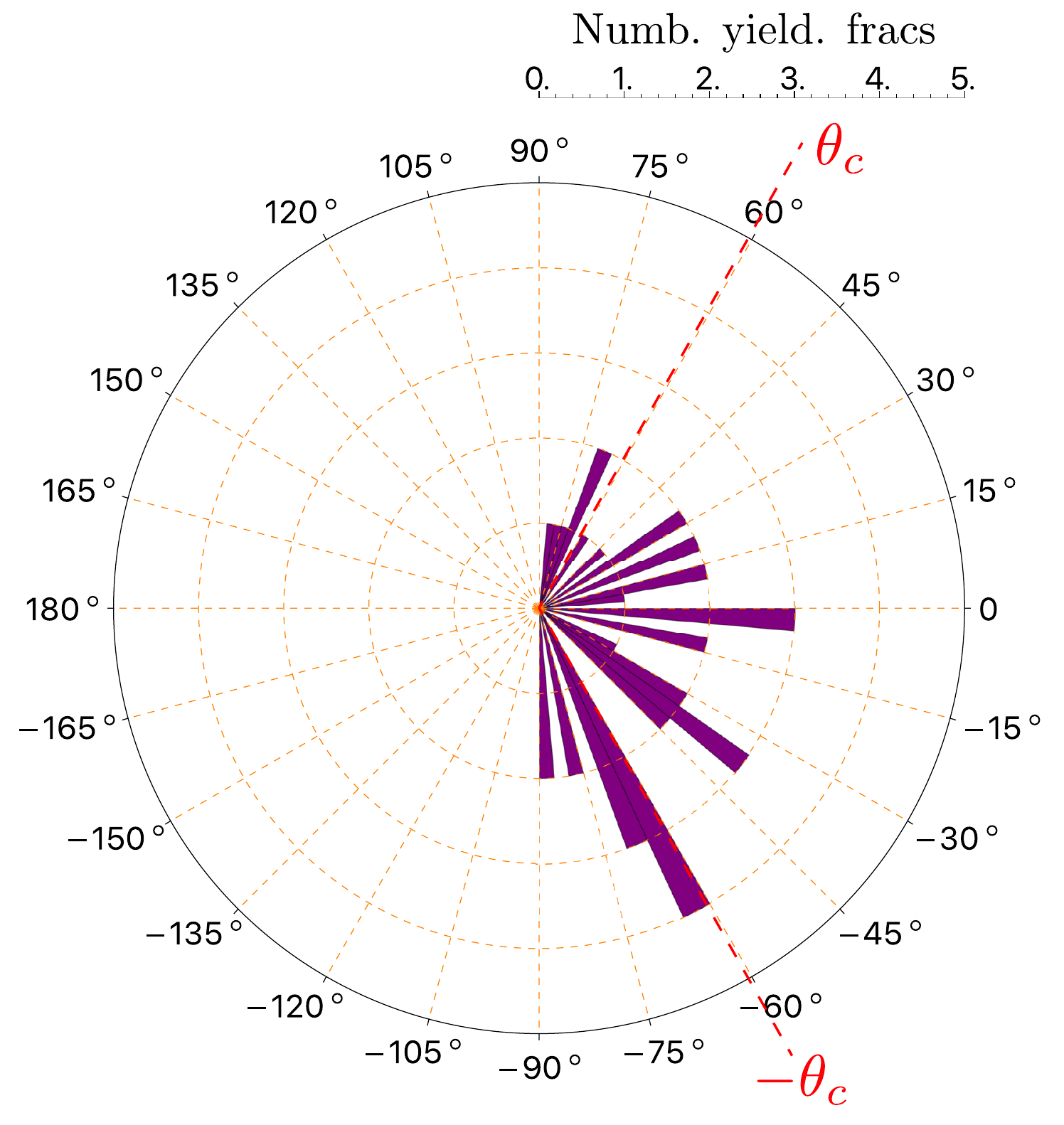}}
\\
   \subfloat[$p=12.21$]{\label{fig:angles_ruptures3}
      \includegraphics[width=.4\textwidth]{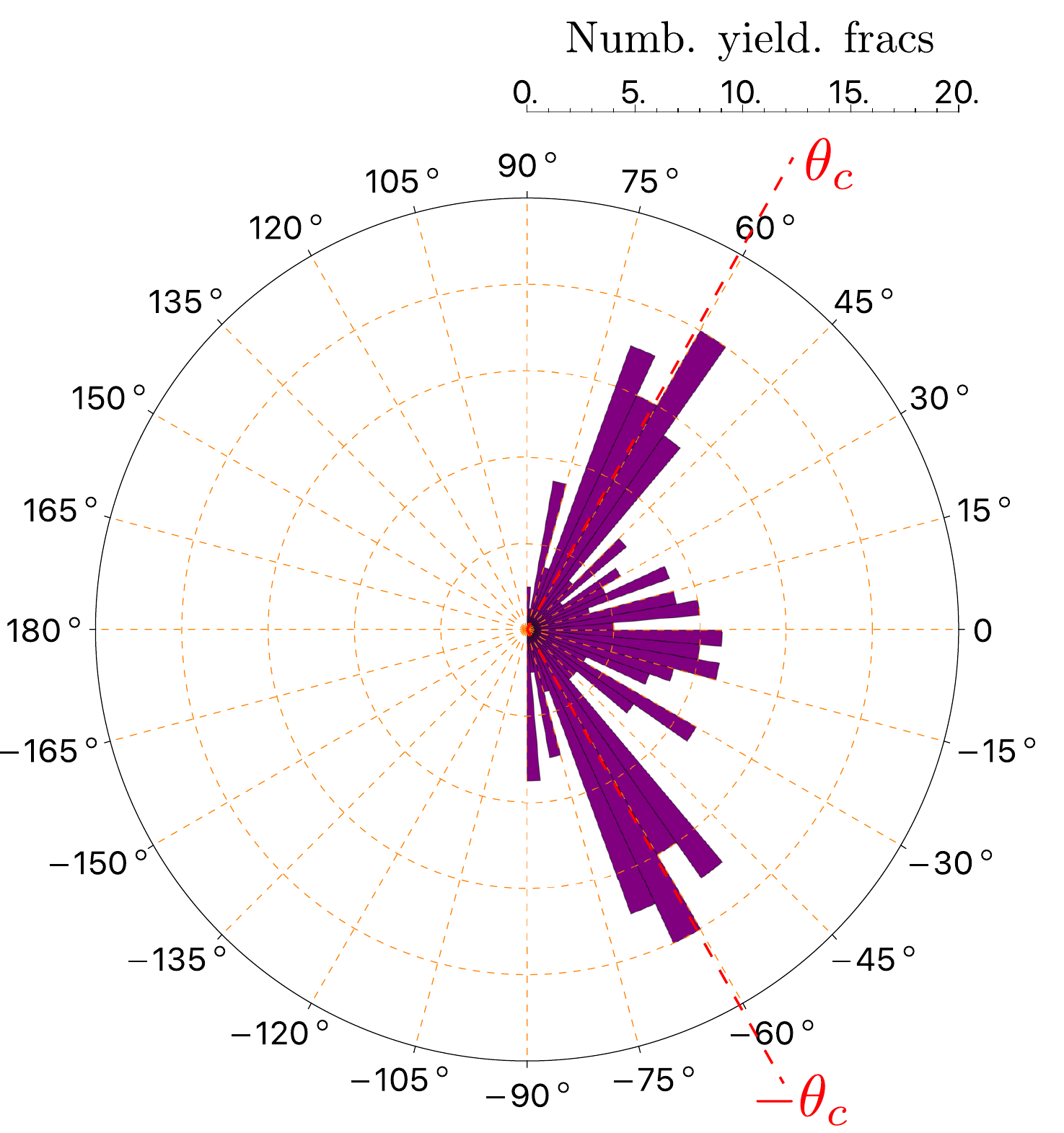}}
~
   \subfloat[$p=26.75$]{\label{fig:angles_ruptures4}
      \includegraphics[width=.4\textwidth]{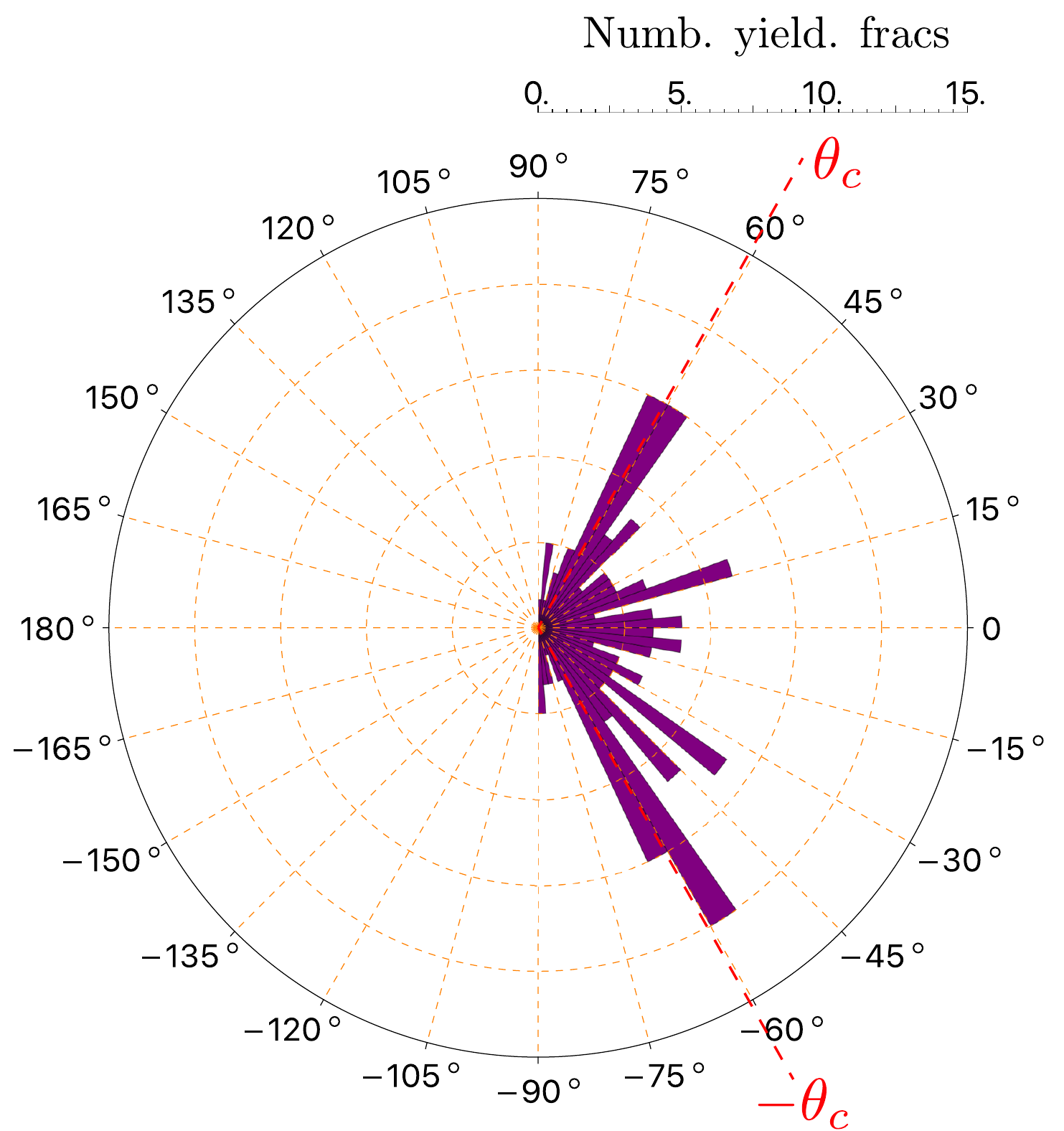}}
   \caption{Angular histograms showing the total number of yielded fractures induced by pressurization, as function of their local orientation $\theta$ for each DFN realization considered. A fracture is considered ``yielded" when at least one of its element has reached the yielding criterion  (\ref{eq:interfacial_law}) throughout pressurization and thus has experienced plastic slip. The dashed red lines denote the critical angle $\pm \theta_c$ (\ref{eq:critical_angle}).}
   \label{fig:angles_ruptures}
\end{figure}

We ran the same numerical experiments (i.e. same DNF realizations, same injection conditions and same locations of injection point), but now with an effective stress anisotropy ratio set to $\kappa = 3$, representative of critically stressed conditions. This implies that all the pre-existing fractures oriented along the critical angle $\theta_c = \pi/4 + \textrm{ArcTan}(f)/2 \simeq 60.5^{\circ}$ are critically stressed and characterized by $\mathcal{T}_c \simeq 0.056$. In other words, they are prompt to fail with little perturbation. Here all the simulations are stopped when the rupture front approached the boundary of the simulation box.\\
\begin{figure}[t!]
\centering
   \subfloat[]{
      \includegraphics[width=.5\textwidth]{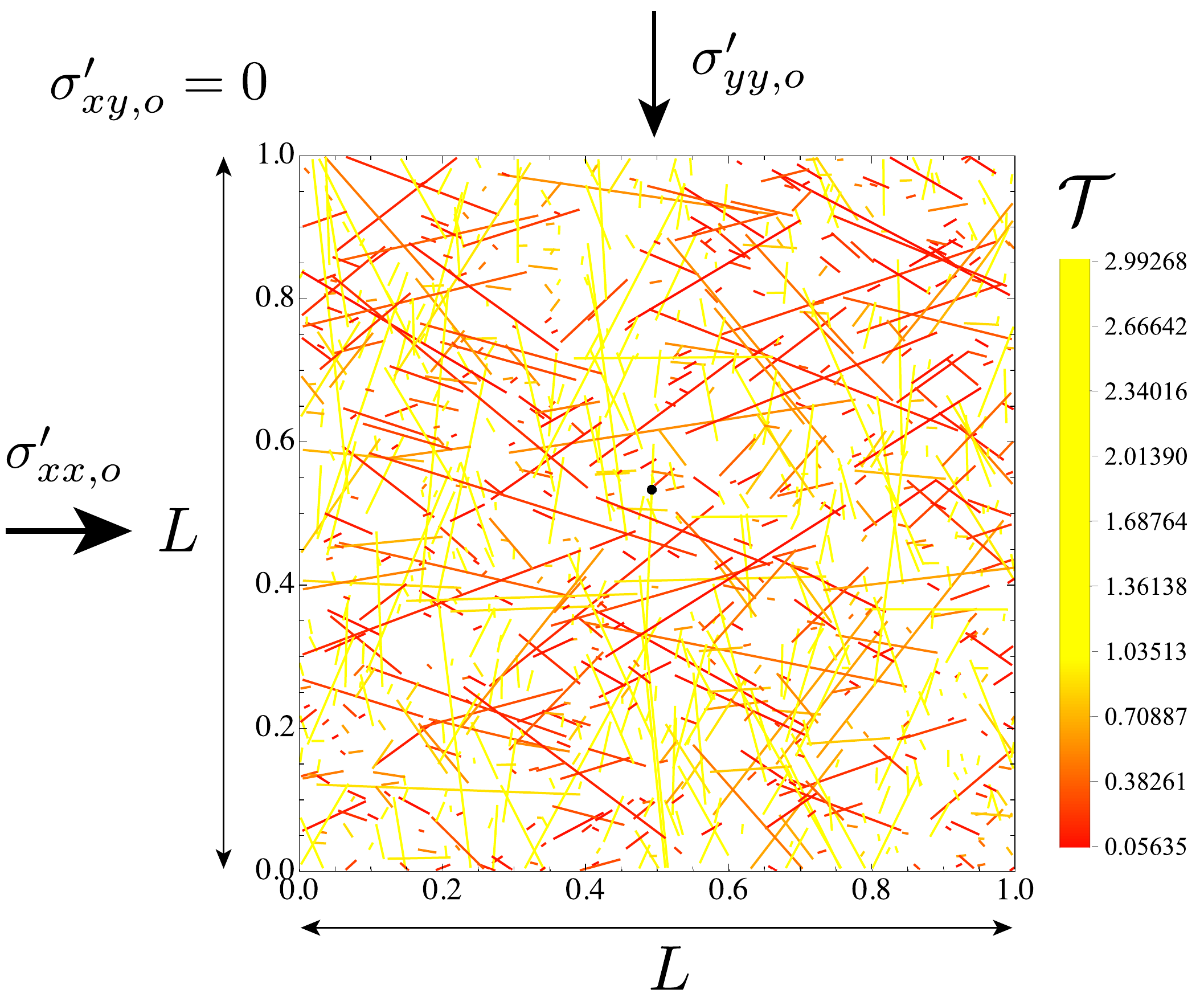}}
~
   \subfloat[]{
      \includegraphics[width=.4\textwidth]{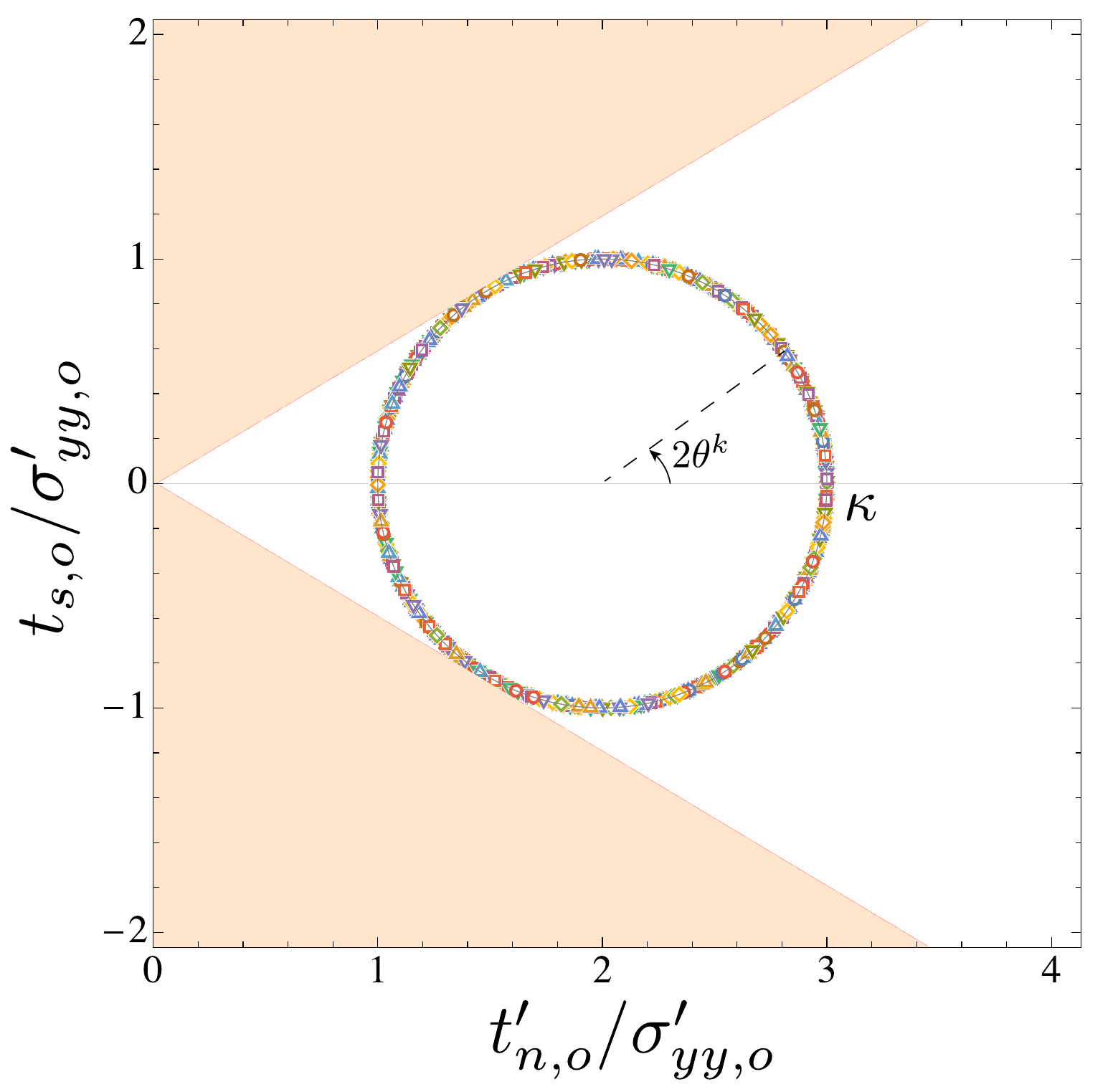}}
      
   \caption{Panel (a): critically stressed Discrete Fracture Network with percolation parameter $p =  12.21$, subjected to an in-situ principal effective stress state. The color of each fracture denotes the corresponding magnitude of the fracture-stress-injection parameter $\mathcal{T}$ prior fluid injection. The black dot, instead, denotes the location of injection point. Panel (b): initial effective stress state in the Mohr-Coulomb plane. Note that the effective tractions are normalized with the minimum far-field principal effective stress $\sigma_{yy,o}^{\prime}$.  
   Along the Mohr-Coulomb circle, whose diameter equals the value of stress anisotropy ratio considered (i.e. $\kappa=2$), all the uniform stress states of each pre-existing fracture are reported. Since all the pre-existing fractures are randomly oriented within the region $L\times L$, the whole circle is uniformly ``covered'' by the fractures' initial stress states.}
   \label{fig:SC2}
\end{figure}
\begin{figure}[t!]
\centering
   \subfloat[$\sqrt{4\alpha t}/L \simeq 0.219 $]{
      \includegraphics[width=.48\textwidth]{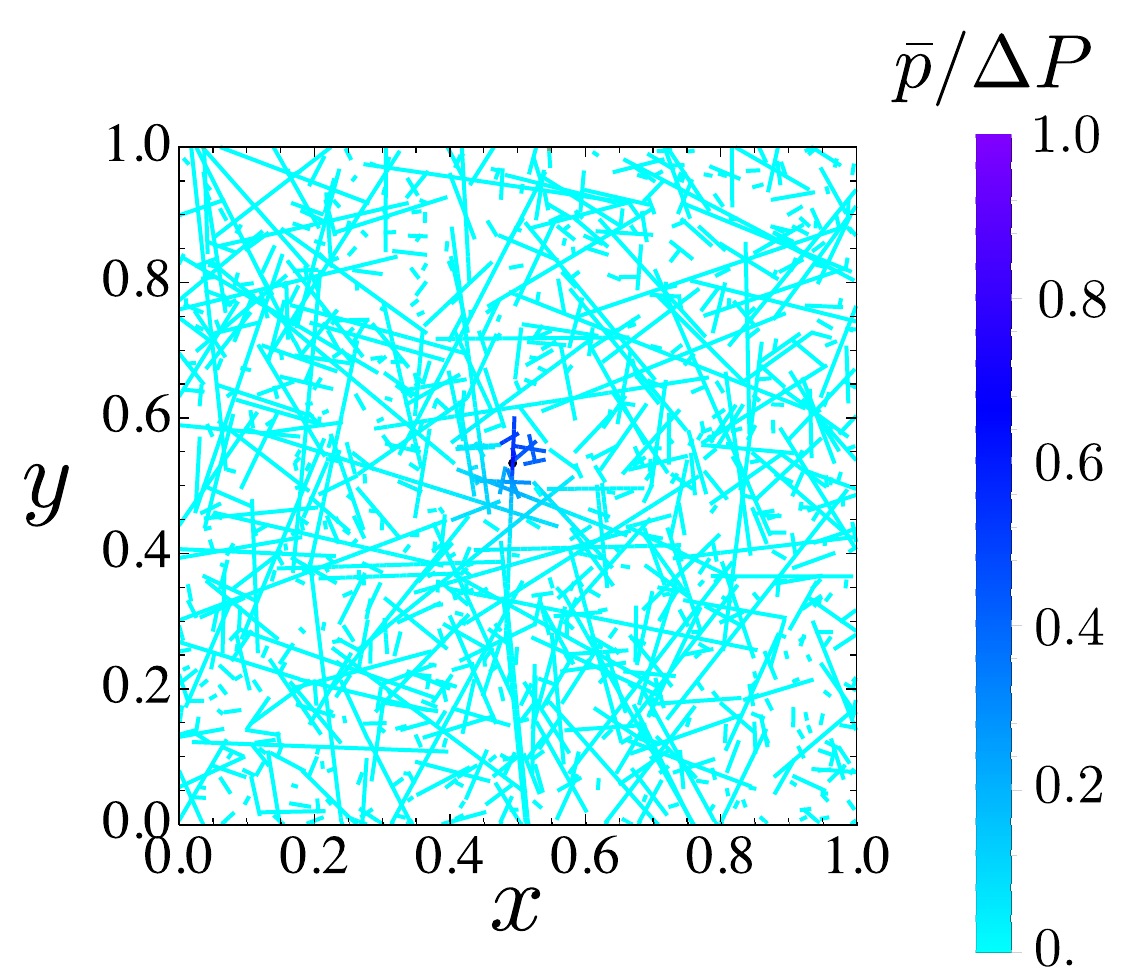}}
~
   \subfloat[$\sqrt{4\alpha t}/L \simeq 0.219 $]{
      \includegraphics[width=.41\textwidth]{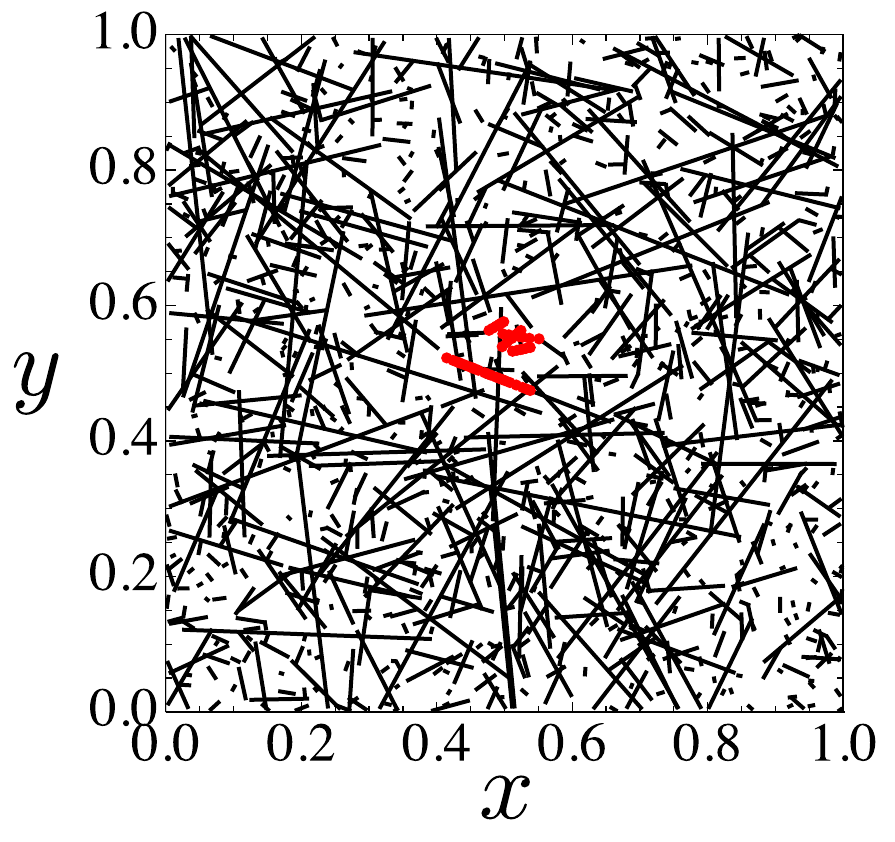}}
\\
   \subfloat[$\sqrt{4\alpha t}/L \simeq 0.596 $]{
      \includegraphics[width=.48\textwidth]{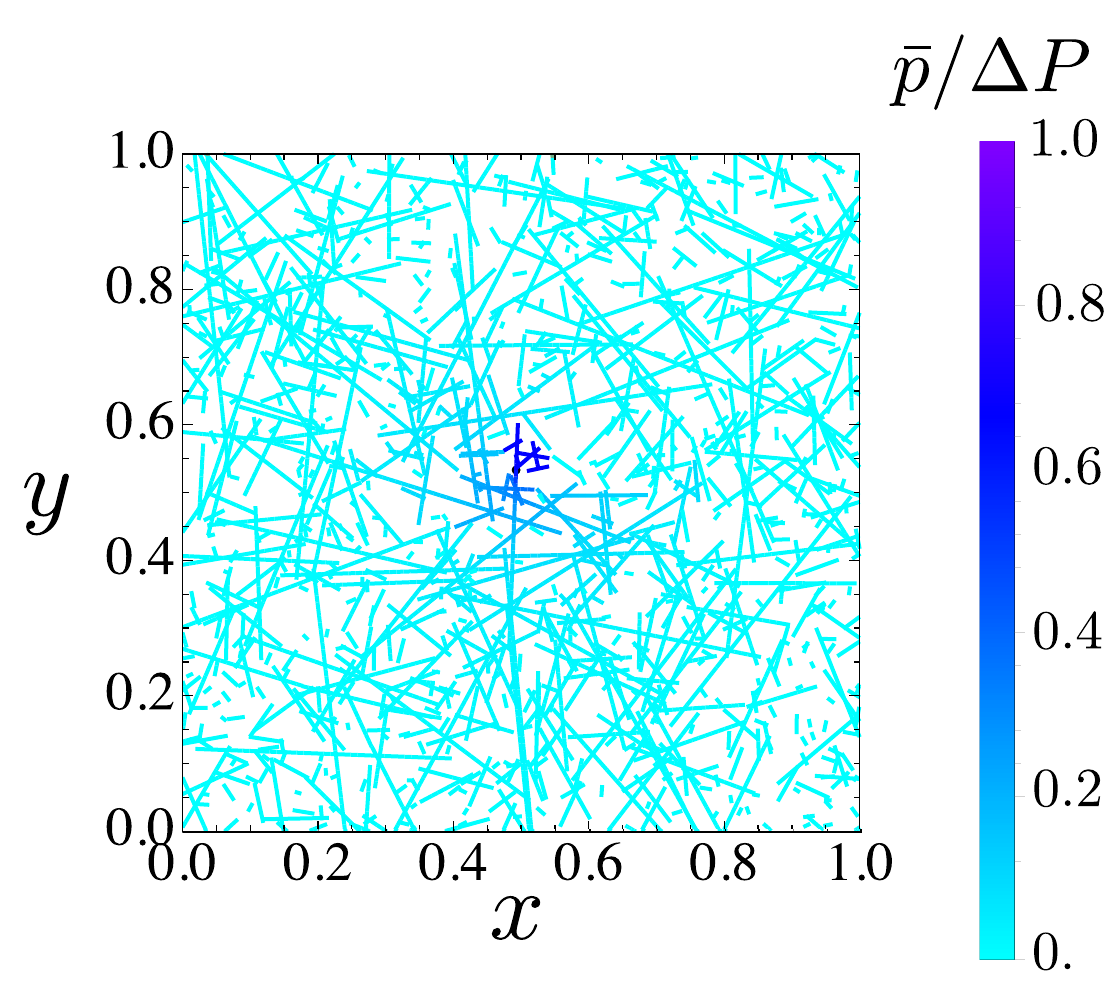}}
~
   \subfloat[$\sqrt{4\alpha t}/L \simeq 0.596 $]{
      \includegraphics[width=.41\textwidth]{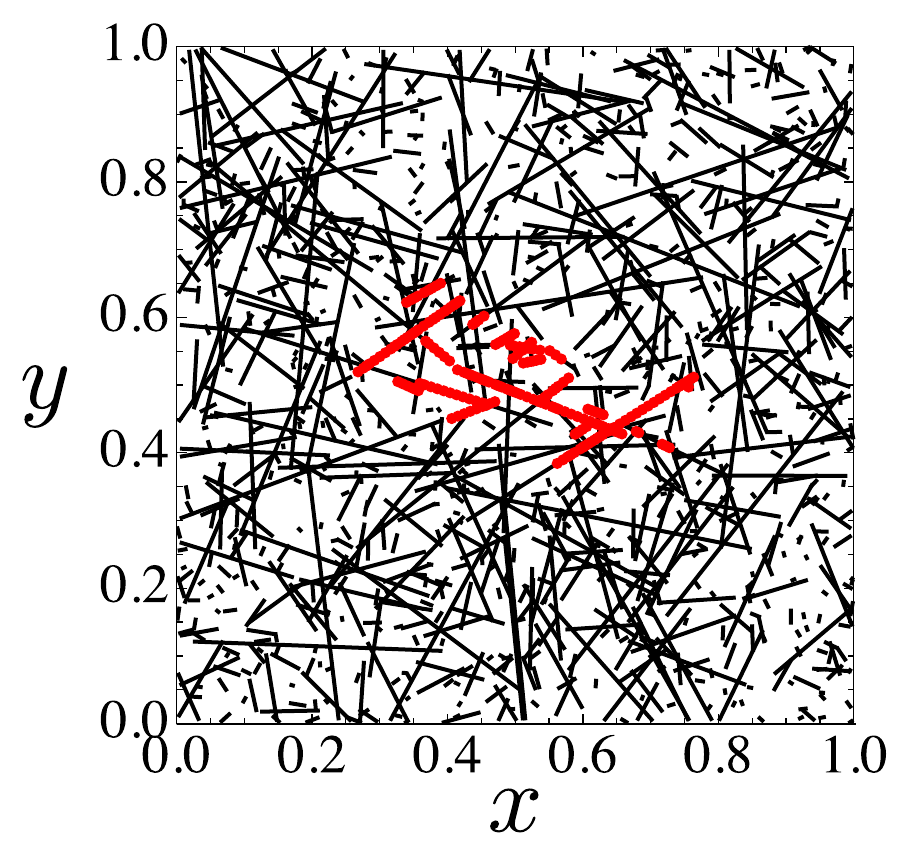}}
\\
   \subfloat[$\sqrt{4\alpha t}/L \simeq 1.021 $]{
      \includegraphics[width=.48\textwidth]{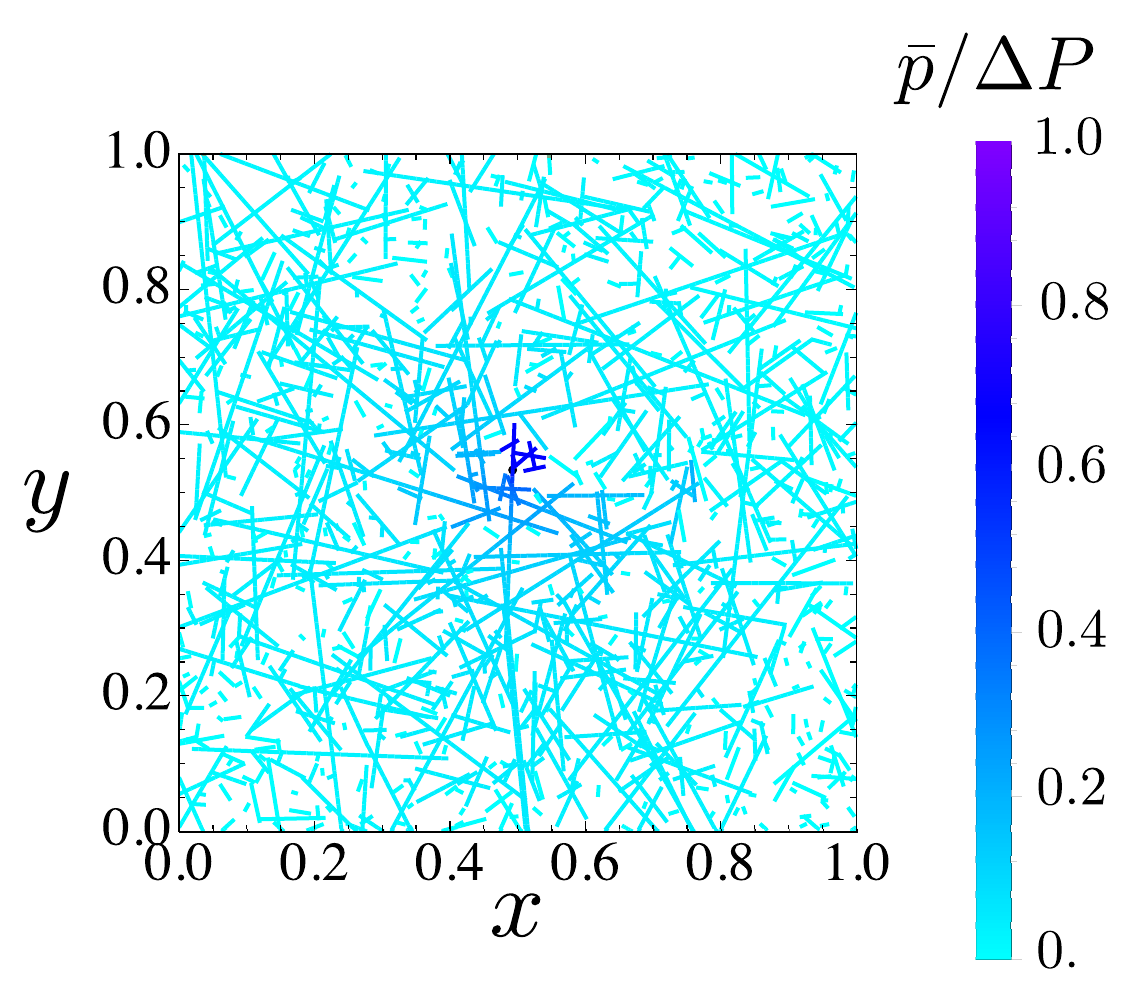}}
~
   \subfloat[$\sqrt{4\alpha t}/L \simeq 1.021 $]{
      \includegraphics[width=.41\textwidth]{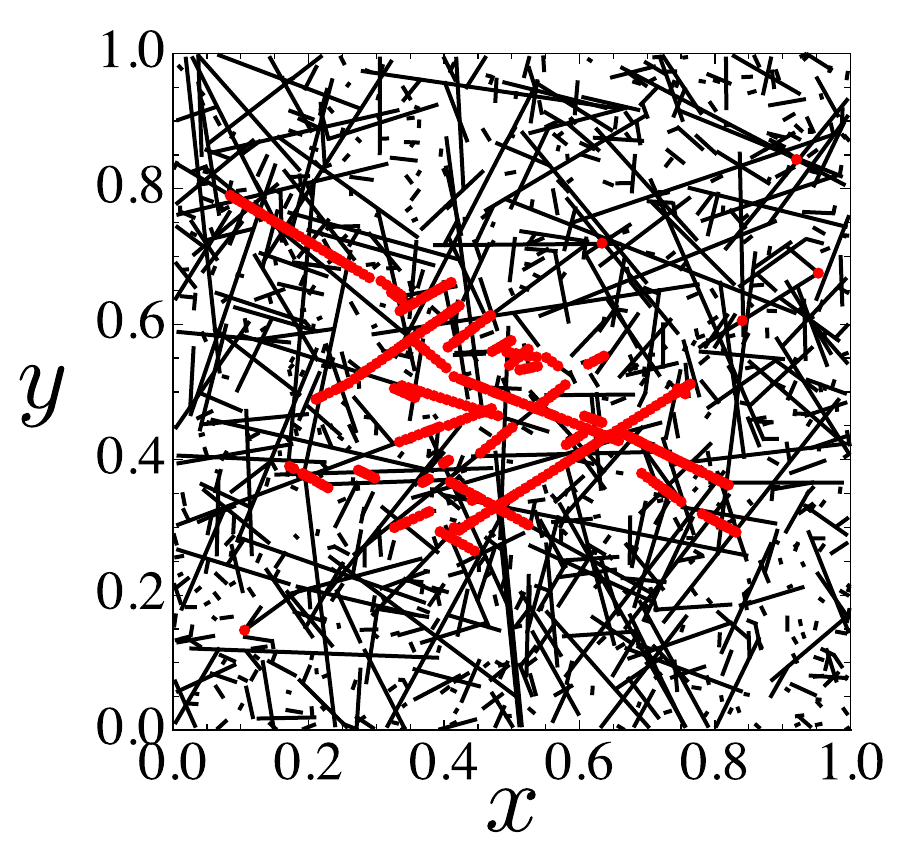}}

   \caption{Evolution of normalized over-pressure $\bar{p}/\Delta P$ (a-c-e) and aseismic rupture extent (denoted by red color - (b-d-f)) along the critically stressed DFN with $p=12.21$, at different normalized time snapshots $\sqrt{4 \alpha t}/L$. Fluid is injected at $\sim\left( 0.493, 0.533 \right)$, in one fracture that is hydraulically connected to many others.}
   \label{fig:overpress&aseismic_patch_CS}
\end{figure}
Restricting to the reference case of injection into the DFN with percolation parameter $p=12.21$, we can observe from Figure \ref{fig:SC2}-a that the fracture in which fluid is injected into is not favourably oriented with respect to far-field stress state: its fracture-stress-injection parameter $\mathcal{T}$ is quite large (albeit the far-field effective stress state is highly critical - see Figure \ref{fig:SC2}-b). Nevertheless, upon fluid injection, pore fluid over-pressure starts to accumulate inside the DFN and activates slip on critically stressed fractures located near injection point (see Figure \ref{fig:overpress&aseismic_patch_CS} at time snapshot $\sqrt{4 \alpha t}/L \simeq 0.219$).   
The activation of these fractures triggers in turn a cascade of other ruptures due to elastic stress interactions between critically stressed pre-existing fractures (denoted by a reddish color in Figure \ref{fig:SC2}-a). The aseismic slipping patch thus propagates outward from injection point  significantly faster than the pressurization front. At normalized time $\sqrt{4 \alpha t}/L \simeq 1.021$, the pore fluid over-pressure zone is still confined to a surrounding of the injection point, while the slipping patch has propagated almost up to the boundaries of the simulation box. 
Their relative position, therefore, is reversed with respect to the case of injection in marginally pressurized fractured rock masses.
Furthermore, such a fast migration of aseismic slipping patch compared to pore fluid diffusion holds regardless whether the DFN is statistically connected or not. 
\begin{figure}[t!]
\centering
\includegraphics[width=0.75\textwidth]{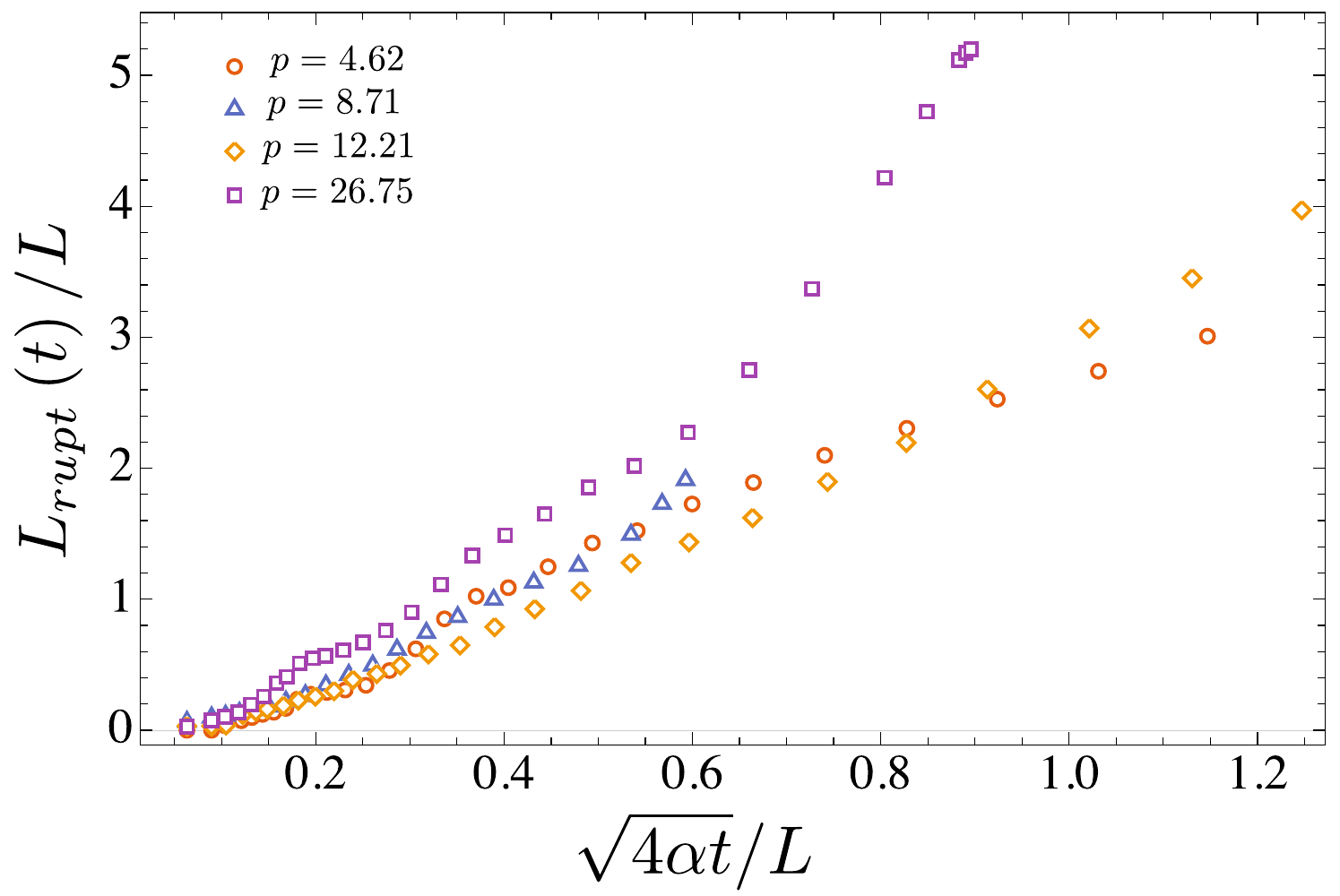}
\caption{Time evolution of normalized total rupture length $L_{rupt}\left(t \right)/L$ for each DFN realization considered (see Appendix \ref{app:appendix1}), under critically stressed conditions. $p$ denotes the percolation parameter (\ref{eq:percolation_parameter}).}
\label{fig:rupture_length_CS}
\end{figure}
Figure \ref{fig:rupture_length_CS} displays the total rupture length as function of pressurization time, for each DFN realization considered. We can  notice that, for each value of percolation $p$, the rupture length increases rapidly and monotonically in time. This suggests that in critically stressed conditions %the percolation degree of a DFN does not influence the hydro-mechanical response as 
the main driving force for the fast aseismic slip propagation is the elastic stress interactions between favourably oriented fractures (i.e. the fractures oriented approximately at $\theta_c \simeq \pm 60.5^{\circ}$ with respect to the direction of $\sigma_{yy,o}^\prime$).
This is further strengthened by looking at the rose plots reported in Figure \ref{fig:angles_ruptures}, which display the angular histograms of the yielded fractures\footnote{Note that one pre-existing fracture is considered ``yielded" if at least one of its finite element has reached the yielding criterion (\ref{eq:interfacial_law}) throughout pressurization and thus has experienced plastic slip.}. 
For each DFN realization, the majority of yielded fractures are oriented approximately along the critical angle $\theta_c$, which, for the value of friction coefficient considered here, is approximately $\pm 60.5^{\circ}$ (\ref{eq:critical_angle}). 

\subsection{Aseismic moment}
\begin{figure}[t!]
\centering
    \subfloat[]{
      \includegraphics[width=.5\textwidth]{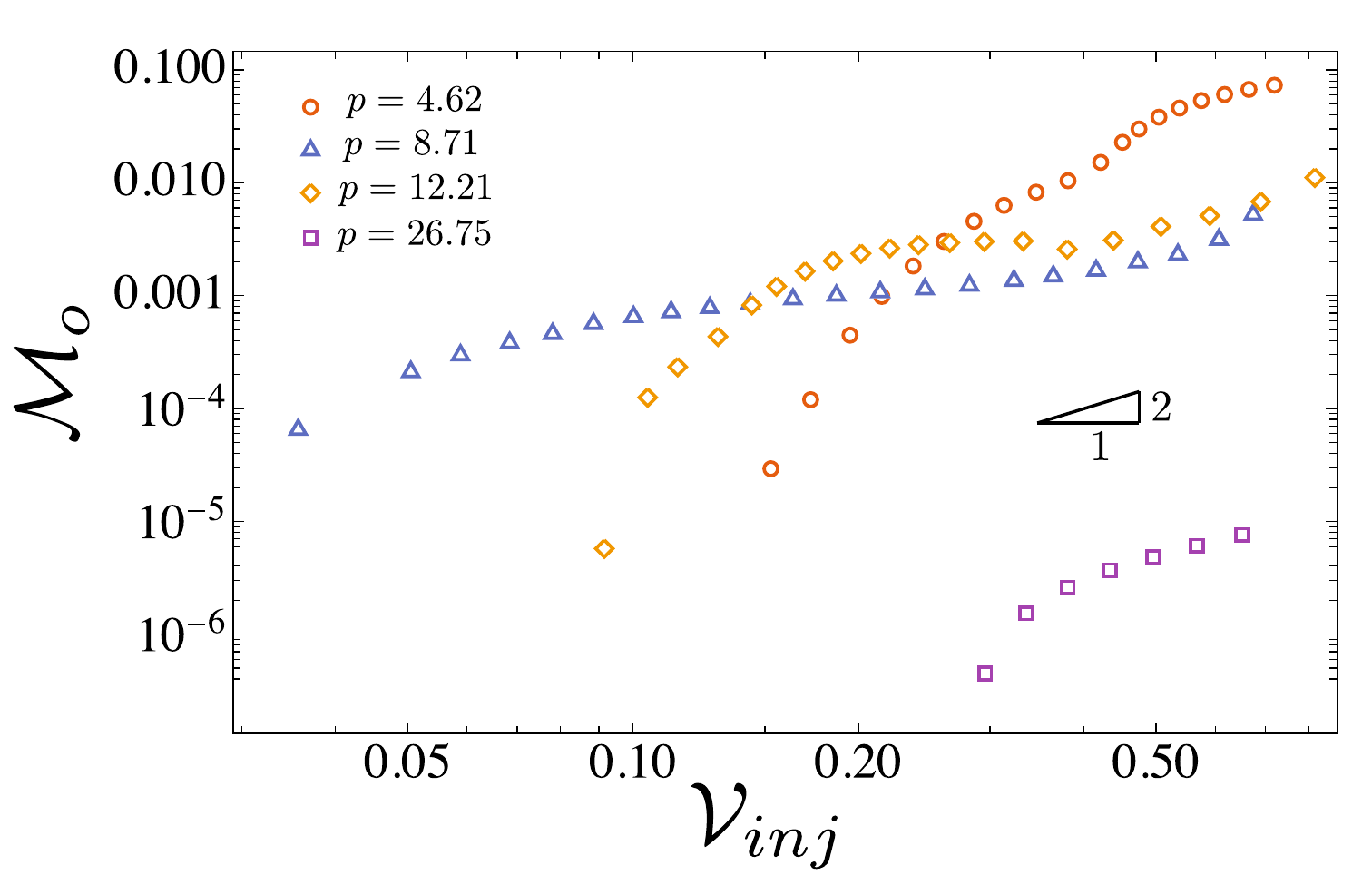}}
~
    \subfloat[]{
      \includegraphics[width=.5\textwidth]{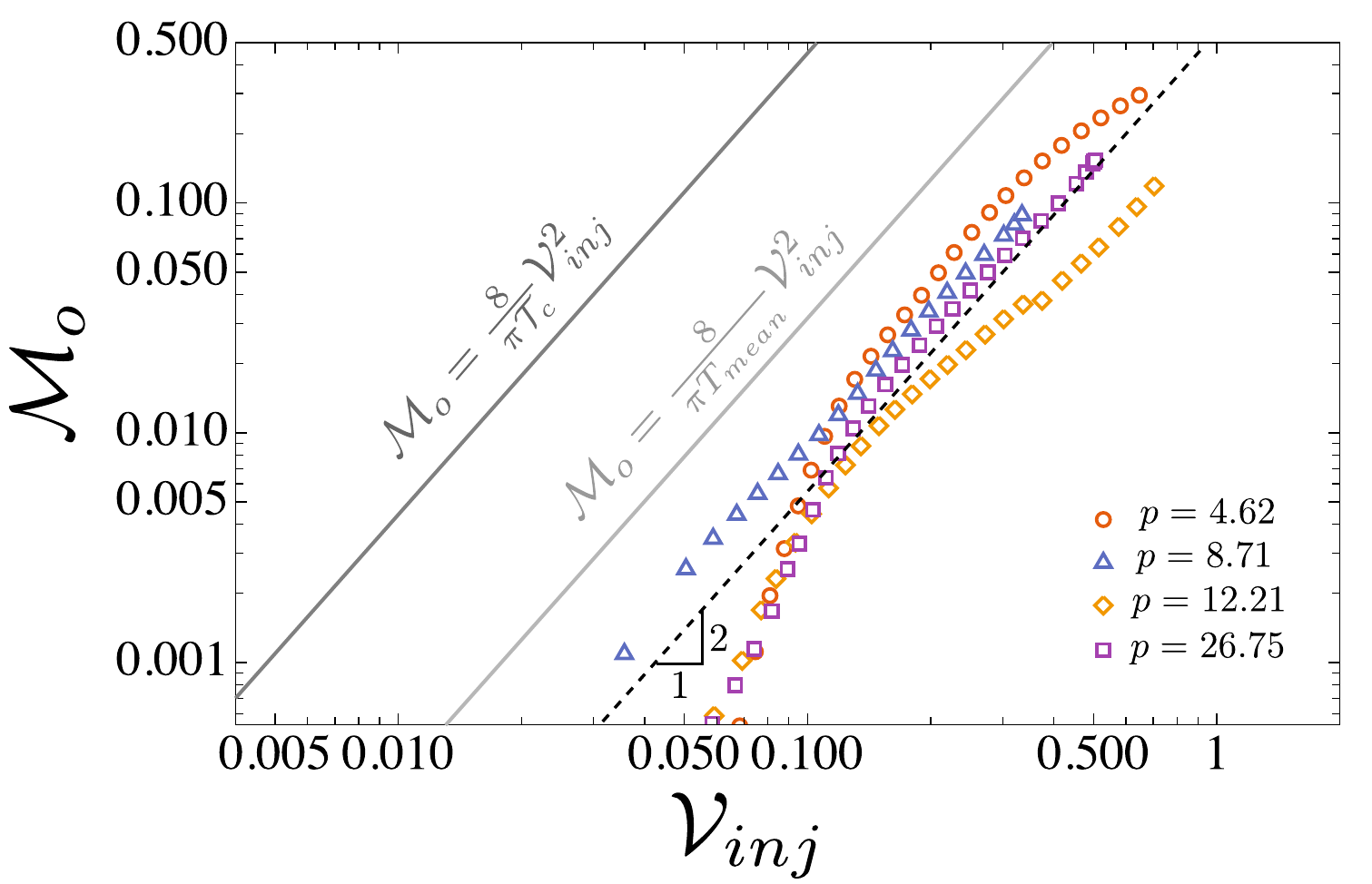}}
\caption{
Log-log plots that display the evolution of normalized aseismic moment with accumulated injected volume, under marginally pressuized conditions (left - $\mathcal{T}_c \simeq 0.528$) and critically stressed conditions (right - $\mathcal{T}_c \simeq 0.056$). The dark gray line represents the analytical  solution (\ref{eq:aseismic_moment_vs_volume_dimensionless})-b valid for a single planar shear fracture with fracture-stress-injection parameter equal to $\mathcal{T}_c \simeq 0.056$, while the light gray line is valid for a single planar shear fracture with fracture-stress-injection parameter equal to the mean value of $\mathcal{T}$ parameters of all the yielded fractures in each realization.}
\label{fig:aseismic_moment_inj_volume}
\end{figure}

Finally, we report how the aseismic moment vary with accumulated injected volume in case fluid is injected in a marginally pressurized or critically stressed DFNs. Referring to the same numerical results presented and discussed in Sections \ref{subsec:MP} and \ref{subsec:CS}, Figure \ref{fig:aseismic_moment_inj_volume} displays the evolution of scaled $M_o(t)$ as function of normalized $V_{inj}(t)$ in log-log plots, for all the DFN realizations considered and for the two values of critical $\mathcal{T}$ parameter: $\mathcal{T}_c \simeq 0.528$ (left - marginally pressurized) and $\mathcal{T}_c \simeq 0.056$ (right - critically stressed). Note that the same scaling has been adopted for both plots.

We can readily observe that the aseismic moment scales to the injected volume as $M_o(t) \propto V_{inj}^2$ in both limiting conditions and regardless the percolation degree of the DFN (see the slopes of each curve that tend to be similar the one of the hypothenuse of the black triangle). In marginally pressurized conditions, however, such a scaling is attained at large injected volumes, with the percolation parameter that affects the factor of proportionality: larger values are obtained at lower percolation degrees due to pore-fluid localization (consistently with the results described in Section \ref{subsec:MP}). In critically stressed conditions, instead, the aseismic moment tends to scale quadratically with injected volume as soon as slip is activated. Moreover, the subsequent fast migration of aseismic slip due to elastic-stress interactions of favourably oriented fractures tends to result in a constant factor of proportionality, implying that there is a family of solutions that does not depend on the percolation parameter (unlike in marginally pressurized conditions). It is interesting to note that such a factor is lower than the one obtained as if the injection would occur in a single planar fracture, with fracture-stress-injection parameter equal to $\mathcal{T}_c$ (see dark gray line) or equal to the mean value of $\mathcal{T}$ parameters of all the yielded fractures in each realization (see light gray line). This suggests that the cascade of ruptures driven by elastic stress transfer leads to a global ``shielding" effect such that the total accumulated slip on a DFN with uniformly distributed and randomly oriented fractures is lower than the corresponding one on a single planar fracture (for the same injection and in-situ conditions).       

\section{Discussions}
\label{sec5}
\subsection{Comparison between injection in a single fracture and in a Discrete Fracture Network}
The problem of injection in a 2-dimensional Discrete Fracture Network with randomly oriented and uniformly distributed fractures described in the previous section shares a lot of similarities with the simpler case of injection in a planar 2D shear fracture (reported in Section \ref{sec3}). Here, we describe them and highlight their implications to injection-induced seismicity.\\
Although the two problems are  different, the mechanisms that govern the evolution of aseismic slip driven by pore-fluid diffusion are the same, with a single dimensionless parameter that govern the overall hydro-mechanical response. 
In the case of injection at constant pressure in a single fracture, we have showed that injection-induced aseismic slip is self-similar in a diffusive manner, with a self-similarity factor that is only function of the fracture-stress-injection parameter $\mathcal{T}$ (\ref{eq:T_parameter}). We have demonstrated that when the $\mathcal{T}$ parameter is large, the aseismic rupture extent lags the pressurization region (marginally pressurized conditions). Vice-versa, when $\mathcal{T}$ parameter is low, slip migrates from injection point much faster than pore-fluid diffusion, with such a quick migration controlled by the stress transfer of the propagating rupture (critically stressed conditions). These scenarios are exactly the same in the case of injection in a marginally pressurized or critically stressed DFN in a global sense, albeit the self-similarity condition is not strictly valid anymore and the governing parameter is the $\mathcal{T}$ parameter evaluated at the critical orientation $\theta_c$ (\ref{eq:critical_angle}). This implies that micro-seismicity in deep fractured reservoirs, commonly observed to correlate well with a pore fluid diffusion process and thus thought to migrate away from injection point proportional to the square-root of time \citep{ShHue97, ShRo02, PaRo03, HaOg05, HaFi12, AlbOy14}, might instead be driven by elastic stress interactions between critically stressed fractures. This would notably be the case when $\mathcal{T}_c$ is low in which aseismic slip always outpaces pore pressure front and may trigger micro-seismicity by stress transfer way ahead of the actual location of the pore-pressure disturbance. 
%Note that these considerations have been also observed in three-dimension, although for a single isolated fracture/fault yet \citep{SaLe22}.  % may be not needed to be repeated here
As a result, the fit of micro-seismicity data (via space-time migration plots)  to infer hydraulic diffusivity $\alpha$ may result in a large over-estimation of the latter. %, as $\lambda\gg 1$ for small value of  $\mathcal{T}$.

A number of efforts have been made to understand how the magnitude of induced seismic events vary with operational parameters, such as the total volume of fluid injected \citep{Mc14, GaAm17} or applied injection conditions: constant injection rate / constant over-pressure \citep{GaGe12} or ramp-up of injection rate commonly present in many injection protocols \citep{CiRi22}.
These studies are indirectly devoted to understand how to constraint the maximum seismic moment and thus reduce the seismic risk. Although they are of great importance for the design of injection operations, in some cases the relevant contribution to injection-induced moment release in fractured reservoirs may be aseismic \citep{McBa18}. It is therefore significant to understand how the aseismic moment evolve in function of operational parameters (such as total volume of fluid injected) when the fractured rock mass is critically stressed or marginally pressurized. 
Our results show that in both limiting regimes the aseismic moment on a DFN scales quadratically to the injected volume, for an injection at constant over-pressure under plane-strain conditions. The same power-law scaling has been found to be valid for a single planar shear fracture subjected to the same injection condition (see Section \ref{subsec2}). 
Finally, it is interesting to note that the DFN response in critically stressed conditions is similar to the one of a single planar fracture: the aseismic moment is not significantly affected by geometrical connectivity and scales as $\propto V_{inj}^2$ with factor of proportionality that is lower than the corresponding one obtained on a single fracture (supposed to be favourably oriented with respect to in-situ stress conditions or supposed to have a fracture-stress-injection parameter $\mathcal{T}$ equal to an average value of all the $\mathcal{T}$ parameters associated with the yielded fractures). A possible explanation for this is the shielding effect between interacting fractures, in a similar way as observed in cases of hydraulic fracturing in naturally fractured rocks (e.g. \citep{WaTe87}).  

\subsection{Modelling assumptions}
Fluid injection in a deep fractured rock mass  is a complex hydro-mechanical problem. 
Large non-linearities arising from the coupling between hydraulic and mechanical deformations make the problem  challenging  from a computational point of view and, at the same time, difficult to get insight into.  
In order to properly explore and understand the basic physical mechanisms that take place at depth upon injection, we have made a number of simplifying assumptions  which nevertheless provide a first order model. Notably, we have assumed that all the pre-existing fractures in the DFN have a uniform hydraulic transmissivity $w_h k_f$ at ambient conditions, although field observations have shown that critically stressed fractures (low $\mathcal{T}$ parameter) act most likely as high permeable fluid conduits compared to not optimally oriented fractures \citep{BaLu83} and that larger apertures are expected on longer fractures (following a sub-linear power law) \citep{Ols03}. 
We can certainly say, however, that an initial heterogeneous distribution of hydraulic transmissivities would not change the DFN hydro-mechanical response in critically stressed conditions, as the driving mechanism for the fast slip propagation is the elastic stress interactions of favourably oriented fractures. In marginally pressurized conditions, instead, it may have an impact due to its effect on pore-fluid diffusion (yet still within the rupture extent that lags the pressurization region). This would occur even in the case where shear-induced dilatancy / compaction is included, such that the fracture hydraulic transmissivities are not only heterogeneous at in-situ conditions, but also evolve in space and time in relation to frictional slip.

A more complex response is surely obtained if the friction coefficient weakens during slip propagation, thus allowing the spontaneous nucleation of dynamic (seismic) ruptures. For instance, in the case of slip weakening friction coefficient and shear-induced dilatancy / compaction of pre-existing fractures, we would expect a faster migration of slip front with respect to fluid front on a critically stressed DFN (especially when run-away ruptures take place), even when dilatant hardening effect is able to quench the dynamic events. This is also valid in the limit case of fluid injection into a critically stressed dilatant fracture/fault that is hydraulically disconnected from the other pre-existing fractures \citep{CiLe19}. %and for which the results of \cite{CiLe19} are well-established. 
On a marginally pressurized DFN, instead, the weakening of friction coefficient with shear deformations may lead to finite-sized dynamic events, even along highly dilatant pre-existing fractures (due to the slow rupture front velocity upon fluid injection). At early pressurization time, the pore-pressure front would always be located ahead the slipping patch front, but it may be outpaced by the rupture front after  nucleation and arrest of such seismic events (when occurring). Obviously, this scenario strongly depends on the percolation parameter, DFN characteristics and injection condition. Further numerical investigations are therefore needed.  

\section{Conclusions}
\label{sec6}
We have studied the propagation of stable frictional slip ruptures in fractured rock masses driven by pore-fluid diffusion. Before exploring  fluid injection in a Discrete Fracture Network with randomly oriented and uniformly distributed fractures, we revisited the problem of injection in a single isolated frictional shear fracture, first investigated by \citet{BaVi19,Vi21}. For constant friction, the stable (aseismic) slip propagates with pore-fluid diffusion in a self-similar manner. The self-similarity factor is function of only one dimensionless parameter $\mathcal{T}$ that identifies two end-member limiting regimes, namely critically stressed ($\mathcal{T}\to0$), associated with fast migration of slip compared to pressurization front, and marginally pressurized ($\mathcal{T}\to1$) regime (reversed scenario).    
We have solved numerically the problem and extensively benchmarked our results against the analytical asymptotic solutions of \citet{BaVi19} and \cite{Vi21}. As demonstrated in Figure \ref{fig:res_viesca1}, our numerical results are very accurate, with a relative error in terms of crack half-length that is always lower than $2\%$ in both critically stressed and marginally pressurized limiting conditions.
Furthermore, we have derived closed form solutions for the aseismic moment evolution with pressurization time and total injected volume in both limiting regimes, and compared them with numerical results obtaining a very good agreement. An interesting finding is that the aseismic moment scales quadratically to the injected volume in both critically stressed and marginally pressurized regime.

Similar considerations and similar limiting regimes have been found to be valid even in the case fluid is injected in one fracture that is hydraulically connected to many others, forming a Discrete Fracture Network (with no preferred orientation). Although the self-similarity between the evolution of rupture extent and pore fluid front does not strictly hold anymore, the hydro-mechanical response is at first order governed by the fracture-stress-injection parameter $\mathcal{T}$ evaluated at the critical orientation $\theta_c = \pm \left(\pi/4 + \textrm{ArcTan}(f)/2 \right)$. % from the min. or max principal stress 

In marginally pressurized conditions ($\mathcal{T}_c \gtrsim 1$), the frictional ruptures are solely contained within the region of pore-pressure disturbance. As a result, in that limit, the percolation number of the DFN affects to first order the response of the medium. 
On the other hand, in critically-stressed conditions ($\mathcal{T}_c\ll 1$), a cascade of ruptures occurs on critically oriented fractures driven by stress transfer at distances largely ahead of the pressurized region. 
As a result, if aseismic slip is the dominant mechanism for the triggering of micro-seismicity, the estimation of hydraulic diffusivity $\alpha$ from the spatio-temporal pattern of micro-seismicity are grossly over-estimated. 
The aseismic moment also scales as $\propto V_{inj}^2$ in both limiting regimes albeit with a different coefficient of proportionality compared to the single fracture case. Indeed, this coefficient of proportionality  clearly depends on the DFN characteristics in the marginally pressurized case (as the rupture tracks the flow and is affected by percolation to first order). In the critically stressed case, this pre-factor appears to be only mildly dependent on the DFN properties. 

From the results presented in this contribution we can conclude that, on a DFN with randomly oriented and uniformly distributed fractures, stable slip propagation activated by an injection at constant over-pressure and driven by pore fluid diffusion is qualitatively similar as if slip would propagate in a single planar fracture under the same injection condition. Future works should combine scaling analysis with numerical simulations to investigate rock masses with non-random fractures orientation (joint sets), quantify the effect introduced by variation of hydraulic properties with slip and the effect of linkage between fractures via the propagation of wing cracks \citep{Jung13}.

%Furthermore, if aseismic slip is the dominant mechanism for the triggering of (micro-)seismicity as in the case of critically stressed fractures rock masses, the estimates of global hydraulic diffusivity $\alpha$ from the spatio-temporal pattern of micro-seismicity should not be done using the classical power law that characterizes pore fluid diffusion processes.  

\section*{Declarations}

\subsection*{CRediT author statement}
\textbf{Federico Ciardo}: Conceptualization, Methodology, Software, Validation, Formal analysis, Writing - Original Draft, Visualization. \textbf{Brice Lecampion}: Conceptualization, Methodology, Software, Validation, Formal analysis, Writing - Review \& Editing.

\subsection*{Competing interests}
The authors have no competing interests to declare that are relevant to the content of this article.

\subsection*{Funding}
F. Ciardo was supported by the Swiss Federal Office of Energy (SFOE) through the ERANET Cofund GEOTHERMICA (Project No. 200320-4001), which is supported by the European Union’s HORIZON 2020 programme for research, technological development and demonstration. B. Lecampion acknowledges funding for the EMOD project (Engineering model for hydraulic stimulation) via a grant (research contract SI/502081-01) and an exploration subsidy (contract number MF-021-GEO-ERK) of the Swiss federal office of energy for the EGS geothermal project in Haute-Sorne, canton of Jura.

\begin{appendices}
\section{DFN realizations}
\label{app:appendix1}

\begin{figure}[t!]
\centering
\includegraphics[width=0.8\textwidth]{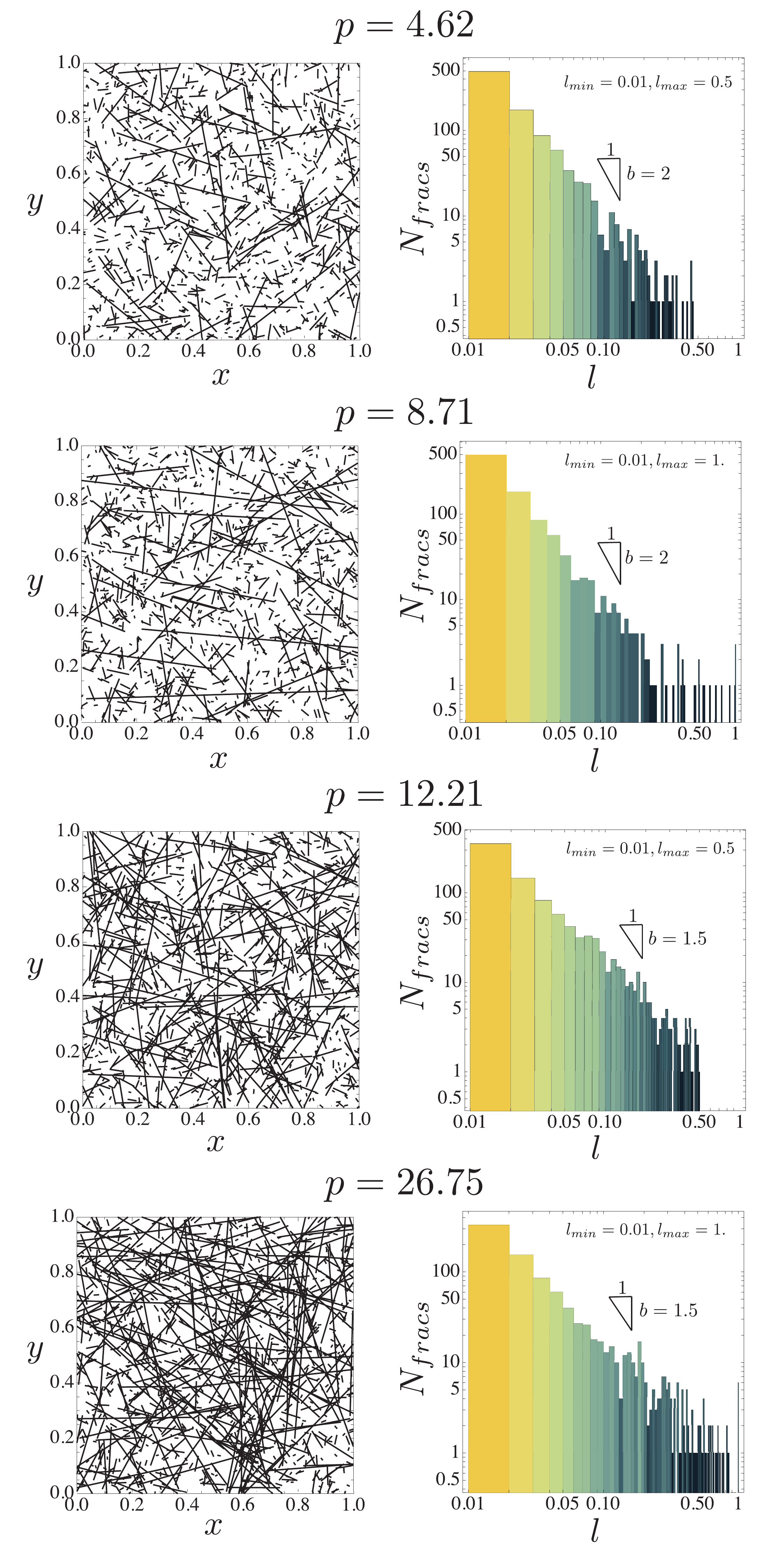}
\caption{Discrete Fracture Networks realizations used in our numerical investigations. The fractures length is sampled from power law distribution (\ref{eq:pdf_lengths}), with fixed minimum length $l_{min} = 0.01 L$ (with $L$ being the characteristic length of the computational domain) and different maximum length $l_{max}$ and power law exponent $b$ in order to obtain different percolation degrees. The left column displays the trace of each DFN, and the right column the histograms of the fractures length.}
\label{fig:DFN_realizations}
\end{figure}

We generated different Discrete Fracture Networks using an open source library called ADFNE \citep{Alg17}. We replaced the original exponential distribution for fracture length with the power law distribution (\ref{eq:pdf_lengths}). We fixed the total number of pre-existing fractures to be 1000 as well as the minimum fracture length $l_{min}$ to be $0.01 L$, being $L$ the characteristic length of the domain area. We then varied the power law exponent $b$ and the maximum fracture length $l_{max}$ in order to generate several DFNs, all characterized by different values of percolation $p$ (see Figure \ref{fig:DFN_realizations}). Note, however, that the fracture lengths in each realization cover two orders of magnitude of the power law distribution (see histograms in Fig. \ref{fig:DFN_realizations}) and that the $b$ values considered fall between 1.3 and 3.5, as suggested by outcrop data (e.g. \citep{BoBo01}).\\
For a given value of $b$ and $l_{max}$, initial and end coordinates of each fracture were obtained.
A pre-processing script in Wolfram Mathematica \citep{Mathematica} was then used to mesh all the generated fractures, with the possibility of i) controlling the mesh element size per fracture, ii) inserting automatically a mesh node at each fracture intersection and finally iii) set a minimum number of finite straight segments per fracture. In all the realizations, the mean mesh size is $0.005 L$ and the minimum number of straight elements per fracture is 5.

\end{appendices}

%%%%%%%%%%%%%%%%%%%%%%%%%%%%%%%%%%%%%%%%%%%%%%%%%%%%
%           BIBLIOGRAPHY        
%%%%%%%%%%%%%%%%%%%%%%%%%%%%%%%%%%%%%%%%%%%%%%%%%%%%
\bibliography{Bibliography.bib}% common bib file

\begin{thebibliography}{}
\providecommand{\doi}[1]{\url{https://doi.org/#1}}
\bibcommenthead

\bibitem [\protect \citeauthoryear {%
{Aki K.}%
}{%
{Aki K.}%
}{%
{\protect \APACyear {1966}}%
}]{%
Aki66}
\APACinsertmetastar {%
Aki66}%
\begin{APACrefauthors}%
{Aki K.}%
\end{APACrefauthors}%
\unskip\
\newblock
\APACrefYearMonthDay{1966}{}{}.
\newblock
{\BBOQ}\APACrefatitle {Generation and {P}ropagation of {G} {Waves} from the
  {N}iigata {E}arthquake of {J}une 16, 1964. {P}art 2. {E}stimation of
  earthquake moment, released energy, and stress-strain drop from the {G} wave
  spectrum.} {Generation and {P}ropagation of {G} {Waves} from the {N}iigata
  {E}arthquake of {J}une 16, 1964. {P}art 2. {E}stimation of earthquake moment,
  released energy, and stress-strain drop from the {G} wave spectrum.}{\BBCQ}
\newblock
\APACjournalVolNumPages{Bulletin of the Earthquake Research
  Institute}{44}{}{73-88}.
\newblock

\newblock

\PrintBackRefs{\CurrentBib}

\bibitem [\protect \citeauthoryear {%
Albaric%
\ \protect \BOthers {.}}{%
Albaric%
\ \protect \BOthers {.}}{%
{\protect \APACyear {2014}}%
}]{%
AlbOy14}
\APACinsertmetastar {%
AlbOy14}%
\begin{APACrefauthors}%
Albaric, J.%
, Oye, V.%
, Langet, N.%
, Hasting, M.%
, Lecomte, I.%
, Iranpour, K.%
\BDBL {}Reid, P.%
\end{APACrefauthors}%
\unskip\
\newblock
\APACrefYearMonthDay{2014}{}{}.
\newblock
{\BBOQ}\APACrefatitle {Monitoring of induced seismicity during the first
  geothermal reservoir stimulation at {P}aralana, {A}ustralia} {Monitoring of
  induced seismicity during the first geothermal reservoir stimulation at
  {P}aralana, {A}ustralia}.{\BBCQ}
\newblock
\APACjournalVolNumPages{Geothermics}{52}{}{120-131}.
\newblock

\newblock

\PrintBackRefs{\CurrentBib}

\bibitem [\protect \citeauthoryear {%
Alghalandis%
}{%
Alghalandis%
}{%
{\protect \APACyear {2017}}%
}]{%
Alg17}
\APACinsertmetastar {%
Alg17}%
\begin{APACrefauthors}%
Alghalandis, Y.F.%
\end{APACrefauthors}%
\unskip\
\newblock
\APACrefYearMonthDay{2017}{}{}.
\newblock
{\BBOQ}\APACrefatitle {{ADFNE}: {O}pen source software for discrete fracture
  network engineering, two and three dimensional applications} {{ADFNE}: {O}pen
  source software for discrete fracture network engineering, two and three
  dimensional applications}.{\BBCQ}
\newblock
\APACjournalVolNumPages{Computers \& Geosciences}{102}{}{1-11}.
\newblock

\newblock

\PrintBackRefs{\CurrentBib}

\bibitem [\protect \citeauthoryear {%
Anders%
\ \BBA {} Wiltschko%
}{%
Anders%
\ \BBA {} Wiltschko%
}{%
{\protect \APACyear {1994}}%
}]{%
AndWil94}
\APACinsertmetastar {%
AndWil94}%
\begin{APACrefauthors}%
Anders, M.H.%
\BCBT {}\ \BBA {} Wiltschko, D.V.%
\end{APACrefauthors}%
\unskip\
\newblock
\APACrefYearMonthDay{1994}{}{}.
\newblock
{\BBOQ}\APACrefatitle {Microfracturing, paleostress and the growth of faults}
  {Microfracturing, paleostress and the growth of faults}.{\BBCQ}
\newblock
\APACjournalVolNumPages{Journal of Structural Geology}{16}{6}{795-815}.
\newblock

\newblock

\PrintBackRefs{\CurrentBib}

\bibitem [\protect \citeauthoryear {%
Bandis%
, Lumsden%
\BCBL {}\ \BBA {} Barton%
}{%
Bandis%
\ \protect \BOthers {.}}{%
{\protect \APACyear {1983}}%
}]{%
BaLu83}
\APACinsertmetastar {%
BaLu83}%
\begin{APACrefauthors}%
Bandis, S.%
, Lumsden, A.%
\BCBL {} Barton, N.%
\end{APACrefauthors}%
\unskip\
\newblock
\APACrefYearMonthDay{1983}{}{}.
\newblock
{\BBOQ}\APACrefatitle {Fundamentals of rock joint deformation} {Fundamentals of
  rock joint deformation}.{\BBCQ}
\newblock
 \APACrefbtitle {International Journal of Rock Mechanics and Mining Sciences \&
  Geomechanics Abstracts} {International journal of rock mechanics and mining
  sciences \& geomechanics abstracts}\ (\BVOL~20, \BPGS\ 249--268).
\PrintBackRefs{\CurrentBib}

\bibitem [\protect \citeauthoryear {%
Bhattacharya%
\ \BBA {} Viesca%
}{%
Bhattacharya%
\ \BBA {} Viesca%
}{%
{\protect \APACyear {2019}}%
}]{%
BaVi19}
\APACinsertmetastar {%
BaVi19}%
\begin{APACrefauthors}%
Bhattacharya, P.%
\BCBT {}\ \BBA {} Viesca, R.C.%
\end{APACrefauthors}%
\unskip\
\newblock
\APACrefYearMonthDay{2019}{}{}.
\newblock
{\BBOQ}\APACrefatitle {Fluid-induced aseismic fault slip outpaces pore-fluid
  migration} {Fluid-induced aseismic fault slip outpaces pore-fluid
  migration}.{\BBCQ}
\newblock
\APACjournalVolNumPages{Science}{364}{6439}{464--468}.
\newblock

\newblock

\PrintBackRefs{\CurrentBib}

\bibitem [\protect \citeauthoryear {%
Bonnet%
\ \protect \BOthers {.}}{%
Bonnet%
\ \protect \BOthers {.}}{%
{\protect \APACyear {2001}}%
}]{%
BoBo01}
\APACinsertmetastar {%
BoBo01}%
\begin{APACrefauthors}%
Bonnet, E.%
, Bour, O.%
, Odling, N.E.%
, Davy, P.%
, Main, I.%
, Cowie, P.%
\BCBL {} Berkowitz, B.%
\end{APACrefauthors}%
\unskip\
\newblock
\APACrefYearMonthDay{2001}{August}{}.
\newblock
{\BBOQ}\APACrefatitle {Scaling of fracture systems in geological media}
  {Scaling of fracture systems in geological media}.{\BBCQ}
\newblock
\APACjournalVolNumPages{Rewiews of Geophysics}{39}{}{347-383}.
\newblock

\newblock

\PrintBackRefs{\CurrentBib}

\bibitem [\protect \citeauthoryear {%
Bour%
\ \BBA {} Davy%
}{%
Bour%
\ \BBA {} Davy%
}{%
{\protect \APACyear {1997}}%
}]{%
BoDa97}
\APACinsertmetastar {%
BoDa97}%
\begin{APACrefauthors}%
Bour, O.%
\BCBT {}\ \BBA {} Davy, P.%
\end{APACrefauthors}%
\unskip\
\newblock
\APACrefYearMonthDay{1997}{}{}.
\newblock
{\BBOQ}\APACrefatitle {Connectivity of random fault networks following a power
  law fault length distribution} {Connectivity of random fault networks
  following a power law fault length distribution}.{\BBCQ}
\newblock
\APACjournalVolNumPages{Water resources research}{33}{7}{1567-1583}.
\newblock

\newblock

\PrintBackRefs{\CurrentBib}

\bibitem [\protect \citeauthoryear {%
Bourouis%
\ \BBA {} Bernard%
}{%
Bourouis%
\ \BBA {} Bernard%
}{%
{\protect \APACyear {2007}}%
}]{%
BoBe07}
\APACinsertmetastar {%
BoBe07}%
\begin{APACrefauthors}%
Bourouis, S.%
\BCBT {}\ \BBA {} Bernard, P.%
\end{APACrefauthors}%
\unskip\
\newblock
\APACrefYearMonthDay{2007}{}{}.
\newblock
{\BBOQ}\APACrefatitle {Evidence for coupled seismic and aseismic fault slip
  during water injection in the geothermal site of {S}oultz ({F}rance), and
  implications for seismogenic transients} {Evidence for coupled seismic and
  aseismic fault slip during water injection in the geothermal site of {S}oultz
  ({F}rance), and implications for seismogenic transients}.{\BBCQ}
\newblock
\APACjournalVolNumPages{Geophys. J. Int.}{169}{}{723-732}.
\newblock

\newblock

\PrintBackRefs{\CurrentBib}

\bibitem [\protect \citeauthoryear {%
Byerlee%
}{%
Byerlee%
}{%
{\protect \APACyear {1978}}%
}]{%
Byerlee78}
\APACinsertmetastar {%
Byerlee78}%
\begin{APACrefauthors}%
Byerlee, J.%
\end{APACrefauthors}%
\unskip\
\newblock
\APACrefYearMonthDay{1978}{}{}.
\newblock
{\BBOQ}\APACrefatitle {Friction of rocks} {Friction of rocks}.{\BBCQ}
\newblock
\APACjournalVolNumPages{Pure and Applied Geophysics}{116}{}{615-626}.
\newblock

\newblock

\PrintBackRefs{\CurrentBib}

\bibitem [\protect \citeauthoryear {%
Cappa%
, Scuderi%
, Collettini%
, Guglielmi%
\BCBL {}\ \BBA {} Avouac%
}{%
Cappa%
\ \protect \BOthers {.}}{%
{\protect \APACyear {2019}}%
}]{%
CaScu19}
\APACinsertmetastar {%
CaScu19}%
\begin{APACrefauthors}%
Cappa, F.%
, Scuderi, M.M.%
, Collettini, C.%
, Guglielmi, Y.%
\BCBL {} Avouac, J\BHBI P.%
\end{APACrefauthors}%
\unskip\
\newblock
\APACrefYearMonthDay{2019}{}{}.
\newblock
{\BBOQ}\APACrefatitle {Stabilization of fault slip by fluid injection in the
  laboratory and in situ} {Stabilization of fault slip by fluid injection in
  the laboratory and in situ}.{\BBCQ}
\newblock
\APACjournalVolNumPages{Science advances}{5}{3}{}.
\newblock

\newblock

\PrintBackRefs{\CurrentBib}

\bibitem [\protect \citeauthoryear {%
Carslaw%
\ \BBA {} Jaeger%
}{%
Carslaw%
\ \BBA {} Jaeger%
}{%
{\protect \APACyear {1959}}%
}]{%
CaJa59}
\APACinsertmetastar {%
CaJa59}%
\begin{APACrefauthors}%
Carslaw, H.S.%
\BCBT {}\ \BBA {} Jaeger, J.C.%
\end{APACrefauthors}%
\unskip\
\newblock
\APACrefYear{1959}.
\newblock
\APACrefbtitle {Conduction of heat in solids} {Conduction of heat in solids}.
\newblock
\APACaddressPublisher{}{Oxford Univ Press}.
\PrintBackRefs{\CurrentBib}

\bibitem [\protect \citeauthoryear {%
Ciardo%
\ \BBA {} Lecampion%
}{%
Ciardo%
\ \BBA {} Lecampion%
}{%
{\protect \APACyear {2019}}%
}]{%
CiLe19}
\APACinsertmetastar {%
CiLe19}%
\begin{APACrefauthors}%
Ciardo, F.%
\BCBT {}\ \BBA {} Lecampion, B.%
\end{APACrefauthors}%
\unskip\
\newblock
\APACrefYearMonthDay{2019}{}{}.
\newblock
{\BBOQ}\APACrefatitle {Effect of dilatancy on the transition from aseismic to
  seismic slip due to fluid injection in a fault} {Effect of dilatancy on the
  transition from aseismic to seismic slip due to fluid injection in a
  fault}.{\BBCQ}
\newblock
\APACjournalVolNumPages{Journal of Geophysical Research: Solid
  Earth}{124}{}{3724-3743}.
\newblock

\newblock

\PrintBackRefs{\CurrentBib}

\bibitem [\protect \citeauthoryear {%
Ciardo%
, Lecampion%
, Fayard%
\BCBL {}\ \BBA {} Chaillat%
}{%
Ciardo%
\ \protect \BOthers {.}}{%
{\protect \APACyear {2020}}%
}]{%
CiLe20}
\APACinsertmetastar {%
CiLe20}%
\begin{APACrefauthors}%
Ciardo, F.%
, Lecampion, B.%
, Fayard, F.%
\BCBL {} Chaillat, S.%
\end{APACrefauthors}%
\unskip\
\newblock
\APACrefYearMonthDay{2020}{}{}.
\newblock
{\BBOQ}\APACrefatitle {A fast boundary element based solver for localized
  inelastic deformations} {A fast boundary element based solver for localized
  inelastic deformations}.{\BBCQ}
\newblock
\APACjournalVolNumPages{Int. J. Numer. Meth. Engng.}{}{}{1-23}.
\newblock

\newblock

\PrintBackRefs{\CurrentBib}

\bibitem [\protect \citeauthoryear {%
Ciardo%
\ \BBA {} Rinaldi%
}{%
Ciardo%
\ \BBA {} Rinaldi%
}{%
{\protect \APACyear {2022}}%
}]{%
CiRi22}
\APACinsertmetastar {%
CiRi22}%
\begin{APACrefauthors}%
Ciardo, F.%
\BCBT {}\ \BBA {} Rinaldi, A.P.%
\end{APACrefauthors}%
\unskip\
\newblock
\APACrefYearMonthDay{2022}{}{}.
\newblock
{\BBOQ}\APACrefatitle {Impact of injection rate ramp-up on nucleation and
  arrest of dynamic fault slip} {Impact of injection rate ramp-up on nucleation
  and arrest of dynamic fault slip}.{\BBCQ}
\newblock
\APACjournalVolNumPages{Geomech. Geophys. Geo-energ. Geo-resour.}{8}{28}{}.
\newblock

\newblock

\PrintBackRefs{\CurrentBib}

\bibitem [\protect \citeauthoryear {%
Cornet%
}{%
Cornet%
}{%
{\protect \APACyear {2016}}%
}]{%
Corn16}
\APACinsertmetastar {%
Corn16}%
\begin{APACrefauthors}%
Cornet, F.H.%
\end{APACrefauthors}%
\unskip\
\newblock
\APACrefYearMonthDay{2016}{}{}.
\newblock
{\BBOQ}\APACrefatitle {Seismic and aseismic motions generated by fluid
  injections} {Seismic and aseismic motions generated by fluid
  injections}.{\BBCQ}
\newblock
\APACjournalVolNumPages{Geomechanics for Energy and the
  Environment}{5}{}{42--54}.
\newblock

\newblock

\PrintBackRefs{\CurrentBib}

\bibitem [\protect \citeauthoryear {%
Cornet%
}{%
Cornet%
}{%
{\protect \APACyear {2021}}%
}]{%
Corn21}
\APACinsertmetastar {%
Corn21}%
\begin{APACrefauthors}%
Cornet, F.H.%
\end{APACrefauthors}%
\unskip\
\newblock
\APACrefYearMonthDay{2021}{June}{}.
\newblock
{\BBOQ}\APACrefatitle {The engineering of safe hydraulic stimulations for {EGS}
  development in hot crystalline rock masses} {The engineering of safe
  hydraulic stimulations for {EGS} development in hot crystalline rock
  masses}.{\BBCQ}
\newblock
\APACjournalVolNumPages{Geomechanics for Energy and the
  Environment}{26}{}{100151}.
\newblock

\newblock

\PrintBackRefs{\CurrentBib}

\bibitem [\protect \citeauthoryear {%
Cornet%
, Helm%
, Poitrenaud%
\BCBL {}\ \BBA {} Etchecopar%
}{%
Cornet%
\ \protect \BOthers {.}}{%
{\protect \APACyear {1997}}%
}]{%
CornHe97}
\APACinsertmetastar {%
CornHe97}%
\begin{APACrefauthors}%
Cornet, F.H.%
, Helm, J.%
, Poitrenaud, H.%
\BCBL {} Etchecopar, A.%
\end{APACrefauthors}%
\unskip\
\newblock
\APACrefYearMonthDay{1997}{}{}.
\newblock
{\BBOQ}\APACrefatitle {Seismic and {A}seismic {S}lips {I}nduced by
  {L}arge-scale {F}luid {I}njections} {Seismic and {A}seismic {S}lips {I}nduced
  by {L}arge-scale {F}luid {I}njections}.{\BBCQ}
\newblock
\APACjournalVolNumPages{Pure Appl. Geophys.}{150}{}{563-583}.
\newblock

\newblock

\PrintBackRefs{\CurrentBib}

\bibitem [\protect \citeauthoryear {%
Davy%
, Bour%
\BCBL {}\ \BBA {} De~Dreuzy%
}{%
Davy%
\ \protect \BOthers {.}}{%
{\protect \APACyear {2006}}%
}]{%
DaBo06}
\APACinsertmetastar {%
DaBo06}%
\begin{APACrefauthors}%
Davy, P.%
, Bour, O.%
\BCBL {} De~Dreuzy, J\BHBI R.%
\end{APACrefauthors}%
\unskip\
\newblock
\APACrefYearMonthDay{2006}{August}{}.
\newblock
\APACrefbtitle {Discrete fracture network for the {F}orsmark site} {Discrete
  fracture network for the {F}orsmark site}\ \APACbVolEdTR{}{\BTR{}}.
\newblock
\APACaddressInstitution{}{Itasca Consulting Group}.
\PrintBackRefs{\CurrentBib}

\bibitem [\protect \citeauthoryear {%
Duboeuf%
\ \protect \BOthers {.}}{%
Duboeuf%
\ \protect \BOthers {.}}{%
{\protect \APACyear {2017}}%
}]{%
DuDe17}
\APACinsertmetastar {%
DuDe17}%
\begin{APACrefauthors}%
Duboeuf, L.%
, De~Barros, L.%
, Cappa, F.%
, Guglielmini, Y.%
, Deschamps, A.%
\BCBL {} Seguy, S.%
\end{APACrefauthors}%
\unskip\
\newblock
\APACrefYearMonthDay{2017}{}{}.
\newblock
{\BBOQ}\APACrefatitle {Aseismic motions drive a sparse seismicity during fluid
  injections into a fractured zone in a carbonate reservoir} {Aseismic motions
  drive a sparse seismicity during fluid injections into a fractured zone in a
  carbonate reservoir}.{\BBCQ}
\newblock
\APACjournalVolNumPages{Journal of Geophysical Research: Solid
  Earth}{122}{}{8285-8304}.
\newblock

\newblock

\PrintBackRefs{\CurrentBib}

\bibitem [\protect \citeauthoryear {%
Ellsworth%
}{%
Ellsworth%
}{%
{\protect \APACyear {2013}}%
}]{%
El13}
\APACinsertmetastar {%
El13}%
\begin{APACrefauthors}%
Ellsworth, W.L.%
\end{APACrefauthors}%
\unskip\
\newblock
\APACrefYearMonthDay{2013}{July}{}.
\newblock
{\BBOQ}\APACrefatitle {Injection-Induced Earthquakes} {Injection-induced
  earthquakes}.{\BBCQ}
\newblock
\APACjournalVolNumPages{Science}{341}{}{}.
\newblock

\newblock

\PrintBackRefs{\CurrentBib}

\bibitem [\protect \citeauthoryear {%
Eyre%
\ \protect \BOthers {.}}{%
Eyre%
\ \protect \BOthers {.}}{%
{\protect \APACyear {2019}}%
}]{%
EyEa19}
\APACinsertmetastar {%
EyEa19}%
\begin{APACrefauthors}%
Eyre, T.%
, Eaton, D.W.%
, Garagash, D.I.%
, Zecevic, M.%
, Venieri, M.%
, Weir, R.%
\BCBL {} Lawton, D.C.%
\end{APACrefauthors}%
\unskip\
\newblock
\APACrefYearMonthDay{2019}{}{}.
\newblock
{\BBOQ}\APACrefatitle {The role of aseismic slip in hydraulic
  fracturing--induced seismicity} {The role of aseismic slip in hydraulic
  fracturing--induced seismicity}.{\BBCQ}
\newblock
\APACjournalVolNumPages{Science advances}{5}{8}{}.
\newblock

\newblock

\PrintBackRefs{\CurrentBib}

\bibitem [\protect \citeauthoryear {%
Faulkner%
\ \protect \BOthers {.}}{%
Faulkner%
\ \protect \BOthers {.}}{%
{\protect \APACyear {2010}}%
}]{%
FauJa10}
\APACinsertmetastar {%
FauJa10}%
\begin{APACrefauthors}%
Faulkner, D.%
, Jackson, C.A.L.%
, Lunn, R.J.%
, Schlische, R.W.%
, Shipton, Z.K.%
, Wibberley, C.A.J.%
\BCBL {} Withjack, M.O.%
\end{APACrefauthors}%
\unskip\
\newblock
\APACrefYearMonthDay{2010}{}{}.
\newblock
{\BBOQ}\APACrefatitle {A review of recent developments concerning the
  structure, mechanics and fluid flow properties of fault zones} {A review of
  recent developments concerning the structure, mechanics and fluid flow
  properties of fault zones}.{\BBCQ}
\newblock
\APACjournalVolNumPages{Journal of Structural Geology}{32}{}{1557-1575}.
\newblock

\newblock

\PrintBackRefs{\CurrentBib}

\bibitem [\protect \citeauthoryear {%
Galis%
, Ampuero%
, Mai%
\BCBL {}\ \BBA {} Cappa%
}{%
Galis%
\ \protect \BOthers {.}}{%
{\protect \APACyear {2017}}%
}]{%
GaAm17}
\APACinsertmetastar {%
GaAm17}%
\begin{APACrefauthors}%
Galis, M.%
, Ampuero, J\BHBI P.%
, Mai, P.M.%
\BCBL {} Cappa, F.%
\end{APACrefauthors}%
\unskip\
\newblock
\APACrefYearMonthDay{2017}{}{}.
\newblock
{\BBOQ}\APACrefatitle {Induced seismicity provides insight into why earthquake
  ruptures stop} {Induced seismicity provides insight into why earthquake
  ruptures stop}.{\BBCQ}
\newblock
\APACjournalVolNumPages{Science advances}{3}{12}{eaap7528}.
\newblock

\newblock

\PrintBackRefs{\CurrentBib}

\bibitem [\protect \citeauthoryear {%
Garagash%
\ \BBA {} Germanovich%
}{%
Garagash%
\ \BBA {} Germanovich%
}{%
{\protect \APACyear {2012}}%
}]{%
GaGe12}
\APACinsertmetastar {%
GaGe12}%
\begin{APACrefauthors}%
Garagash, D.I.%
\BCBT {}\ \BBA {} Germanovich, L.N.%
\end{APACrefauthors}%
\unskip\
\newblock
\APACrefYearMonthDay{2012}{}{}.
\newblock
{\BBOQ}\APACrefatitle {Nucleation and arrest of dynamic slip on a pressurized
  fault} {Nucleation and arrest of dynamic slip on a pressurized fault}.{\BBCQ}
\newblock
\APACjournalVolNumPages{Journal of Geophysical Research}{117}{B10310}{}.
\newblock

\newblock

\PrintBackRefs{\CurrentBib}

\bibitem [\protect \citeauthoryear {%
Gebbia%
}{%
Gebbia%
}{%
{\protect \APACyear {1891}}%
}]{%
Ge1891}
\APACinsertmetastar {%
Ge1891}%
\begin{APACrefauthors}%
Gebbia, M.%
\end{APACrefauthors}%
\unskip\
\newblock
\APACrefYearMonthDay{1891}{}{}.
\newblock
{\BBOQ}\APACrefatitle {Formule fondamentali della statica dei corpi elastici}
  {Formule fondamentali della statica dei corpi elastici}.{\BBCQ}
\newblock
\APACjournalVolNumPages{Rend. Circ. Mat. di Palermo}{5}{}{320-323}.
\newblock

\newblock

\PrintBackRefs{\CurrentBib}

\bibitem [\protect \citeauthoryear {%
Guglielmini%
, Cappa%
, Avouac%
, Henry%
\BCBL {}\ \BBA {} Elsworth%
}{%
Guglielmini%
\ \protect \BOthers {.}}{%
{\protect \APACyear {2015}}%
}]{%
GuCa15}
\APACinsertmetastar {%
GuCa15}%
\begin{APACrefauthors}%
Guglielmini, Y.%
, Cappa, F.%
, Avouac, J\BHBI P.%
, Henry, P.%
\BCBL {} Elsworth, D.%
\end{APACrefauthors}%
\unskip\
\newblock
\APACrefYearMonthDay{2015}{June}{}.
\newblock
{\BBOQ}\APACrefatitle {Seismicity triggered by fluid injection-induced aseismic
  slip} {Seismicity triggered by fluid injection-induced aseismic slip}.{\BBCQ}
\newblock
\APACjournalVolNumPages{Science}{348}{}{}.
\newblock

\newblock

\PrintBackRefs{\CurrentBib}

\bibitem [\protect \citeauthoryear {%
Hainzl%
, Fischer%
\BCBL {}\ \BBA {} Dahm%
}{%
Hainzl%
\ \protect \BOthers {.}}{%
{\protect \APACyear {2012}}%
}]{%
HaFi12}
\APACinsertmetastar {%
HaFi12}%
\begin{APACrefauthors}%
Hainzl, S.%
, Fischer, T.%
\BCBL {} Dahm, T.%
\end{APACrefauthors}%
\unskip\
\newblock
\APACrefYearMonthDay{2012}{}{}.
\newblock
{\BBOQ}\APACrefatitle {Seismicity{-}based estimation of the driving fluid
  pressure in the case of swarm activity in {W}estern {B}ohemia}
  {Seismicity{-}based estimation of the driving fluid pressure in the case of
  swarm activity in {W}estern {B}ohemia}.{\BBCQ}
\newblock
\APACjournalVolNumPages{Geophys. J. Int.}{191}{271-281}{}.
\newblock

\newblock

\PrintBackRefs{\CurrentBib}

\bibitem [\protect \citeauthoryear {%
Hainzl%
\ \BBA {} Ogata%
}{%
Hainzl%
\ \BBA {} Ogata%
}{%
{\protect \APACyear {2005}}%
}]{%
HaOg05}
\APACinsertmetastar {%
HaOg05}%
\begin{APACrefauthors}%
Hainzl, S.%
\BCBT {}\ \BBA {} Ogata, Y.%
\end{APACrefauthors}%
\unskip\
\newblock
\APACrefYearMonthDay{2005}{}{}.
\newblock
{\BBOQ}\APACrefatitle {Detecting fluid signals in seismicity data through
  statistical earthquake modeling} {Detecting fluid signals in seismicity data
  through statistical earthquake modeling}.{\BBCQ}
\newblock
\APACjournalVolNumPages{Journal of Geophysical Research: Solid Earth}{110}{}{}.
\newblock

\newblock

\PrintBackRefs{\CurrentBib}

\bibitem [\protect \citeauthoryear {%
Hamilton%
\ \BBA {} Meehan%
}{%
Hamilton%
\ \BBA {} Meehan%
}{%
{\protect \APACyear {1971}}%
}]{%
HaMe71}
\APACinsertmetastar {%
HaMe71}%
\begin{APACrefauthors}%
Hamilton, D.H.%
\BCBT {}\ \BBA {} Meehan, R.L.%
\end{APACrefauthors}%
\unskip\
\newblock
\APACrefYearMonthDay{1971}{}{}.
\newblock
{\BBOQ}\APACrefatitle {Ground rupture in the {B}aldwin {H}ills} {Ground rupture
  in the {B}aldwin {H}ills}.{\BBCQ}
\newblock
\APACjournalVolNumPages{Science}{172}{3981}{333--344}.
\newblock

\newblock

\PrintBackRefs{\CurrentBib}

\bibitem [\protect \citeauthoryear {%
Hatton%
, Main%
\BCBL {}\ \BBA {} Meredith%
}{%
Hatton%
\ \protect \BOthers {.}}{%
{\protect \APACyear {1994}}%
}]{%
HaMa94}
\APACinsertmetastar {%
HaMa94}%
\begin{APACrefauthors}%
Hatton, C.G.%
, Main, I.G.%
\BCBL {} Meredith, P.G.%
\end{APACrefauthors}%
\unskip\
\newblock
\APACrefYearMonthDay{1994}{January}{}.
\newblock
{\BBOQ}\APACrefatitle {Non-universal scaling of fracture length and opening
  displacement} {Non-universal scaling of fracture length and opening
  displacement}.{\BBCQ}
\newblock
\APACjournalVolNumPages{Nature}{367}{}{}.
\newblock

\newblock

\PrintBackRefs{\CurrentBib}

\bibitem [\protect \citeauthoryear {%
Healy%
, Rubey%
, Griggs%
\BCBL {}\ \BBA {} Raleigh%
}{%
Healy%
\ \protect \BOthers {.}}{%
{\protect \APACyear {1968}}%
}]{%
HeRu68}
\APACinsertmetastar {%
HeRu68}%
\begin{APACrefauthors}%
Healy, J.H.%
, Rubey, W.W.%
, Griggs, D.T.%
\BCBL {} Raleigh, C.B.%
\end{APACrefauthors}%
\unskip\
\newblock
\APACrefYearMonthDay{1968}{}{}.
\newblock
{\BBOQ}\APACrefatitle {The {D}enver {E}arthquakes} {The {D}enver
  {E}arthquakes}.{\BBCQ}
\newblock
\APACjournalVolNumPages{Science}{161}{3848}{1301--1310}.
\newblock

\newblock

\PrintBackRefs{\CurrentBib}

\bibitem [\protect \citeauthoryear {%
Hill%
, Kelly%
, Dai%
\BCBL {}\ \BBA {} Korsunsky%
}{%
Hill%
\ \protect \BOthers {.}}{%
{\protect \APACyear {1996}}%
}]{%
HiKe96}
\APACinsertmetastar {%
HiKe96}%
\begin{APACrefauthors}%
Hill, D.A.%
, Kelly, P.A.%
, Dai, D.N.%
\BCBL {} Korsunsky, A.M.%
\end{APACrefauthors}%
\unskip\
\newblock
\APACrefYear{1996}.
\newblock
\APACrefbtitle {Solution of {C}rack {P}roblems: the {D}istributed {D}islocation
  {T}echnique} {Solution of {C}rack {P}roblems: the {D}istributed {D}islocation
  {T}echnique}.
\newblock
\APACaddressPublisher{}{{Kluwer Academic Publishers}}.
\PrintBackRefs{\CurrentBib}

\bibitem [\protect \citeauthoryear {%
Jung%
}{%
Jung%
}{%
{\protect \APACyear {2013}}%
}]{%
Jung13}
\APACinsertmetastar {%
Jung13}%
\begin{APACrefauthors}%
Jung, R.%
\end{APACrefauthors}%
\unskip\
\newblock
\APACrefYearMonthDay{2013}{}{}.
\newblock
{\BBOQ}\APACrefatitle {{EGS - Goodbye or Back to the Future 95}} {{EGS -
  Goodbye or Back to the Future 95}}.{\BBCQ}
\newblock
 \APACrefbtitle {Effective and Sustainable Hydraulic Fracturing.} {Effective
  and sustainable hydraulic fracturing.}
\PrintBackRefs{\CurrentBib}

\bibitem [\protect \citeauthoryear {%
Keranen%
\ \BBA {} Weingarten%
}{%
Keranen%
\ \BBA {} Weingarten%
}{%
{\protect \APACyear {2018}}%
}]{%
KeWe18}
\APACinsertmetastar {%
KeWe18}%
\begin{APACrefauthors}%
Keranen, K.M.%
\BCBT {}\ \BBA {} Weingarten, M.%
\end{APACrefauthors}%
\unskip\
\newblock
\APACrefYearMonthDay{2018}{}{}.
\newblock
{\BBOQ}\APACrefatitle {Induced seismicity} {Induced seismicity}.{\BBCQ}
\newblock
\APACjournalVolNumPages{Annual Review of Earth and Planetary
  Sciences}{46}{1}{149--174}.
\newblock

\newblock

\PrintBackRefs{\CurrentBib}

\bibitem [\protect \citeauthoryear {%
Kranz%
}{%
Kranz%
}{%
{\protect \APACyear {1983}}%
}]{%
Kra83}
\APACinsertmetastar {%
Kra83}%
\begin{APACrefauthors}%
Kranz, R.L.%
\end{APACrefauthors}%
\unskip\
\newblock
\APACrefYearMonthDay{1983}{}{}.
\newblock
{\BBOQ}\APACrefatitle {Microcracks in rocks: a review} {Microcracks in rocks: a
  review}.{\BBCQ}
\newblock
\APACjournalVolNumPages{Tectonophysics}{100}{}{449-480}.
\newblock

\newblock

\PrintBackRefs{\CurrentBib}

\bibitem [\protect \citeauthoryear {%
Lei%
\ \BBA {} Gao%
}{%
Lei%
\ \BBA {} Gao%
}{%
{\protect \APACyear {2018}}%
}]{%
LeiGao18}
\APACinsertmetastar {%
LeiGao18}%
\begin{APACrefauthors}%
Lei, Q.%
\BCBT {}\ \BBA {} Gao, K.%
\end{APACrefauthors}%
\unskip\
\newblock
\APACrefYearMonthDay{2018}{}{}.
\newblock
{\BBOQ}\APACrefatitle {Correlation between fracture network properties and
  stress variability in geological media} {Correlation between fracture network
  properties and stress variability in geological media}.{\BBCQ}
\newblock
\APACjournalVolNumPages{Geophysical Research Letters}{45}{}{}.
\newblock

\newblock

\PrintBackRefs{\CurrentBib}

\bibitem [\protect \citeauthoryear {%
Lei%
, Latham%
\BCBL {}\ \BBA {} Tsang%
}{%
Lei%
\ \protect \BOthers {.}}{%
{\protect \APACyear {2017}}%
}]{%
LeLa17}
\APACinsertmetastar {%
LeLa17}%
\begin{APACrefauthors}%
Lei, Q.%
, Latham, J\BHBI P.%
\BCBL {} Tsang, C\BHBI F.%
\end{APACrefauthors}%
\unskip\
\newblock
\APACrefYearMonthDay{2017}{}{}.
\newblock
{\BBOQ}\APACrefatitle {The use of discrete fracture networks for modelling
  coupled geomechanical and hydrological behaviour of fractured rocks} {The use
  of discrete fracture networks for modelling coupled geomechanical and
  hydrological behaviour of fractured rocks}.{\BBCQ}
\newblock
\APACjournalVolNumPages{Computers and Geotechnics}{85}{}{151--176}.
\newblock

\newblock

\PrintBackRefs{\CurrentBib}

\bibitem [\protect \citeauthoryear {%
Lei%
\ \BBA {} Wang%
}{%
Lei%
\ \BBA {} Wang%
}{%
{\protect \APACyear {2016}}%
}]{%
LeiWa16}
\APACinsertmetastar {%
LeiWa16}%
\begin{APACrefauthors}%
Lei, Q.%
\BCBT {}\ \BBA {} Wang, X.%
\end{APACrefauthors}%
\unskip\
\newblock
\APACrefYearMonthDay{2016}{}{}.
\newblock
{\BBOQ}\APACrefatitle {Tectonic interpretation of the connectivity of a
  multiscale fracture system in limestone} {Tectonic interpretation of the
  connectivity of a multiscale fracture system in limestone}.{\BBCQ}
\newblock
\APACjournalVolNumPages{Geophysical Research Letters}{43}{}{1551-1558}.
\newblock

\newblock

\PrintBackRefs{\CurrentBib}

\bibitem [\protect \citeauthoryear {%
McGarr%
}{%
McGarr%
}{%
{\protect \APACyear {2014}}%
}]{%
Mc14}
\APACinsertmetastar {%
Mc14}%
\begin{APACrefauthors}%
McGarr, A.%
\end{APACrefauthors}%
\unskip\
\newblock
\APACrefYearMonthDay{2014}{}{}.
\newblock
{\BBOQ}\APACrefatitle {Maximum magnitude earthquakes induced by fluid
  injection} {Maximum magnitude earthquakes induced by fluid injection}.{\BBCQ}
\newblock
\APACjournalVolNumPages{Journal of Geophysical Research: Solid
  Earth}{119}{2}{1008-1019}.
\newblock

\newblock

\PrintBackRefs{\CurrentBib}

\bibitem [\protect \citeauthoryear {%
McGarr%
\ \BBA {} Barbour%
}{%
McGarr%
\ \BBA {} Barbour%
}{%
{\protect \APACyear {2018}}%
}]{%
McBa18}
\APACinsertmetastar {%
McBa18}%
\begin{APACrefauthors}%
McGarr, A.%
\BCBT {}\ \BBA {} Barbour, A.J.%
\end{APACrefauthors}%
\unskip\
\newblock
\APACrefYearMonthDay{2018}{}{}.
\newblock
{\BBOQ}\APACrefatitle {Injection-induced moment release can also be aseismic}
  {Injection-induced moment release can also be aseismic}.{\BBCQ}
\newblock
\APACjournalVolNumPages{Geophys. Res. Letters}{45}{}{5344-5351}.
\newblock

\newblock

\PrintBackRefs{\CurrentBib}

\bibitem [\protect \citeauthoryear {%
Mogilevskaya%
}{%
Mogilevskaya%
}{%
{\protect \APACyear {2014}}%
}]{%
Mogi2014}
\APACinsertmetastar {%
Mogi2014}%
\begin{APACrefauthors}%
Mogilevskaya, S.G.%
\end{APACrefauthors}%
\unskip\
\newblock
\APACrefYearMonthDay{2014}{}{}.
\newblock
{\BBOQ}\APACrefatitle {Lost in translation: Crack problems in different
  languages} {Lost in translation: Crack problems in different
  languages}.{\BBCQ}
\newblock
\APACjournalVolNumPages{International Journal of Solids and
  Structures}{51}{25}{4492--4503}.
\newblock

\newblock

\PrintBackRefs{\CurrentBib}

\bibitem [\protect \citeauthoryear {%
Olson%
}{%
Olson%
}{%
{\protect \APACyear {2003}}%
}]{%
Ols03}
\APACinsertmetastar {%
Ols03}%
\begin{APACrefauthors}%
Olson, J.E.%
\end{APACrefauthors}%
\unskip\
\newblock
\APACrefYearMonthDay{2003}{}{}.
\newblock
{\BBOQ}\APACrefatitle {Sublinear scaling of fracture aperture versus length:
  {A}n exception or the rule?} {Sublinear scaling of fracture aperture versus
  length: {A}n exception or the rule?}{\BBCQ}
\newblock
\APACjournalVolNumPages{Journal of Geophysical Research}{108}{B9}{}.
\newblock

\newblock

\PrintBackRefs{\CurrentBib}

\bibitem [\protect \citeauthoryear {%
Parotidis%
, Rothert%
\BCBL {}\ \BBA {} Shapiro%
}{%
Parotidis%
\ \protect \BOthers {.}}{%
{\protect \APACyear {2003}}%
}]{%
PaRo03}
\APACinsertmetastar {%
PaRo03}%
\begin{APACrefauthors}%
Parotidis, M.%
, Rothert, E.%
\BCBL {} Shapiro, S.A.%
\end{APACrefauthors}%
\unskip\
\newblock
\APACrefYearMonthDay{2003}{}{}.
\newblock
{\BBOQ}\APACrefatitle {Pore{-}pressure diffusion: {A} possible triggering
  mechanism for the earthquake swarms 2000 in {V}ogtland{/NW}{-}{B}ohemia{,}
  central {E}urope} {Pore{-}pressure diffusion: {A} possible triggering
  mechanism for the earthquake swarms 2000 in {V}ogtland{/NW}{-}{B}ohemia{,}
  central {E}urope}.{\BBCQ}
\newblock
\APACjournalVolNumPages{Geophysical Research Letters}{30}{}{}.
\newblock

\newblock

\PrintBackRefs{\CurrentBib}

\bibitem [\protect \citeauthoryear {%
Quarteroni%
, Sacco%
\BCBL {}\ \BBA {} Saleri%
}{%
Quarteroni%
\ \protect \BOthers {.}}{%
{\protect \APACyear {2000}}%
}]{%
QuaSa2000}
\APACinsertmetastar {%
QuaSa2000}%
\begin{APACrefauthors}%
Quarteroni, A.%
, Sacco, R.%
\BCBL {} Saleri, F.%
\end{APACrefauthors}%
\unskip\
\newblock
\APACrefYear{2000}.
\newblock
\APACrefbtitle {Numerical mathematics} {Numerical mathematics}\ (T.~in~applied
  mathematics, \BED{}).
\newblock
\APACaddressPublisher{}{Springer}.
\PrintBackRefs{\CurrentBib}

\bibitem [\protect \citeauthoryear {%
S{\'a}ez%
, Lecampion%
, Bhattacharya%
\BCBL {}\ \BBA {} Viesca%
}{%
S{\'a}ez%
\ \protect \BOthers {.}}{%
{\protect \APACyear {2022}}%
}]{%
SaLe22}
\APACinsertmetastar {%
SaLe22}%
\begin{APACrefauthors}%
S{\'a}ez, A.%
, Lecampion, B.%
, Bhattacharya, P.%
\BCBL {} Viesca, R.C.%
\end{APACrefauthors}%
\unskip\
\newblock
\APACrefYearMonthDay{2022}{}{}.
\newblock
{\BBOQ}\APACrefatitle {Three-dimensional fluid-driven stable frictional
  ruptures} {Three-dimensional fluid-driven stable frictional ruptures}.{\BBCQ}
\newblock
\APACjournalVolNumPages{J. Mech. Phys. Solids}{160}{}{}.
\newblock

\newblock

\PrintBackRefs{\CurrentBib}

\bibitem [\protect \citeauthoryear {%
Scotti%
\ \BBA {} Cornet%
}{%
Scotti%
\ \BBA {} Cornet%
}{%
{\protect \APACyear {1994}}%
}]{%
ScCo94}
\APACinsertmetastar {%
ScCo94}%
\begin{APACrefauthors}%
Scotti, O.%
\BCBT {}\ \BBA {} Cornet, F.H.%
\end{APACrefauthors}%
\unskip\
\newblock
\APACrefYearMonthDay{1994}{}{}.
\newblock
{\BBOQ}\APACrefatitle {In {S}itu {E}vidence for {F}luid-{I}nduced {A}seismic
  {S}lip {E}vents {A}long {F}ault {Z}ones} {In {S}itu {E}vidence for
  {F}luid-{I}nduced {A}seismic {S}lip {E}vents {A}long {F}ault {Z}ones}.{\BBCQ}
\newblock
\APACjournalVolNumPages{Int. J. Rock Mech. Min. Sci. \& Geom.
  Abstr.}{31}{4}{347-358}.
\newblock

\newblock

\PrintBackRefs{\CurrentBib}

\bibitem [\protect \citeauthoryear {%
Scuderi%
\ \BBA {} Collettini%
}{%
Scuderi%
\ \BBA {} Collettini%
}{%
{\protect \APACyear {2016}}%
}]{%
ScCo16}
\APACinsertmetastar {%
ScCo16}%
\begin{APACrefauthors}%
Scuderi, M.M.%
\BCBT {}\ \BBA {} Collettini, C.%
\end{APACrefauthors}%
\unskip\
\newblock
\APACrefYearMonthDay{2016}{}{}.
\newblock
{\BBOQ}\APACrefatitle {The role of fluid pressure in induced vs. triggered
  seismicity: insights from rock deformation experiments on carbonates} {The
  role of fluid pressure in induced vs. triggered seismicity: insights from
  rock deformation experiments on carbonates}.{\BBCQ}
\newblock
\APACjournalVolNumPages{Scientific Reports}{6}{24852}{}.
\newblock

\newblock

\PrintBackRefs{\CurrentBib}

\bibitem [\protect \citeauthoryear {%
Shapiro%
, Huenges%
\BCBL {}\ \BBA {} Borm%
}{%
Shapiro%
\ \protect \BOthers {.}}{%
{\protect \APACyear {1997}}%
}]{%
ShHue97}
\APACinsertmetastar {%
ShHue97}%
\begin{APACrefauthors}%
Shapiro, S.A.%
, Huenges, E.%
\BCBL {} Borm, G.%
\end{APACrefauthors}%
\unskip\
\newblock
\APACrefYearMonthDay{1997}{}{}.
\newblock
{\BBOQ}\APACrefatitle {Estimating the crust permeability from
  fluid-injection-induced seismic emission at the {KTB} site} {Estimating the
  crust permeability from fluid-injection-induced seismic emission at the {KTB}
  site}.{\BBCQ}
\newblock
\APACjournalVolNumPages{Geophys. J. Int.}{131}{}{F15-F18}.
\newblock

\newblock

\PrintBackRefs{\CurrentBib}

\bibitem [\protect \citeauthoryear {%
Shapiro%
, Rothert%
, Rath%
\BCBL {}\ \BBA {} Rindschwentner%
}{%
Shapiro%
\ \protect \BOthers {.}}{%
{\protect \APACyear {2002}}%
}]{%
ShRo02}
\APACinsertmetastar {%
ShRo02}%
\begin{APACrefauthors}%
Shapiro, S.A.%
, Rothert, E.%
, Rath, V.%
\BCBL {} Rindschwentner, J.%
\end{APACrefauthors}%
\unskip\
\newblock
\APACrefYearMonthDay{2002}{}{}.
\newblock
{\BBOQ}\APACrefatitle {Characterization of fluid transport properties of
  reservoirs using induced seismicity} {Characterization of fluid transport
  properties of reservoirs using induced seismicity}.{\BBCQ}
\newblock
\APACjournalVolNumPages{Geophysics}{67}{1}{212-220}.
\newblock

\newblock

\PrintBackRefs{\CurrentBib}

\bibitem [\protect \citeauthoryear {%
S{\"o}derlind%
\ \BBA {} Wang%
}{%
S{\"o}derlind%
\ \BBA {} Wang%
}{%
{\protect \APACyear {2003}}%
}]{%
SoWa03}
\APACinsertmetastar {%
SoWa03}%
\begin{APACrefauthors}%
S{\"o}derlind, G.%
\BCBT {}\ \BBA {} Wang, L.%
\end{APACrefauthors}%
\unskip\
\newblock
\APACrefYearMonthDay{2003}{}{}.
\newblock
{\BBOQ}\APACrefatitle {Adaptive time-stepping and computational stability}
  {Adaptive time-stepping and computational stability}.{\BBCQ}
\newblock
\APACjournalVolNumPages{Journal of Computational and Applied
  Mathematics}{185}{}{225-243}.
\newblock

\newblock

\PrintBackRefs{\CurrentBib}

\bibitem [\protect \citeauthoryear {%
Sornette%
, Davy%
\BCBL {}\ \BBA {} Sornette%
}{%
Sornette%
\ \protect \BOthers {.}}{%
{\protect \APACyear {1993}}%
}]{%
SorDa93}
\APACinsertmetastar {%
SorDa93}%
\begin{APACrefauthors}%
Sornette, A.%
, Davy, P.%
\BCBL {} Sornette, D.%
\end{APACrefauthors}%
\unskip\
\newblock
\APACrefYearMonthDay{1993}{}{}.
\newblock
{\BBOQ}\APACrefatitle {Fault growth in brittle-ductile experiments and the
  {M}echanics of continental collisions} {Fault growth in brittle-ductile
  experiments and the {M}echanics of continental collisions}.{\BBCQ}
\newblock
\APACjournalVolNumPages{Journal of Geophysical Research}{98}{N7}{12111-12139}.
\newblock

\newblock

\PrintBackRefs{\CurrentBib}

\bibitem [\protect \citeauthoryear {%
Viesca%
}{%
Viesca%
}{%
{\protect \APACyear {2021}}%
}]{%
Vi21}
\APACinsertmetastar {%
Vi21}%
\begin{APACrefauthors}%
Viesca, R.%
\end{APACrefauthors}%
\unskip\
\newblock
\APACrefYearMonthDay{2021}{}{}.
\newblock
{\BBOQ}\APACrefatitle {Self-similar fault slip in response to fluid injection}
  {Self-similar fault slip in response to fluid injection}.{\BBCQ}
\newblock
\APACjournalVolNumPages{Journal of Fluid Mechanics}{928}{A29}{}.
\newblock

\newblock

\PrintBackRefs{\CurrentBib}

\bibitem [\protect \citeauthoryear {%
Walmann%
, Malthe-S{\o}renssen%
, Feder%
, J{\o}ssang%
\BCBL {}\ \BBA {} Maekin%
}{%
Walmann%
\ \protect \BOthers {.}}{%
{\protect \APACyear {1996}}%
}]{%
WalSor96}
\APACinsertmetastar {%
WalSor96}%
\begin{APACrefauthors}%
Walmann, T.%
, Malthe-S{\o}renssen, A.%
, Feder, J.%
, J{\o}ssang, T.%
\BCBL {} Maekin, P.%
\end{APACrefauthors}%
\unskip\
\newblock
\APACrefYearMonthDay{1996}{}{}.
\newblock
{\BBOQ}\APACrefatitle {Scaling relations for the {L}enghts and {W}idths of
  {F}ractures} {Scaling relations for the {L}enghts and {W}idths of
  {F}ractures}.{\BBCQ}
\newblock
\APACjournalVolNumPages{Phys. Rev. Letters}{77}{27}{}.
\newblock

\newblock

\PrintBackRefs{\CurrentBib}

\bibitem [\protect \citeauthoryear {%
Warpinski%
\ \BBA {} Teufel%
}{%
Warpinski%
\ \BBA {} Teufel%
}{%
{\protect \APACyear {1987}}%
}]{%
WaTe87}
\APACinsertmetastar {%
WaTe87}%
\begin{APACrefauthors}%
Warpinski, N.R.%
\BCBT {}\ \BBA {} Teufel, L.W.%
\end{APACrefauthors}%
\unskip\
\newblock
\APACrefYearMonthDay{1987}{}{}.
\newblock
{\BBOQ}\APACrefatitle {Influence of geologic discontinuities on hydraulic
  fracture propagation} {Influence of geologic discontinuities on hydraulic
  fracture propagation}.{\BBCQ}
\newblock
\APACjournalVolNumPages{Journal of Petroleum Technology}{39}{02}{209--220}.
\newblock

\newblock

\PrintBackRefs{\CurrentBib}

\bibitem [\protect \citeauthoryear {%
Wibberley%
, Yielding%
\BCBL {}\ \BBA {} Di~Toro%
}{%
Wibberley%
\ \protect \BOthers {.}}{%
{\protect \APACyear {2008}}%
}]{%
Wibberley20085}
\APACinsertmetastar {%
Wibberley20085}%
\begin{APACrefauthors}%
Wibberley, C.A.J.%
, Yielding, G.%
\BCBL {} Di~Toro, G.%
\end{APACrefauthors}%
\unskip\
\newblock
\APACrefYearMonthDay{2008}{}{}.
\newblock
{\BBOQ}\APACrefatitle {Recent advances in the understanding of fault zone
  internal structure: A review} {Recent advances in the understanding of fault
  zone internal structure: A review}.{\BBCQ}
\newblock
\APACjournalVolNumPages{Geological Society Special Publication}{299}{}{5-33}.
\newblock

\newblock

\PrintBackRefs{\CurrentBib}

\bibitem [\protect \citeauthoryear {%
Witherspoon%
\ \BBA {} Gale%
}{%
Witherspoon%
\ \BBA {} Gale%
}{%
{\protect \APACyear {1977}}%
}]{%
WhGa77}
\APACinsertmetastar {%
WhGa77}%
\begin{APACrefauthors}%
Witherspoon, P.A.%
\BCBT {}\ \BBA {} Gale, J.E.%
\end{APACrefauthors}%
\unskip\
\newblock
\APACrefYearMonthDay{1977}{}{}.
\newblock
{\BBOQ}\APACrefatitle {Mechanical and hydraulic properties of rocks related to
  induced seismicity} {Mechanical and hydraulic properties of rocks related to
  induced seismicity}.{\BBCQ}
\newblock
\APACjournalVolNumPages{Engineering Geology}{11}{1}{23--55}.
\newblock

\newblock

\PrintBackRefs{\CurrentBib}

\bibitem [\protect \citeauthoryear {%
{Wolfram Research, Inc.}%
}{%
{Wolfram Research, Inc.}%
}{%
{\protect \APACyear {2022}}%
}]{%
Mathematica}
\APACinsertmetastar {%
Mathematica}%
\begin{APACrefauthors}%
{Wolfram Research, Inc.}%
\end{APACrefauthors}%
\unskip\
\newblock
\APACrefYearMonthDay{2022}{}{}.
\newblock
\APACrefbtitle {Mathematica, {V}ersion 13.1.} {Mathematica, {V}ersion 13.1.}
\newblock
\begin{APACrefURL} {https://www.wolfram.com/mathematica} \end{APACrefURL}
\newblock
\APACrefnote{Champaign, IL, 2022}
\PrintBackRefs{\CurrentBib}

\bibitem [\protect \citeauthoryear {%
Zoback%
\ \BBA {} Harjes%
}{%
Zoback%
\ \BBA {} Harjes%
}{%
{\protect \APACyear {1997}}%
}]{%
ZoHa97}
\APACinsertmetastar {%
ZoHa97}%
\begin{APACrefauthors}%
Zoback, M.%
\BCBT {}\ \BBA {} Harjes, H\BHBI P.%
\end{APACrefauthors}%
\unskip\
\newblock
\APACrefYearMonthDay{1997}{}{}.
\newblock
{\BBOQ}\APACrefatitle {Injection-induced earthquakes and crustal stress at 9 km
  depth at the {KTB} deep drilling site, {G}ermany} {Injection-induced
  earthquakes and crustal stress at 9 km depth at the {KTB} deep drilling site,
  {G}ermany}.{\BBCQ}
\newblock
\APACjournalVolNumPages{Journal of Geophysical Research: Solid
  Earth}{102}{B8}{18477-18491}.
\newblock

\newblock

\PrintBackRefs{\CurrentBib}

\end{thebibliography}

\end{document}